%% file: Draft_36_AER.tex
\documentclass[12pt]{article}

\usepackage{natbib}
\usepackage{breakcites}

\usepackage{setspace}
\usepackage{datetime}
\usepackage{marginnote}
\usepackage{enumitem}
\usepackage{rotating}
\usepackage{fancyvrb}
\usepackage{pdfpages}
\usepackage{float}
\usepackage{url}
\usepackage{indentfirst}
\usepackage{xcolor}
\usepackage{letltxmacro}
\usepackage{xparse}
\usepackage{needspace}
\usepackage[multiple]{footmisc}
\usepackage{afterpage}

\dimen\footins=.7\textheight        

\usepackage{amsmath,amsfonts,amssymb,amsthm}
\allowdisplaybreaks

\usepackage{graphicx}
\usepackage{epstopdf}
\DeclareGraphicsExtensions{.pdf,.jpg}

\usepackage[caption=false]{subfig}
\usepackage[font=footnotesize,labelfont=bf,labelsep=space]{caption}

\usepackage[bb=libus]{mathalpha}
\usepackage{wasysym}
\usepackage{nicefrac}
\usepackage{faktor}
\usepackage{accents}

\usepackage{tikz}
\usetikzlibrary{arrows,petri,topaths}
\usepackage{tkz-graph}
\usepackage{tkz-berge}

\usepackage{stackengine}
\usepackage{verbatim}
\usepackage{array}
\usepackage{placeins}

\usepackage{longtable}
\usepackage{booktabs}

\usepackage{etoolbox}
\usepackage{chngcntr}
\usepackage{apptools}

\makeatletter
\newenvironment{canonicalgame}{%
  \par\addvspace{\thm@preskip}%
  \trivlist\item[]\itshape
}{%
  \endtrivlist\addvspace{\thm@postskip}%
}
\makeatother

\newcommand{\tabsec}[1]{%
  \addlinespace[6pt]%
  \multicolumn{2}{@{}l@{}}{\textit{#1}}\\*[2pt]%
}

\IfFileExists{versionPO.sty}{%
  \usepackage{versionPO}%
  \excludeversion{notes} 
  \includeversion{links} 
}{%
  \newif\ifnoteson\notesonfalse
  \newif\iflinkson\linksontrue
  \newcommand{\ifnotes}[2]{\ifnoteson ##1\else ##2\fi}
  \newcommand{\iflinks}[2]{\iflinkson ##1\else ##2\fi}

  \newcommand{\includeversion}[1]{}
  \newcommand{\excludeversion}[1]{}
}

\ifnotes{%
  \usepackage[paperwidth=10in,top=1in,bottom=1in,left=1in,right=2.5in]{geometry}%
  \usepackage[textwidth=1.4in,shadow,colorinlistoftodos]{todonotes}%
}{%
  \usepackage[paperwidth=9.75in,top=1in,bottom=1in,left=1.25in,right=1.25in]{geometry}%
  \usepackage[disable]{todonotes}%
}

\makeatletter\let\chapter\@undefined\makeatother 

\usepackage[hidelinks]{hyperref}
\iflinks{}{\hypersetup{draft=true}}

\newdateformat{mydate}{\monthname[\THEMONTH] \THEYEAR}

\setlist[itemize,2]{label=\(\circ\)}
\setlist[description]{%
  topsep=5pt,
  itemsep=5pt,
  font={\slshape\normalfont},
}

\theoremstyle{plain}
\newtheorem{theorem}{Theorem}[section]

\newtheorem{conjecture}[theorem]{Conjecture}
\newtheorem{hypothesis}[theorem]{Hypothesis}
\newtheorem{question}[theorem]{Question}
\newtheorem{corollary}[theorem]{Corollary}

\newtheorem{definition}{Definition}
\newtheorem{example}[theorem]{Example}

\newtheorem{lemma}{Lemma}[section]

\newtheorem{proposition}[theorem]{Proposition}
\newtheorem{remark}[theorem]{Remark}

\newtheorem{assumption}{Assumption}

\AtAppendix{\counterwithin{lemma}{section}}
\AtAppendix{\counterwithin{assumption}{section}}

\makeatletter
\def\overUnderArrow{\@ifnextchar[\overUnderArrow@i{\overUnderArrow@i[]}}
\def\overUnderArrow@i[#1]#2#3{
  \ifx\relax#1\relax
    \array[b]{c}\overset{\text{#2}}{\uparrow}\\#3\endarray
  \else\ifx\relax#2\relax
    \array[t]{c}#3\\\underset{\text{#1}}{\downarrow}\endarray
  \else
    \array{c}\overset{\text{#2}}{\uparrow}\\#3\\\underset{\text{#1}}{\downarrow}\endarray
  \fi\fi
}
\makeatother

\makeatletter
\newcommand\varitem[1]{\item[\textbf{A\arabic{enumi}\rlap{$#1$}.}]%
  \edef\@currentlabel{A\arabic{enumi}{$#1$}}}
\makeatother

\newcounter{footequation}
\renewcommand{\thefootequation}{F.\arabic{footequation}}

\makeatletter
\LetLtxMacro\latex@@footnote\footnote
\RenewDocumentCommand{\footnote}{om}{%
  \begingroup
  \let\c@equation\c@footequation
  \let\theequation\thefootequation
  \IfValueTF{#1}{%
    \latex@@footnote[#1]{#2}%
  }{%
    \latex@@footnote{#2}%
  }%
  \endgroup
}
\makeatother

\makeatletter
\patchcmd{\env@cases}{1.2}{0.72}{}{}
\makeatother

\newcommand{\R}{{\mathbb{R}}}

\newcommand\TikCircle[1][0.6]{\tikz[baseline=-2.9pt]{\draw[blue, very thick](0,0)circle[radius=#1mm];}}
\newcommand\TikDot[1][0.6]{\tikz[baseline=-2.9pt]{\draw[red,fill=red](0,0)circle[radius=#1mm];}}

\newcommand{\Cov}{{\operatorname{\mathbb{Cov}}}}

\DeclareMathOperator*{\argmax}{argmax}

\mathchardef\mhyphen="2D

\begin{document}

\setlist{noitemsep} 
\onehalfspacing 


\title{Arrow-Debreu Meets Kyle: \\ Price Discovery Across Derivatives} 


\author{Christian Keller \and Michael C. Tseng\thanks{Christian Keller is from the Department of Mathematics, University of Central Florida.
Michael Tseng is from the Department of Economics, University of Central Florida.
We thank Roberto Burguet, Alain Chaboud, Zhenzhen Fan, Michael Gallmeyer, Sergei Glebkin, Vladimir Gatchev, Vijay Krishna, Melody Lo,
Neil Pearson,
Christo Pirinsky, and Marzena Rostek for valuable feedback and discussions on earlier drafts.
We also thank participants at the Northern Finance Association Meeting, the HEC-McGill Winter Finance Workshop, and 
the Econometric Society World Congress for comments and questions.
All remaining errors are our own. 
Corresponding author: Michael Tseng. Email: michael.tseng@ucf.edu. Address: 
Department of Economics,
College of Business,
University of Central Florida,
12744 Pegasus Dr., Orlando, Florida 32816-1400.
}
}



\date{
\mydate\today
}

\renewcommand{\thefootnote}{\fnsymbol{footnote}}

\singlespacing

\maketitle

\vspace{0.5cm}

\noindent \rule{6.5in}{1pt}

{\flushleft \bf Abstract}

We study price discovery in a model where an informed agent has arbitrary private information about future state probabilities and trades state-contingent claims. 
The model unifies key elements of \cite{arrow1954existence} and \cite{kyle1985continuous}.
When the claims are options, the informed agent has arbitrary information about the underlying asset's payoff distribution and trades option portfolios. 
Our setting provides the first equilibrium framework that encompasses long-standing option-market practices and regularities, including common trading strategies and the volatility smile across strikes.

\vspace{0.5cm}

\noindent 

\medskip
\vspace{0.5cm}


\thispagestyle{empty}

\clearpage

\onehalfspacing
\setcounter{footnote}{0}
\renewcommand{\thefootnote}{\arabic{footnote}}
\setcounter{page}{1}

\section{Introduction}

A fundamental economic function of prices is to aggregate and convey information. In its most basic form, this informational role is price discovery about the mean---when informed agents trade to exploit private knowledge 
about an asset’s expected payoff, that knowledge is impounded into its price.  
But private information is rarely confined to the mean: it often concerns broader aspects of the payoff distribution (e.g., volatility, skewness, tail risk), and derivatives are the natural instruments for exploiting such information.
Indeed, since \cite{ross1976options}, it has been understood that derivatives allow traders to target particular aspects of the payoff distribution, and these trades are routine in practice (see \cite{hull2003options}).
The price discovery implication is immediate---derivative prices must impound broad information about the payoff distribution beyond the mean.

Yet the literature lacks a basic model that articulates how derivative markets incorporate diverse information signals into prices.
This theoretical gap reflects the modeling challenges posed by nonlinear derivative payoffs, the interdependence across their trades and prices, and the resulting complex cross-market information dynamics.
We address this gap by unifying two cornerstone frameworks: \cite{arrow1954existence}, the foundation of contingent-claim pricing, and \cite{kyle1985continuous}, a canonical model of informed trading.
Our unified model yields equilibrium characterizations of informed demand, cross-market price impact, and the joint informativeness of the price cross-section. 

In our model, an informed trader has private information about the probabilities of future states and trades state-contingent claims.
Equivalently, he has information about the payoffs of Arrow-Debreu (AD) securities and trades in the AD markets. 
We place no restrictions on the probability distribution across states or the form of private information.

Since any derivative payoff---such as options---can be replicated by a portfolio of AD claims, our general AD formulation immediately yields derivative-market specializations that capture the structural linkages across particular instruments. 
In the options specialization, the informed trader has arbitrary private information about the underlying asset’s payoff distribution and trades options. 
Our results can then be read directly as characterizing how option markets aggregate that information. 
While the same translation applies to other derivative classes, we focus on options as the most empirically salient application, after establishing the core mechanisms in the AD formulation.

When trading is restricted to the underlying asset only, our model reduces to the single-asset price discovery benchmark of \cite{kyle1985continuous}. 
There, the informed trader’s private signal is the asset’s expected payoff, and equilibrium pricing is linear. 
The slope of the inverse supply curve, Kyle’s lambda, is the price impact: the marginal price response to demand as trades reveal information about the asset’s expected payoff. 
Kyle’s lambda increases with the private signal’s intensity (its prior variation): higher information intensity makes order flow more informative, so price adjusts more sharply.

In our more general AD setting, by contrast, private information concerns the entire state distribution.
Each private signal is thus a payoff-distribution ``story'' that can describe any scenario, far beyond the mean.
In this environment, trades are jointly informative---a trade in one security can move prices across markets as it reveals information about the payoffs of other securities.
We show that equilibrium cross-price impact is proportional to payoff covariance across signals, generalizing Kyle’s lambda to an (infinite-dimensional) payoff-covariance matrix that governs cross-market price impact. 
The intuition is immediate: cross-price impact is high precisely when two securities have similar payoff profiles across information stories---so demand for one is informative about the other.

Viewed through the options lens, the cross-impact characterization provides a structural framework for a basic empirical question: what information, precisely, do option trades and prices contain? 
A large empirical literature documents that option prices and order flow embed information about the underlying payoff distribution, but absent an equilibrium account of how information propagates across option markets, 
the mapping from observables to latent information dimensions necessarily depends on the proxies used.
We characterize cross-option information flow by mapping cross-price impact to general features of the underlying payoff distribution---volatility, skewness, tail risk, etc. 
In doing so, we consolidate prior empirical findings and generate novel, testable predictions.

While cross-price impact is the local (marginal) footprint of learning from order flow, the corresponding global object is the joint system of supply schedules for all AD securities---an asymmetric-information pricing kernel that 
prices all contingent-claims, and hence derivatives.\footnote{This pricing kernel is a high-dimensional, nonlinear object. It maps cross-market order-flow to the cross-section 
of state prices; its marginal quantity---the price impact kernel---is already an infinite-dimensional linear operator.}
The extensive derivative-pricing literature since \cite{black1973pricing} has largely focused on hedging and replication under complete information. 
Yet hedging is clearly not the only motive for trading derivatives. We show how informed trading explains observed price behavior that hedging alone does not account for. 
In particular, the asymmetric-information pricing kernel endogenously generates the volatility smile across option strikes, a well-documented empirical regularity of option-implied volatilities.
By contrast, reduced-form frameworks descended from Black-Scholes can only fit the smile by \textit{ex post} calibration, an exercise that offers no structural explanation---it treats a market outcome implied by option prices as an exogenous input to be fitted.\footnote{
A large financial econometrics literature fits the volatility smile/surface by modifying components of Black-Scholes-type models.
Prominent approaches include the jump-diffusion models (\cite{merton1976option}, \cite{bates1991crash}), 
stochastic volatility models (\cite{heston1993closed}, \cite{bates1996jumps}, \cite{duffie2000transform}, \cite{britten2000option}, \cite{bates2000post}, \cite{ait2007maximum}, \cite{christoffersen2010volatility}, \cite{ait2021implied}),
and local volatility models (\cite{dupire1994pricing}, \cite{berestycki2002asymptotics}, and \cite{carr2012explicit}).}

Our model also yields a simple, practical expression of equilibrium informed demand. 
The informed trader trades on his private signal via a long-short contingent-claim portfolio: he goes long the signal-implied payoff distribution and shorts those implied by alternative signals.
In the options specialization, this translates to concrete multi-leg trades that target the chosen aspects of the underlying payoff distribution; for example, volatility information induces the \emph{straddle}, a workhorse volatility strategy. 
More broadly, equilibrium informed demand rationalizes familiar, long-standing option strategies---going beyond the existing (mostly single-contract) models that are silent on observed multi-option trading practices. 
This equilibrium portfolio characterization provides a constructive guide for designing and empirically interpreting option portfolios that load on particular informational dimensions.

Finally, we ask the natural question: in a complete market, how much information is impounded into the cross-section of prices?
We show that aggregate price informativeness is summarized by a single scalar measure.
Notably, this measure is invariant to the possible payoff distributions and to noise trading intensities across markets.
In other words, while contingent claim prices can reflect a wide range of payoff-distribution ``stories,'' aggregate price informativeness does not depend on the specific details of those stories. 
This invariance parallels textbook complete-markets risk sharing: just as only aggregate risk matters, only the \emph{aggregate dimension of uncertainty}—how many distinct stories the market must distinguish, not their details—matters for the informativeness of the pricing kernel.

\paragraph{Related Literature}

\cite{kyle1985continuous} is the seminal framework for single-asset price discovery.
Subsequent work extends this framework to settings with multiple insiders and information dispersed across securities---e.g., \cite{holden1992longlived,caballe1994imperfect,foster1996strategic,back2000imperfect}.
All of these models impose a linear-Gaussian structure---joint normality of payoffs and signals---leaving price discovery across claims with nonlinear payoffs beyond their scope.
\cite{rochet1994insider} relaxes normality but remains in a single-asset setting.
We carry the Kyle price-discovery logic to Arrow-Debreu markets.
The Arrow-Debreu pricing kernel is generically infinite-dimensional---even with finitely many primitive assets---whereas the linear-Gaussian setting is finite-dimensional by construction. 
Accordingly, our departure from that setting calls for both new conceptual insights and methodological tools.

A large empirical literature documents that option prices and order flow embed broad information about the underlying payoff distribution.
Regarding the mean, signed option order imbalance and relative call-put pricing predict subsequent stock returns (\cite{pan2006information,CremersWeinbaum2010}), and options contribute materially to price discovery alongside the underlying (\cite{chakravarty2004informed,Muravyev2016}).
Beyond the mean, demand pressure shifts the implied-volatility surface (\cite{BollenWhaley2004}), option-implied moments forecast realized volatility and tail risk (\cite{Jiang2005,BollerslevTodorov2011}), and straddle-based evidence 
points to meaningful volatility mispricing (\cite{ni2008volatility,goyal2009cross}).
More broadly, option order flow is linked to information about firm fundamentals and downside risk (\cite{cao2005informational,Roll2009,augustin2019informed}).

Despite this robust evidence of joint price formation across option contracts, existing informed-trading models (e.g., \cite{back1993asymmetric,biais1994insider,brennan1996information,Easley1998,collin2021informed}) remain confined to single-option settings 
and therefore cannot speak to the cross-contract mechanisms at the core of these empirical findings---cross-strike dynamics, implied-volatility-surface formation, and multi-option trades such as the straddle.
Our model yields cross-option equilibrium restrictions that consolidate and sharpen this body of evidence while generating additional testable predictions.
For instance, we identify cross-strike volume imbalance as the equilibrium channel behind the return predictability of signed option volume documented by \cite{pan2006information}, and provide the first equilibrium rationale for the volatility-forecasting power of straddle
demand observed in  \cite{ni2008volatility} and \cite{goyal2009cross}.

There have also been theoretical studies of informed trading across multiple securities in competitive (non-strategic) settings, such as \cite{admati1985noisy}, \cite{malamud2015noisy}, and \cite{chabakauri2022multi}.
Without strategic order flow, these models cannot generate the derivatives trading strategies used in practice.\footnote{This mirrors the familiar distinction between \cite{kyle1985continuous} and \cite{grossman1980impossibility} settings in the single-asset case.}
Similarly, the insight that maps payoff characteristics to observable cross-price impact and cross-strike regularities is possible only in a strategic setting.
Moreover, \cite{admati1985noisy} and \cite{chabakauri2022multi} impose restrictive parametric assumptions on the payoff distribution, while \cite{malamud2015noisy} relies on properties intrinsic to the continuum.
By contrast, we impose no restrictions on the payoff distribution, and our results do not depend on the discrete–continuous modeling choice. 

To foreground the economic intuition, we adopt the discrete formulation in this paper.
At the level of intuition, a continuous setup---such as the Gaussian prior-payoff environment of \cite{kyle1985continuous}---is well approximated by a discrete signal-state grid.
A rigorous treatment of the continuous formulation is provided in the companion paper \cite{kellertsengmath}, which addresses the substantive technical considerations it entails.

Empirical work consistently finds that access to derivatives markets reduces information asymmetry and improves price efficiency---evidence spans futures (e.g., \cite{GarbadeSilber1983,Chan1992}), credit derivatives (e.g., \cite{BlancoBrennanMarsh2005,LongstaffMithalNeis2005,AcharyaJohnson2007}), 
and options (e.g., \cite{Hu2018}).
Theoretical foundations for this efficiency channel, however, remain limited.
Our model isolates a simple and intuitive mechanism: once derivatives are tradable, an informed trader is no longer restricted to generally suboptimal linear demand and can instead express his information through targeted state-contingent positions.
In other words, we show that the linear equilibrium in \cite{kyle1985continuous} is not robust to the introduction of derivatives.\footnote{See Example~\ref{example: Kyle with options}.}
Derivatives thus expand the information that prices can aggregate beyond what the underlying asset alone can support.

The remainder of the paper is organized as follows.
Section~\ref{sec: main intuition} conveys the core intuition of our results in a simplified setting. 
Section~\ref{sec: model outline} presents the general model.
Section~\ref{sec: Market Maker's Pricing Kernel} derives the market maker’s sufficient statistic and pricing kernel.
Section~\ref{sec: Informed Trader's Problem} characterizes the informed trader's cross-market price impact.
Section~\ref{sec: Equilibrium, Finite $S$} then establishes equilibrium via a canonical reduction of the general trading game. 
Section~\ref{sec: discussion} analyzes price discovery across contingent claims---the informed demand, price impact, and information efficiency of prices.
Section~\ref{sec: discussion subsection} discusses empirical implications.
Section~\ref{sec: conclusion} concludes.  
Technical assumptions and proofs are provided in the Appendix.

The reduction of Section~\ref{sec: Equilibrium, Finite $S$} yields a key insight that, despite full nonparametric Arrow--Debreu generality, equilibrium is pinned down by a single endogenous constant.
Readers who prefer a high-level overview without the formal developments can find the intuition distilled in Section~\ref{sec: e.g. binary signal}, which links directly back to the simplified environment of Section~\ref{sec: main intuition}.

\begin{figure}[htbp!]
\centering 
\subfloat[$x$-axis: states, $y$-axis: probability]{
\includegraphics[width = 7cm, height = 3cm, trim=.41cm .42cm .39cm .39cm, clip]{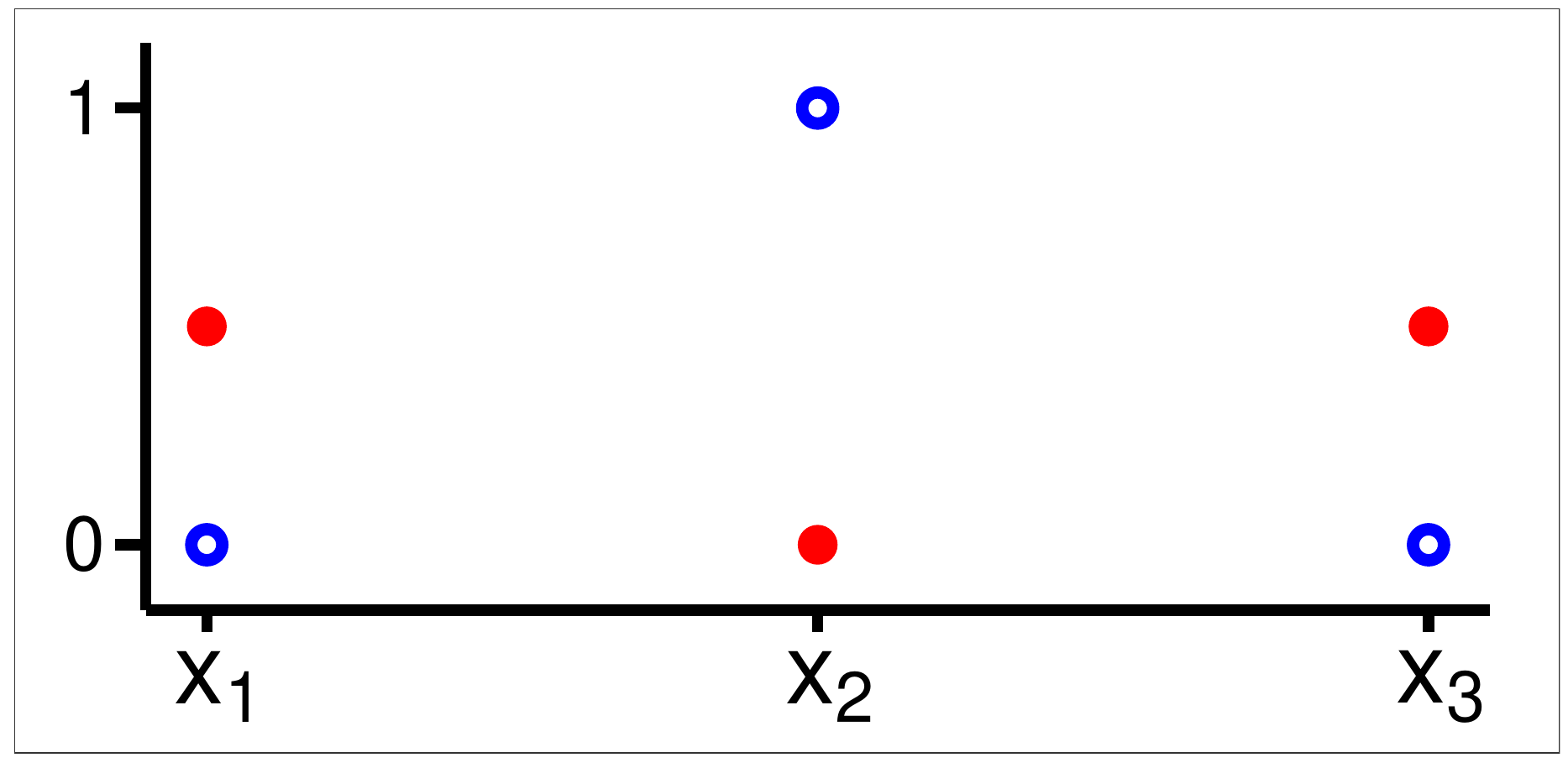} \label{fig: toy model payoff}
}
\hfil
\subfloat[$x$-axis: insider portfolio size $\alpha$, $y$-axis: expected profit]{
\includegraphics[width = 7cm, height = 3cm]{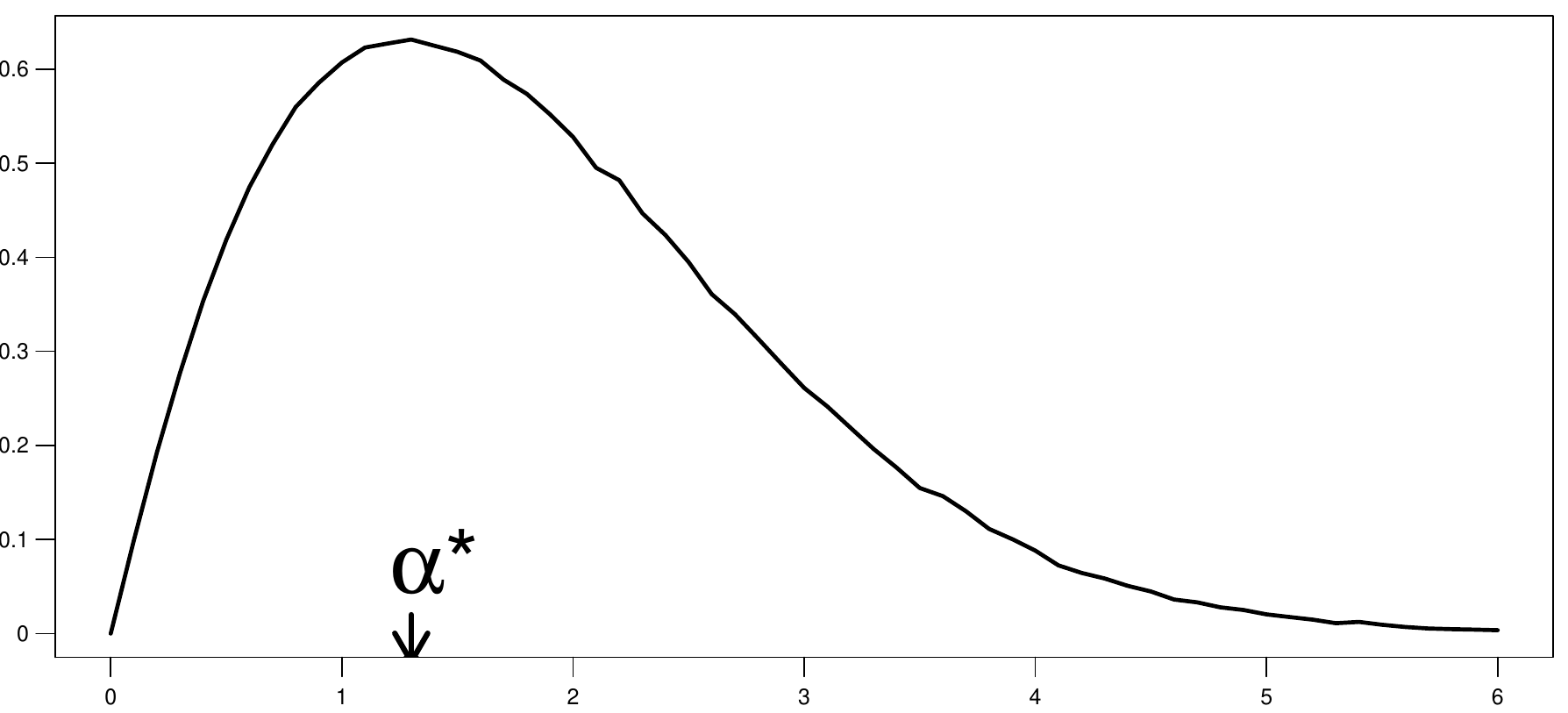} \label{fig: toy model insider profit}
}

\caption{
{\footnotesize
{\bf Simplified Setting}\\
(a) {\bf Conditional State Probabilities.}  Solid dots $\TikDot$ represent signal $s_1$ (high volatility), and hollow dots $\TikCircle$ represent signal $s_2$ (low volatility).\\
(b) {\bf Insider Expected Profit Curve.} The insider's optimal portfolio size {\bf $\alpha^*$} is the endogenous constant characterizing equilibrium.
}
}
\label{fig: toy model}
\end{figure}

\section{Basic Intuition}
\label{sec: main intuition}

We start with a simplified setting that encapsulates the core intuition. Although deliberately stripped-down, it maps directly onto canonical ideas in asset pricing and price formation under asymmetric information. 
The informed trader observes a full state-probability vector, making AD securities the natural primitive claims---their prices are state prices.
This toy setting provides a first look at informed trading across AD markets.
In the general model of Section~\ref{sec: model outline}, we will develop the robustness of the intuition illustrated here---with no restrictions on the number of securities, payoff distributions, or the nature of private information.

There are two risk-neutral agents: the insider and the market maker.
At $t=0$, the agents know that there are three possible $t=1$ states, denoted $x_i$ for $i = 1, 2, 3$, and trade the corresponding AD securities. 

At $t=0$, the insider privately observes a signal $s\in\{s_1,s_2\}$. Conditional on $s_1$, the distribution over $t=1$ states is $\big(\tfrac12,0,\tfrac12\big)$; conditional on $s_2$, it is $(0,1,0)$. 
See Figure~\ref{fig: toy model payoff}. The market maker's prior $\pi_0$ is uniform over signals.

After observing his private signal, the insider submits his demand (market orders) for AD securities to maximize his expected utility at $t=1$.
Noise traders trade for exogenous reasons such as liquidity needs; their trades across the AD markets are $\varepsilon_i \!\! \stackrel{i.i.d.}{\sim} \!\! \mathcal{N}(0,1)$ for $i = 1, 2, 3$.
The market maker acts as a competitive liquidity provider.

The market maker receives the combined order flow $\omega = (\omega_1, \omega_2, \omega_3)$ of the insider and noise traders across markets, updates his prior, and executes the orders at his zero-profit prices $P_i$, $i = 1, 2, 3$.\footnote{The numeraire is the consumption good.} 
If the market maker's posterior on signals is $( \pi_1(s_1 \vert \omega), \pi_1(s_2 \vert \omega) )$, the AD prices are then his posterior means of the security payoffs, 
$$
(P_1, P_2, P_3) = (\frac12 \pi_1(s_1 \vert \omega), \,  \pi_1(s_2 \vert \omega), \, \frac12 \pi_1(s_1 \vert \omega)).
$$

An intuitive trading strategy for the insider, conditional on observing $s_1$, is to buy securities $x_1$ and $x_3$  (where payoffs are positive) and sell security $x_2$ (where payoff is zero) for 
$\alpha$ shares each.\footnote{There are no leverage or short-selling constraints.}
Then, conditional on the insider observing $s_1$, the order flow across markets received by the market maker is 
$$
\omega = (\omega_1, \omega_2, \omega_3), \; \mbox{ where } \; \omega_1 = \alpha + \varepsilon_1, \; \omega_2 = - \alpha + \varepsilon_2, \; \omega_3 = \alpha + \varepsilon_3.
$$ 
Similarly, conditional on $s_2$, the insider buys security $x_2$ and sells securities $x_1$ and $x_3$  for $\alpha$ shares each.

Given the insider's trading strategy, the market maker's posterior likelihood ratio,
conditional on order flow $\omega$, is 
\begin{equation}
\label{eqn: posterior likelihood ratio in toy model}
\frac{\pi_1(s_1 \!\mid\! \omega)}{\pi_1(s_2 \!\mid\! \omega)}
= \frac{\exp\!\left(\alpha(\omega_1-\omega_2+\omega_3)-\tfrac{3}{2}\alpha^2\right)}
        {\exp\!\left(-\alpha(\omega_1-\omega_2+\omega_3)-\tfrac{3}{2}\alpha^2\right)}
= \exp\!\left(2\alpha\Delta(\omega)\right),
\end{equation}
where 
$\Delta(\omega) = \omega_1+\omega_3-\omega_2$ is the \emph{cross-market order imbalance}.
Therefore, his zero-profit AD prices are
\begin{equation}
\label{eqn: AD price in toy model}
(P_1,P_2,P_3)
=
\left(
\frac{\tfrac12 \exp(2\alpha \Delta)}{\exp(2\alpha \Delta)+1},
\,
\frac{1}{\exp(2\alpha \Delta)+1},
\,
\frac{\tfrac12 \exp(2\alpha \Delta)}{\exp(2\alpha \Delta)+1}
\right),
\;
\text{with }\Delta=\Delta(\omega).
\end{equation}

\paragraph{Market Maker's Sufficient Statistic}
Equation~\eqref{eqn: AD price in toy model} shows that the market maker does not learn from $(\omega_1,\omega_2,\omega_3)$ separately.
Instead, his sufficient statistic is the cross-market order imbalance $\Delta(\omega)$ for securities $x_1$ and $x_3$ relative to $x_2$.
When $\Delta(\omega)$ is high, order flow resembles the insider's long--short trades under $s_1$, so the market maker tilts 
beliefs toward $s_1$, raising $P_1$ and $P_3$ and lowering $P_2$ (and conversely when $\Delta(\omega)$ is low). This joint inference generates \emph{cross-price impact}: trades in one security move prices of other securities.

\paragraph{Cross-Price Impact}
Cross-price impact is positive when two securities are informational complements: learning that one security has a high payoff raises the market maker's belief that the other security also pays.
Securities $x_1$ and $x_3$ are complements: their payoff covariance is $\tfrac18$, so their cross-price impact is positive.
By contrast, the other two pairs have payoff covariance $-\tfrac18$, resulting in negative cross-price impact.
Payoff covariances determine the extent to which order flow for one security reveals information about others, and thus determine price impact across markets.
In the sections that follow, we bring out this intuition under the general model.

\paragraph{Insider Portfolio Choice}
The insider, conditional on, say, $s_1$, scales his buy-sell (long-short) portfolio by $\alpha$ to maximize expected profit 
$$
\max_{\alpha \geq 0} \, \mathbb{E}[\alpha \, ( \frac12 + \frac12 - P_1 + P_2 - P_3 )]. 
$$
Figure~\ref{fig: toy model insider profit} illustrates the insider's expected profit as a function of  $\alpha$.
If $\alpha \! = \! 0$, the insider's zero demand results in zero profit, and the AD prices are $(\frac14, \frac12, \frac14)$; since order flow contains no information, the market maker sets the prices based on his prior.
If $\alpha \! \rightarrow \! \infty$, the market maker's posterior probability of $s_1$ approaches one, which causes the AD prices to converge to the true distribution $(\frac12, 0, \frac12)$, driving the insider’s profit to zero due to full information revelation.

In choosing the scale $\alpha$ of his long-short portfolio, the insider trades off the marginal gains from exploiting the true payoff distribution against the marginal trading cost. 
That cost has two components. First, there is the cost of transacting at prevailing prices, absent any price impact. 
Second, there is an informational component: larger trades make order flow more informative for the market maker, which steepens cross-market price impact and erodes the insider's information rent. 
The optimal trading scale $\alpha^*$ balances these margins (see Figure~\ref{fig: toy model insider profit}).  
Equilibrium is pinned down by this single endogenous constant $\alpha^*$. \emph{Nontrivially, the same one-parameter equilibrium structure carries over to the general model (see Section~\ref{sec: e.g. binary signal} for intuition)}.
Allowing correlated noise trades, varying noise intensity, and non-uniform priors has no material effect on the core mechanisms, either in this simplified setting or in the general model.%
{%
  \setlength{\skip\footins}{6pt plus 2pt minus 2pt}%
  \setlength{\footnotesep}{4pt}%
  \interfootnotelinepenalty=0%
  \long\def\ThreeParaFootnote{%
    {\bf Correlated Noise Trades}
    Suppose noise trades are correlated (holding their marginal variances fixed). This changes the {\em pattern} of uninformed flow across claims but not {\em marginal} price impact.
    This is because the market maker filters out the predictable noise correlation and prices only the residual surprise.
    For example, if noise tends to co-move {\em opposite} to payoff co-movement, the insider would hide in the offsetting flow and trade more aggressively.
    Anticipating this, the market maker shifts supply schedules but the marginal price responses are unchanged.
    This mirrors the single-asset case, where biased noise shifts the level of the supply schedule but not Kyle’s lambda.
    \label{fn: correlated noise}%

    \par\medskip
    {\bf Varying Noise Intensity}
    Suppose noise intensity differs across markets, with $\epsilon_i\sim\mathcal N(0,\sigma_i^2)$ for $i=1,2,3$.
    Then the market maker’s order-imbalance sufficient statistic becomes
    $\Delta(\omega)=\frac{\omega_1}{\sigma_1^2}+\frac{\omega_3}{\sigma_3^2}-\frac{\omega_2}{\sigma_2^2}$.
    This is the standard signal-to-noise intuition---flow in a noisier market (larger $\sigma_i$) receives less weight in inference.
    Therefore, trades in noisier contracts have lower price impact both within and across markets.
    \label{fn: varying noise intensity}%

    \par\medskip
    {\bf General Prior}
    With a general prior $(\pi_0(s_1),\pi_0(s_2))$, the posterior likelihood ratio in \eqref{eqn: posterior likelihood ratio in toy model} is scaled by the prior odds:
    {\setlength{\abovedisplayskip}{3pt}\setlength{\belowdisplayskip}{3pt}%
    \[
      \frac{\pi_1(s_1\mid \omega)}{\pi_1(s_2\mid \omega)}
      =\exp\!\big(2\alpha\,\Delta(\omega)\big)\cdot \frac{\pi_0(s_1)}{\pi_0(s_2)},
      \qquad \Delta(\omega) = \omega_1+\omega_3-\omega_2.
    \]
    }%
    A higher $\pi_0(s_1)$ means the same order imbalance produces a larger posterior tilt toward the $s_1$ story.
    In the straddle example, if the market is already biased toward high volatility, co-buying the straddle put--call pair is more readily read as volatility demand, so the put and call are repriced more in tandem (higher put--call cross-impact).
    \label{fn: general prior}%
  }%
  \footnote{\ThreeParaFootnote}%
}%

\paragraph{Information Role of Derivatives}
If fewer than all three AD securities are available for trade, both the insider's profit and the information efficiency of prices decline. 
Here, information efficiency is naturally measured by the posterior probability weight that the market maker assigns to the true signal. 
Derivative markets allow informed traders' orders to better express (and therefore reveal) their information, thereby facilitating its incorporation into prices. 
Therefore, derivatives do more than just add leverage: they change what information can be traded and how that information propagates across markets.
This result provides an equilibrium foundation for a long-held intuition that derivatives concentrate informed trading in directions that are difficult to trade in the underlying alone, 
and it is consistent with empirical evidence that option markets contain incremental information about underlying returns and higher moments. 

\paragraph{Example: Straddle}
A key implication is not merely that derivative trades reflect information, but that information is transmitted through \emph{cross-market} price impact.
We demonstrate this mechanism in the case of volatility. 
Interpret the three states $x_1<x_2<x_3$ as the possible $t=1$ realizations of an underlying asset price, and normalize so that the $t=0$ price is $x_2$.
Then $s_1$ is the ``high volatility'' signal and $s_2$ the ``low volatility'' signal.
The signal $s_1$ reveals that the asset price will move, but without a specified direction, whereas $s_2$ reveals that there will be no price movement.
Security $x_1$ is a put-like claim (it pays when the asset price falls), while $x_3$ is a call-like claim (it pays when the asset price rises).
If the insider observes $s_1$, he buys the put-call pair $(x_1,x_3)$---a ``straddle''---to capitalize on price movement regardless of direction.
In real option markets, the straddle is a standard volatility strategy (see \cite{hull2003options}) and is widely discussed in the empirical literature.\footnote{See \cite{coval2001expected,pan2006information,ni2008volatility,driessen2009price,goyal2009cross}.}
This strategy, along with other common option strategies, will be derived as an equilibrium portfolio in the general model.

Under the high volatility signal, the insider therefore loads on both put and call, so buy pressure in either leg is interpreted as evidence favoring high volatility, which pushes up the other leg as well.
Consequently, the cross-price impact between the put-call pair is informative about future volatility.
In the general model, we systematically extend this insight to actual options: cross-option price impact reveals information about the underlying payoff distribution, yielding empirical 
predictions for how option prices and implied-volatilities adjust as price discovery unfolds (see Section~\ref{sec: discussion subsection}).

\section{General Model}
\label{sec: model outline}

As in Section~\ref{sec: main intuition}, there are two risk-neutral agents, the insider and the market maker.
At $t=0$, the insider observes a signal that reveals the probability distribution over the set of possible $t=1$ states, $X = \{x_1, x_2, \cdots \}$ (countably infinite, with finite $X$ included as a special case).
The market maker has a Bayesian prior over the possible signals. At $t=0$, there is a complete market of AD securities for $t=1$ states.\footnote{The model can be adapted to the incomplete market case by partitioning the state space. 
In this paper, we focus on the complete market case.}
There is a risk-free asset in perfectly elastic supply at a risk-free rate of zero.

After observing his private signal at $t=0$, the insider submits his demand for AD securities to maximize his expected utility at $t=1$.
The market maker then receives the combined order flow of the insider and noise traders across AD markets and executes the orders at his zero-profit prices.

The set of possible signals is $S = \{s_1, \cdots, s_K\}$, with prior probabilities $\pi_0(s_k)$, $1 \leq k \leq K$.       
Conditional on a signal $s_k$, the probabilities of $t=1$ states are given by $\eta(\,\cdot\,  \vert s_k ) \colon X \rightarrow [0, \infty)$.
After observing $s_k$, the insider 
chooses a portfolio $W(\,\cdot\, \vert s_k) \colon X \rightarrow \mathbb{R}$ where $W(x_i \vert s_k)$ is his order for security $x_i$ conditional on $s_k$.

Noise trader orders $\varepsilon_i  \stackrel{d}{\sim} \mathcal{N}(0,\sigma_i^2)$, $i \geq 1$, are normally distributed with mean zero in each market and uncorrelated across markets. 
When the insider chooses portfolio $W \colon X \rightarrow \mathbb{R}$, the order flow received by the market maker is a stochastic sequence $\omega = (\omega_i)$ where  
$$
\omega_i = W(x_i) + \varepsilon_i \quad \mbox{for each market $i$.}
$$

The market maker holds a belief $\widetilde{W}(\,\cdot\, \vert \,\cdot\, ) \colon X \times S \rightarrow \mathbb{R}$ about the insider's trading strategy.
Based on this belief and the received order flow $\omega$, the market maker updates his prior $\pi_0$ to the posterior $\pi_1(s_k \vert \omega \kern 0.045em ; \kern 0.045em \widetilde{W})$, $1 \leq k \leq K$, over signals. 
His zero-profit price for each security $x_i$ is his posterior mean of its payoff $\eta(x_i \vert \, \cdot \,)$, 
\begin{equation}
\label{eqn: security x_i price}
\underbrace{P(x_i \vert \omega \kern 0.045em ;  \widetilde{W})}_\text{security $x_i$ price} = \sum_{k} \! \eta(x_i \vert s_k) \, \pi_1(s_k \vert \omega \kern 0.045em ; \kern 0.045em \widetilde{W})
\quad \mbox{for each market $i$.}
\end{equation}

Conditional on observing $s_k$ and given market maker belief $\widetilde{W}$, the insider's AD portfolio choice problem is
\begin{equation}
\label{eqn: informed trader's problem informal}
\max_{ W \colon X \rightarrow \mathbb{R}  } \mathbb{E}^{\mathbb{P}_{\scriptscriptstyle W}} \!\left[ \sum_{i} ( \eta(x_i \vert s_k) -  P(x_i \vert \omega \kern 0.045em ; \widetilde{W}) ) \cdot W(x_i) \right] \equiv  \max\limits_{ W \colon X \rightarrow \mathbb{R} } J(W \vert s_k ; \widetilde{W}). 
\end{equation}
Here, the expectation $\mathbb{E}^{\mathbb{P}_{\scriptscriptstyle W}}[\,\cdot\,]$ is taken over order flow $\omega$  
under its distribution $\mathbb{P}_{\!\scriptscriptstyle W}$ induced by $W$.
The functional $J(\, \cdot \, \vert s_k ; \widetilde{W})$, as defined in \eqref{eqn: informed trader's problem informal}, is the insider's expected utility functional conditional on $s_k$ and given market maker belief $\widetilde{W}$.

\begin{remark}
We place no restrictions on $\eta(\,\cdot\, \vert \,\cdot\, )$, i.e., no restrictions on the probability distributions across states or the nature of private information.
\end{remark}

In equilibrium, the insider’s optimal trading strategy given the market maker's pricing rule $P(\,\cdot\, \vert \,\cdot\, \kern 0.045em ; \kern 0.045em W^*)$ based on the latter's 
belief $W^*$ coincides with $W^*$.
That is, conditional on each $s_k$, the insider's optimal portfolio is $W^*(\,\cdot\, \vert s_k)$, thus confirming the market maker's belief.

\begin{definition}
\label{def: equilibrium}
A (Perfect Bayesian) \textbf{equilibrium} in our model is a trading strategy $W^*(\,\cdot\, \vert \,\cdot\,) \colon X \times S \rightarrow \mathbb{R}$ such that, conditional on each signal $s_k$,
$$
W^*(\,\cdot\,\vert s_k) \in \argmax\limits_{ W \colon X \rightarrow \mathbb{R} }  J(W \vert s_k ; W^*).
$$
\end{definition}

\subsection*{Single-Asset Special Case (\cite{kyle1985continuous})}
\label{sec: single asset case}

Reducing $X$ to a singleton yields the single-asset special case of our model.
In this case, the insider's information $\eta(\,\cdot\,\vert s) : X \rightarrow [0, \infty)$ can be collapsed to just the signal $s$ that parameterizes the asset's expected payoff at $t=1$. 
This is the (static) \cite{kyle1985continuous} setting.

The insider’s strategy reduces to $W \colon S \rightarrow \mathbb{R}$, where he submits a market order $W(s)$ for the asset after observing $s \in S$. 
When the insider submits order $W$, the market maker receives the combined order $\omega  =  W  +  \varepsilon$, where $\varepsilon$ is the noise trader order.

The market maker has a belief $\widetilde{W} \colon S \rightarrow \mathbb{R}$ about the insider's strategy and, based on this belief, 
forms his posterior regarding the asset value conditional on $\omega$. 
His zero-profit price $P(\omega \kern 0.045em ; \widetilde{W})$ for the asset is his posterior mean. 
The insider’s problem, conditional on observing $s$ and given market maker belief $\widetilde{W}$, is
$$
\max_{ W \in \mathbb{R} } \, \mathbb{E}^{\mathbb{P}_{\scriptscriptstyle W}} [ ( s - P(\omega \kern 0.045em ; \widetilde{W}) )\cdot W ]
$$
where the expectation $\mathbb{E}^{ \mathbb{P}_{\scriptscriptstyle W}  }[\,\cdot\,]$ is taken over order flow $\omega$ with respect to its distribution $\mathbb{P}_{\!\scriptscriptstyle W}$ induced by his order $W$.

An equilibrium is a trading strategy $W^*(\,\cdot\,)$ such that, conditional on each possible asset value $s$,
\begin{equation*}
\label{eqn: static Kyle equil}
W^*(s) \in \argmax_{ W \in \mathbb{R} } \, \mathbb{E}^{\mathbb{P}_{\scriptscriptstyle W}} [ ( s - P(\omega \kern 0.045em ; W^*) ) \cdot W ].
\end{equation*} 
This is Definition~1 of \cite{kyle1985continuous} and a special case of our Definition~\ref{def: equilibrium} when 
there are no derivative markets.\footnote{In Definition~1 of \cite{kyle1985continuous}, his Equation~(2.2)—$\tilde{p}(X, P) = E\{\tilde{v} | \tilde{x} + \tilde{u} \}$ where $\tilde{x} = X(\tilde{v})$—expresses the same equilibrium fixed-point property in his notation: 
the insider’s strategy $X(\,\cdot\,)$ confirms the belief that underlies the market maker's pricing rule $P(\,\cdot\,)$.}

Assume the asset value follows a normal prior distribution $\pi_0 \! \stackrel{d}{\sim} \! \mathcal{N}(v_0, \sigma_v^2)$ with prior mean $v_0$ and variance $\sigma_v^2$, and that noise trades follow $\varepsilon \! \stackrel{d}{\sim} \! \mathcal{N}(0, \sigma_{\varepsilon}^2)$.
(As noted in the introduction, our results are fully robust to whether the signal and state spaces are discrete or continuous---intuitively, this joint normal setup can be approximated by a discrete grid.)
Under these assumptions, there exists a linear equilibrium  
\begin{equation}
\label{eqn: static Kyle solution pre}
W^*(s) = \beta (s - v_0) \, \mbox{ and } \, P(\omega \kern 0.045em ; W^*) = v_0 + \lambda \omega
\end{equation}
where
\begin{equation}
\label{eqn: static Kyle solution}
\beta = \frac{\sigma_{\varepsilon}}{\sigma_v} \, \mbox{ and ({\em Kyle's lambda}) } \lambda = \frac{\sigma_v}{2 \sigma_{\varepsilon}}.
\end{equation} 
This is Theorem 1 of \cite{kyle1985continuous}.

In this single-asset special case, the price impact is the slope $\lambda$ of the market maker's inverse supply curve $P(Q) = v_0 + \lambda Q$, as specified by Equations~\eqref{eqn: static Kyle solution pre} and \eqref{eqn: static Kyle solution}.
Price impact is proportional to the (noise-adjusted) signal variation $\frac{\sigma_v}{\sigma_{\varepsilon}}$.
Higher signal variation leads to greater informed demand variation across signals, making order flow more informative and resulting in a higher price impact.
In other words, price impact is determined by the insider's \emph{information intensity}.
As suggested in Section~\ref{sec: main intuition}, this intuition generalizes across AD markets.

\subsection*{Options Formulation}

The AD formulation can be readily specialized to option markets.
Suppose the states $x_1<x_2<\cdots$ are the possible $t=1$ prices of an underlying asset.
Let $x_{i_0}$ denote the market maker's prior mean of the $t=1$ price at $t=0$.
Then for any state-contingent portfolio $W(\,\cdot\,)$ there exist holdings $(a_i)$ such that
\begin{align}
\label{eqn: Breeden-Litzenberger}
W(x_j)
&= a_{i_0}\,(x_j-x_{i_0})
  + \sum_{i<i_0} a_i\,(x_j-x_i)_-
  + \sum_{i>i_0} a_i\,(x_j-x_i)_+,
\qquad j=1,2,\ldots.
\end{align}
This is the standard (discrete) Breeden--Litzenberger formula (see \cite{back2010asset}, Exercise 3.5).
Thus, $W(\,\cdot\,)$ can be implemented by holding $a_{i_0}$ shares of the underlying, $a_i$ puts at strikes $x_i<x_{i_0}$, and $a_i$ calls at strikes $x_i>x_{i_0}$.
So the AD securities can be replaced by the underlying plus a menu of options.

In this options representation, the insider observes $s_k$ and hence the signal-implied payoff distribution $\eta(\,\cdot\,\mid s_k)$ for the underlying, and submits orders $(a_i)$ across the underlying and options.
In each market $i$, aggregate order flow is
\[
\omega_i = a_i + \varepsilon_i,
\quad  \mbox{ with noise trades } \varepsilon_i \stackrel{d}{\sim} \mathcal{N}(0,\sigma_i^2).
\]
The market maker prices contracts conditional on $\omega=(\omega_i)$, and the rest of the model proceeds exactly as in the AD formulation.
Conversely, all results derived in the AD formulation specialize immediately to the options setting.

\paragraph{Options in the \cite{kyle1985continuous} Framework}

The options formulation relates to the single-asset case simply as follows:

\begin{proposition}
\label{prop: options}
$\;$
\nopagebreak

(i) If signals shift the payoff distribution only through its mean, and trading is restricted to the underlying (no options), then our model collapses to the \cite{kyle1985continuous} setting.

(ii) If all signal-implied payoff distributions share the same mean, then the insider does not trade the underlying at all. In other words, exploiting information beyond the expected payoff requires options.
\end{proposition}

Proposition~\ref{prop: options}(i) is a knife-edge case---once options are available, the linear equilibrium~\eqref{eqn: static Kyle solution pre} in \cite{kyle1985continuous} is not robust.
Options let the insider concentrate on the strikes where his signal matters and avoid low-probability tail exposure, so equilibrium informed demand is inherently nonlinear.
We will illustrate this explicitly in Example~\ref{example: Kyle with options}.

\section{Cross-Market Inference}
\label{sec: Market Maker's Pricing Kernel}

The market maker’s inference problem is infinite-dimensional: as noisy order flow arrives across many claims, he must infer the informed portfolio behind it, which is unrestricted.
Making the underlying economic intuition precise therefore requires going beyond the textbook finite-dimensional Bayesian formalism.
Once we do, the takeaway is simple. His sufficient statistic is the degree to which order flow aligns with the insider-demand portfolio he expects under each signal---a direct generalization of the cross-market imbalance $\Delta(\omega)$ of Section~\ref{sec: main intuition}.

\subsection{The Posterior and Pricing Kernel}
\label{sec: Posterior and Pricing Kernel}
 
Formally, order flow is a (random) sequence $\omega=(\omega_i)_{i=1,2,\ldots}$ taking values in the space $\Omega$ of countable sequences. 
Under the market maker's belief $W(\,\cdot\,\vert\,\cdot\,)$ about how the insider trades, conditional on signal $s_k$ the observed order flow is
\begin{equation}
\label{eqn: dY s x}
\omega_i \,=\, W(x_i \vert s_k) + \varepsilon_i,
\qquad i=1,2,\ldots,
\end{equation}
where $\varepsilon_i \stackrel{d}{\sim} \mathcal{N}(0,\sigma_i^2)$ are noise trades.

The market maker treats $s_k$ as the latent parameter, with prior $\pi_0(s_k)$ for $k=1,\ldots,K$, and updates his belief based on the realized order flow $\omega$.
Let $\mathbb{P}_{\scriptscriptstyle W(\,\cdot\,\vert s_k)}$ denote the likelihood of $\omega$ under \eqref{eqn: dY s x} when the insider’s mean order is $W(\,\cdot\,\vert s_k)$, and let $\mathbb{P}_{\! \scriptscriptstyle 0}$ denote the noise-only likelihood (i.e., $W\equiv 0$).
Since the noise-only likelihood does not depend on signal, the market maker can summarize the likelihood of $\omega$ conditional on $s_k$ via the likelihood ratio\footnote{Formally, $\mathbb{P}_{\scriptscriptstyle W(\,\cdot\,\vert s_k)}$ and $\mathbb{P}_{\! \scriptscriptstyle 0}$ are probability measures on $\Omega$. The likelihood ratio~\eqref{eqn: Radon-Nikodym derivative} is their 
Radon-Nikodym derivative.}

\begin{equation}
\label{eqn: Radon-Nikodym derivative}
\frac{ \mathbb{P}_{\scriptscriptstyle W(\,\cdot\,\vert s_k)} (\omega) }{ \mathbb{P}_{\! \scriptscriptstyle 0} (\omega) }
\,=\,
\exp\!\left(
\sum_i \frac{ W(x_i \vert s_k)\,\omega_i}{\sigma_i^2}
-\frac12 \sum_i \frac{ W(x_i \vert s_k)^2}{\sigma_i^2}
\right).
\end{equation}


By Bayes’ Rule, after observing order flow $\omega$ the market maker’s posterior belief over $s_k\in S$ is then
\begin{equation}
\label{eqn: candidate expression for posterior}
\pi_1(s_k \mid \omega \kern 0.045em ; \widetilde{W})
\ \propto\
\underbrace{\exp\!\left(
\sum_i \frac{ \widetilde{W}(x_i \mid s_k)\,\omega_i}{\sigma_i^2}
-\frac12 \sum_i \frac{ \widetilde{W}(x_i \mid s_k)^2}{\sigma_i^2}
\right)}_{\text{{\sl likelihood of $\omega$ conditional on $s_k$}}}
\cdot
\underbrace{\pi_0(s_k)\,\vphantom{\exp\!\left(
\sum_i \frac{ \widetilde{W}(x_i \mid s_k)\,\omega_i}{\sigma_i^2}
-\frac12 \sum_i \frac{ \widetilde{W}(x_i \mid s_k)^2}{\sigma_i^2}
\right)}}_{\text{{\sl prior}}},
\qquad s_k\in S.
\end{equation}
\noindent
with the proportionality constant $C(\omega)$ normalizing the expression to sum to one over $S$.
Given this posterior, the market maker prices AD claims by the zero-profit condition~\eqref{eqn: security x_i price}.

\begin{samepage}
\begin{theorem} 
\label{thm: Market Maker's Posterior}
\nopagebreak
Suppose the market maker holds belief $\widetilde{W}(\,\cdot\, \vert \,\cdot\,)$ about insider demand and receives aggregate order flow $\omega$.

\smallskip
\noindent
(i) His posterior distribution over signals is the $\pi_1(s_k \mid \omega \kern 0.045em ; \widetilde{W})$ in \eqref{eqn: candidate expression for posterior}.  

\smallskip
\noindent
(ii) His posterior $\pi_1(s_k \mid \omega \kern 0.045em ; \widetilde{W})$ depends on $\omega$ only through the noise-adjusted scores
\begin{equation}
\label{eqn:  projection coefficient profile}
\left( \sum\limits_i \frac{ \widetilde{W}(x_i \vert s_k) \,\omega_i }{\sigma_i^2}  \right)_{s_k \in S} \in \mathbb{R}^K,
\end{equation}
which measure how closely the realized order flow aligns with the expected signal-contingent demand profiles.
These scores therefore form the sufficient statistic for his inference.\\
\medskip
\end{theorem}
\end{samepage}

\subsection{Sufficient Statistic}
\label{sec: sufficient statistics}

Recall the three-state illustration in Section~\ref{sec: main intuition}. 
There, the market maker does not treat $(\omega_1,\omega_2,\omega_3)$ as three standalone order imbalances. 
What matters for inference is not whether one particular contract has buy pressure, but whether the order flow across contracts resembles a recognizable informed portfolio. 
In that toy volatility interpretation, he effectively asks: does the joint order flow look more like the portfolio he expects under high volatility (a straddle) or under low volatility? 
The cross-market imbalance $\Delta(\omega)$ of \eqref{eqn: AD price in toy model} summarizes exactly this ``fit.''

Theorem~\ref{thm: Market Maker's Posterior} makes this intuition precise under arbitrary information structures.
For each signal $s_k$, the market maker computes a noise-adjusted alignment score
\begin{equation}
\label{eqn:  projection coefficient}
\sum_i \frac{\widetilde{W}(x_i \vert s_k)\,\omega_i}{\sigma_i^2},
\end{equation}
which measures how well the realized order flow matches the demand profile he expects conditional on $s_k$.
A high score means: ``the order flow looks like the portfolio I would expect if traders were acting on $s_k$.''

His posterior is therefore determined by the profile of alignment scores~\eqref{eqn:  projection coefficient profile}---one for each information type.
The weights $1/\sigma_i^2$ are the signal-to-noise adjustment: trades in noisier markets (larger $\sigma_i$) are less informative and therefore receive less weight in inference.
At the extreme, if $\sigma_i\to\infty$ for all $i$, order flow becomes pure noise, the alignment scores shrink to zero, and the posterior reduces to the prior.

When $X$ is a singleton, Theorem~\ref{thm: Market Maker's Posterior} reduces to the familiar Kyle single-asset inference reviewed in Section~\ref{sec: single asset case}: the sufficient statistic~\eqref{eqn:  projection coefficient profile} becomes a single scalar, 
a scaled version of net signed order flow $\omega$.
Accordingly, buy pressure makes high asset value more likely and raises the market maker’s posterior mean (the price), while sell pressure does the opposite.

We also define the analogous \emph{impact exposure} by replacing order flow $\omega$ with a candidate insider portfolio $W$ in the alignment score~\eqref{eqn:  projection coefficient}.

\begin{definition}
\label{def: overlap measures}

$\;$
\nopagebreak
Given the market maker's belief $\widetilde{W}(\,\cdot\,\vert s_k)$ about insider demand conditional on $s_k$:

\smallskip
\noindent
(i) The noise-adjusted \textbf{alignment score} of order flow $\omega$ with $\widetilde{W}(\,\cdot\, \vert s_k)$ is
\begin{equation}
\label{eqn: MM overlap measure}
\Pi_{mm}(\omega, s_k ; \widetilde{W}) \;=\; \sum_i \frac{\widetilde{W}(x_i \vert s_k)\,\omega_i}{\sigma_i^2} \, \mbox{ (as in \eqref{eqn:  projection coefficient})}.
\end{equation}

\smallskip
\noindent
(ii) The noise-adjusted \textbf{impact exposure} of an insider portfolio $W(\,\cdot\,)$ to $\widetilde{W}(\,\cdot\,\vert s_k)$ is
\begin{equation}
\label{eqn: insider overlap measure}
\Pi_{insider}(W, s_k ; \widetilde{W}) \;=\; \sum_i \frac{\widetilde{W}(x_i \vert s_k)\,W(x_i)}{\sigma_i^2}.
\end{equation}
\end{definition}

By Theorem~\ref{thm: Market Maker's Posterior}, the market maker's inference from order flow $\omega$ is fully summarized by the alignment-score profile 
$\Pi_{mm}(\omega, \,\cdot\, ;\widetilde{W}) = \big(\Pi_{mm}(\omega,s_k;\widetilde{W})\big)_{s_k\in S}$.
On the other hand, the impact-exposure profile $\Pi_{insider}(W,\,\cdot\,;\widetilde{W})$ measures how a contemplated portfolio $W$ loads on the market maker’s conjectured informed-demand template signal by signal. 
In Section~\ref{sec: Informed Trader's Problem}, we will show that this profile of signal-wise loadings governs the cross-market price impact of $W$.

\section{Cross-Market Price Impact}
\label{sec: Informed Trader's Problem}

Just as the market maker’s inference is infinite-dimensional, so is the insider’s portfolio choice. 
After observing his signal, the insider chooses an unrestricted portfolio $W(\,\cdot\,)$ of traded claims. 
Because there is no restriction on $W$, there is likewise no restriction on the pattern of order flow the market maker might rationalize under his belief. 
Prices are then set by how the realized order flow aligns with these conjectured patterns (see Theorem~\ref{thm: Market Maker's Posterior}). 
As a result, changing $W$ changes the alignment of aggregate order flow with each signal, and therefore can move prices across all claims. 
Characterizing this price-impact requires taking marginal perturbations of the infinite-dimensional choice variable $W(\,\cdot\,)$.

Given market maker belief $\widetilde{W}$ and signal $s_k$, the insider’s portfolio choice problem~\eqref{eqn: informed trader's problem informal} can be written as
\begin{align}
\max\limits_{ {\scriptscriptstyle W(\cdot) }} \, J(W \vert s_k; \widetilde{W})    &= \max\limits_{  {\scriptscriptstyle W(\,\cdot\,)} }  \,   \underbrace{ \left( \sum_i W(x_i) \eta(x_i \vert s_k) \right) }_\text{expected payoff}  - \underbrace{ \left(  \sum_i W(x_i) \overline{P}(x_i, W; \widetilde{W}) \right)  }_\text{expected execution cost}. 
\label{eqn: informed trader's problem}
\end{align}
Here,
\[
\overline{P}(x_i, W; \widetilde{W}) \equiv \mathbb{E}^{\mathbb{P}_{\!\scriptscriptstyle W}}
\!\left[\, P(x_i \vert \omega \kern 0.045em ; \widetilde{W}) \,\right]
\]
is the expected execution price of claim $x_i$ when the insider submits portfolio $W$.
The expectation is taken under $\mathbb{P}_{\!\scriptscriptstyle W}$, the probability distribution over realized order flow $\omega$ generated by $W$. 
Let $\partial \! \kern 0.08em J(v; W)$ denote the insider’s marginal gain from adding a small portfolio $v$ to $W$.\footnote{Formally, this first-order change is the G\^ateaux derivative of $J(\,\cdot\,\vert s_k;\widetilde{W})$ in \eqref{eqn: informed trader's problem} at $W$ in direction $v$.}

\subsection{First-Order Condition}
\label{subsec: FOC and Arrow-Debreu-Kyle Decomposition}

Optimal trading equalizes marginal benefits and marginal costs at the level of \emph{portfolios}.
For any small trade package $v$, the insider compares its marginal payoff to its marginal execution cost.
Beyond paying prevailing execution prices, adding $v$ can move prices across claims by changing how aggregate order flow loads on the market maker’s conjectured informed trades across signals.
Theorem~\ref{thm: informed trader 1} shows that this cross-market effect enters through the \emph{impact exposure} $\Pi_{insider}(v,\,\cdot\,;\widetilde{W})$ of the trade package $v$, as defined in Definition~\ref{def: overlap measures}(ii).

\begin{theorem}
\label{thm: informed trader 1} \textup{(Insider FOC)}
\nopagebreak

For any portfolio $W$ and any marginal trade package $v$, the insider's marginal gain decomposes as
\[
\underbrace{\partial \! \kern 0.08em J(v;W)\vphantom{\Bigl(\partial \! \kern 0.08em J_{\scriptscriptstyle \! AD}(v;W)+\partial \! \kern 0.08em J_{\scriptscriptstyle \! K}(v;W)\Bigr)}}_{\text{marginal gain}}
=
\underbrace{\partial \! \kern 0.08em J_{p}(v;W)\vphantom{\Bigl(\partial \! \kern 0.08em J_{\scriptscriptstyle \! AD}(v;W)+\partial \! \kern 0.08em J_{\scriptscriptstyle \! K}(v;W)\Bigr)}}_{\text{marginal payoff}}
-
\underbrace{\Bigl(\partial \! \kern 0.08em J_{\scriptscriptstyle \! AD}(v;W)
      + \partial \! \kern 0.08em J_{\scriptscriptstyle \! K}(v;W)\Bigr)}_{\text{marginal execution cost}},
\]
where
\begin{align}
\underbrace{\partial \! \kern 0.08em J_p(v ; W)}_{\text{marginal payoff}}
&=  \sum_i v(x_i)\,\eta(x_i \vert s_k),
\label{eqn: marginal payoff}\\[0.25em]
\underbrace{\partial \! \kern 0.08em J_{\scriptscriptstyle \! AD}(v ; W)}_{\text{prevailing execution-price}}
&= \sum_i v(x_i)\,\overline{P}(x_i, W; \widetilde{W}),
\label{eqn: AD term}\\[0.25em]
\underbrace{\partial \! \kern 0.08em J_{\scriptscriptstyle \! K}(v; W)}_{\text{cross-market price-impact}}
&= \sum_i W(x_i)\;
\mathbb{E}^{\mathbb{P}_{\scriptscriptstyle W}}
\!\Bigl[
  \underbrace{\Cov\!\Bigl(
    \eta(x_i \vert \,\cdot\,),
    \Pi_{insider}(v, \,\cdot\, ; \widetilde{W})
    \,\big|\, \omega
  \Bigr)}_{\substack{\text{price impact of }v\text{ on }x_i\\ \text{conditional on realized order flow }\omega}}
\Bigr].
\label{eqn: Kyle term in FOC}
\end{align}

\noindent
Therefore, an optimal portfolio $W$ must satisfy the first-order condition
\begin{equation}
\label{eqn: insider FOC}
 \partial \! \kern 0.08em J_p(v; W)
 \;=\;
 \partial \! \kern 0.08em J_{\scriptscriptstyle \! AD}(v; W)
 \;+\;
 \partial \! \kern 0.08em J_{\scriptscriptstyle \! K}(v; W)
 \quad \mbox{for all marginal trade packages $v$.}
\end{equation}

\end{theorem}

The first-order condition~\eqref{eqn: insider FOC} says that, at an optimal portfolio $W$, no marginal trade $v$ yields a positive gain: its payoff is exactly offset by its execution cost.
This execution cost comprises, first, the expected execution-price term~\eqref{eqn: AD term} and, second, the cross-market price-impact term~\eqref{eqn: Kyle term in FOC}.

\begin{itemize}
\item \textbf{Marginal Payoff $\partial \! \kern 0.08em J_p(v;W)$}
\textup{\eqref{eqn: marginal payoff}.}
This is the expected payoff of the incremental trade package $v$ under the signal-implied state probabilities $\eta(\,\cdot\,\mid s_k)$.
For example, if $v$ is an options package across strikes, then $\partial \! \kern 0.08em J_p$ is the valuation of that package 
under the trader’s private distributional view (volatility, skew, tail risk, etc.).

\item \textbf{AD Term $\partial \! \kern 0.08em J_{\scriptscriptstyle \! AD}(v;W)$}
\textup{\eqref{eqn: AD term}.}
This is the baseline execution bill: what it costs to add $v$ at the prevailing
execution prices $\overline{P}(\,\cdot\,, W; \widetilde{W})$.

\item \textbf{Price-Impact Term $\partial \! \kern 0.08em J_{\scriptscriptstyle \! K}(v;W)$}
\textup{\eqref{eqn: Kyle term in FOC}.}
This is the extra information-leakage slippage. Adding $v$ can reprice the entire cross-section of claims through its impact exposure $\Pi_{insider}(v,\,\cdot\,;\widetilde{W})$—its loading on what the market maker conjectures informed trades to be for each signal 
(i.e., how much \emph{more} $v$ makes observed order flow resemble each information story). The price of claim $x_i$ moves precisely when this loading co-moves with $x_i$’s payoff profile across information stories.
For example, buying wing options can depress prices of near-to-ATM options if the market maker reads the package as volatility demand and reprices the entire strip against the trader.
\end{itemize}

\paragraph{Single-Contract Cross Impact}
The general price-impact term $\partial \! \kern 0.08em J_{\scriptscriptstyle \! K}(v;W)$ is written for an arbitrary trade package $v$.
The primitive empirical object is the cross-impact kernel---how signed order flow in one claim moves the price of another.
When $v$ consists of one share of claim $x_j$, its impact exposure collapses to $\Pi_{insider}(v,\,\cdot\,;\widetilde{W})=\widetilde{W}(x_j\mid \,\cdot\,)/\sigma_j^{2}$. 
In words, order flow in $x_j$ pushes beliefs toward whichever story predicts heavier insider demand for $x_j$, adjusted for noise.
Corollary~\ref{cor: def of price impact} characterizes the resulting cross-impact kernel.

\begin{corollary}
\label{cor: def of price impact}
\nopagebreak

Under the market maker’s belief $\,\widetilde{W}$, the price impact of marginal order flow for $x_j$ on the price of another claim $x_i$, given realized aggregate order flow $\omega$, is
\begin{equation}
\label{eqn: conditional price impact y on x}
\frac{\partial}{\partial W(x_j)} P(x_i, \omega \kern 0.045em ; \widetilde{W}) =  \frac{1}{\sigma_j^2} \cdot \Cov \left( \eta(x_i \vert \,\cdot\, ), \widetilde{W}(x_j \vert \,\cdot\,) \, \bigg| \, \omega \right).
\end{equation}
Averaging \eqref{eqn: conditional price impact y on x} over the distribution of realized order flow gives the expected cross-impact
\begin{equation}
\label{eqn: price impact y on x}
\frac{\partial}{\partial W(x_j)} \overline{P}(x_i, W; \widetilde{W})
= \mathbb{E}^{\mathbb{P}_{\scriptscriptstyle W}} \! \kern -0.08em \Bigl[ \frac{\partial}{\partial W(x_j)} P(x_i, \omega \kern 0.045em ; \widetilde{W}) \Bigr].
\end{equation}

\end{corollary}

The intuition behind Corollary~\ref{cor: def of price impact} is the same information-leakage slippage embedded in the general price-impact term $\partial \! \kern 0.08em J_{\scriptscriptstyle\!K}(v;W)$, now in its sharpest single-contract form.
A marginal trade in $x_j$ moves the price of $x_i$ \emph{to the extent that the information stories the market maker reads into $x_j$ order flow correlate with the stories that drive $x_i$’s payoff}.
This is exactly what the posterior covariance in \eqref{eqn: conditional price impact y on x} captures:
\[
\Cov \!\left( \eta(x_i \vert \,\cdot\, ),\, \widetilde{W}(x_j \vert \,\cdot\,) \,\bigg|\, \omega \right).
\]
$\widetilde{W}(x_j\mid\,\cdot\,)$ is the market maker’s story-by-story predicted informed demand for $x_j$, and $\eta(x_i\mid\,\cdot\,)$ is  $x_i$’s corresponding payoff profile across those same stories.

\subsection{No-Arbitrage}
\label{sec: no arb}

No-arbitrage in our setting admits an intuitive characterization: \emph{any portfolio that generates no price impact cannot yield a positive payoff for the insider.} 
Given market maker belief $\widetilde{W}$, we call a portfolio $W$ a \emph{zero--price-impact portfolio} if its impact exposure~\eqref{eqn: insider overlap measure} vanishes,
\[
\Pi_{insider}(W,\,\cdot\,;\widetilde{W})=0.
\]
These are ``stealth'' portfolios---they look like noise trades to the market maker (i.e., they are orthogonal to the market maker's belief about informed demand), so scaling them does not tilt the market maker's posterior and therefore does not move prices.
If such a zero-impact portfolio yielded a nonzero payoff to the insider, the insider could lever it up without moving prices, generating an arbitrarily large payoff. This is Theorem~\ref{thm: informed trader 2}.

\begin{theorem}
\label{thm: informed trader 2}
\textup{(No free lunch with zero price impact)}
\nopagebreak
Under market maker belief $\widetilde{W}$, an optimal insider portfolio can exist only if every zero price impact
portfolio yields zero expected payoff to the insider. Otherwise, the insider can scale a zero-impact trading direction without
moving prices and obtain an arbitrarily large expected payoff.
\end{theorem}

\section{Equilibrium}
\label{sec: Equilibrium, Finite $S$}

Up to this point we have solved each side of the market taking the other as given: the market maker’s pricing kernel as a mapping from cross-market order flow to the cross-section of claim prices (Section~\ref{sec: Market Maker's Pricing Kernel}), 
and the insider’s optimal portfolio in the same rich claim space (Section~\ref{sec: Informed Trader's Problem}). 
In equilibrium, the conjectured ``informed-demand template'' that drives pricing must coincide with the insider’s optimal trades under that same pricing kernel---a joint fixed point of two high-dimensional problems.
A key insight is that the trading game is isomorphic to a canonical game that is \emph{invariant} to the information structure and noise intensities. 
This sets up a striking reduction---equilibrium, an infinite-dimensional fixed point \emph{a priori}, is characterized by a single endogenous constant.
(The economic intuition underlying this reduction is robust whether the signal and state spaces are discrete or continuous; a formal treatment of the continuous case can be found in \cite{kellertsengmath}.)

\begin{assumption}
\label{assumtion: finite S} {\bf (w.l.o.g.)} 
$\;$
\nopagebreak
The market maker has a uniform prior on the signal space $S$.
\end{assumption}

For expositional convenience, we maintain Assumption~\ref{assumtion: finite S} in this section.
This is without loss of generality and serves only to streamline the derivations.
Starting from any prior $\pi_0$, we can rescale payoffs and work {\em as if} the prior were uniform on $S$ by replacing $\eta(\,\cdot\,\vert s_k)$ with
$\frac{\pi_0(s_k)}{1/|S|}\,\eta(\,\cdot\,\vert s_k)$ for each $s_k\in S$. 
Once we characterize equilibrium under the uniform prior, comparative statics for arbitrary priors follow immediately. 
For instance, if two securities tend to pay off together under some $s_k$, then putting more prior weight on $s_k$ makes their co-movement in order flow more diagnostic, so their equilibrium cross-price impact increases with $\pi_0(s_k)$.
This generalizes from the toy model of Section~\ref{sec: main intuition}; see footnote~\ref{fn: general prior}.

\subsection{Canonical Game}
\label{sec: finite dim reduction}

In equilibrium, the portfolios that yield zero expected payoff to the insider must be exactly the portfolios that generate zero price impact. 
Zero payoff implies zero price impact because equilibrium beliefs are correct (Definition~\ref{def: equilibrium}).  
Conversely, by Theorem~\ref{thm: informed trader 2}, a zero-impact portfolio with nonzero payoff would be a stealth arbitrage that the insider could scale without moving prices.

This coincidence translates to a restriction on the relevant trading subspace.
The insider’s expected payoff is driven only by how $W$ loads on the signal-implied payoff distributions $\{\eta(\,\cdot\,\mid s_k)\}_k$. 
Any portfolio $W$ that is orthogonal to each distribution—i.e., $\sum_i W(x_i)\eta(x_i\mid s_k)=0$ for all $k$—has zero expected payoff under every signal.  
Hence, without loss, equilibrium informed demand can be taken to lie in the linear span of $\{\eta(\,\cdot\,\mid s_k)\}_k$.
Conversely, to rule out stealth arbitrage, the market maker's (noise-adjusted) equilibrium ``informed-demand templates'' must cover every payoff-relevant direction—otherwise the insider could hide profits in a trading direction the market maker cannot see.
That is, each $\eta(\,\cdot\,\mid s_k)$ must lie in the span of $\{W^*(\,\cdot\,\mid s_\ell)/\sigma_{\,\cdot\,}^2\}_\ell$.
We record these two spanning restrictions in Proposition~\ref{cor: finite S 1}.

\Needspace{4\baselineskip}
\begin{proposition}
\label{cor: finite S 1}
\nopagebreak
$\;$\\
\indent (i) The insider's equilibrium demand $W^*(\,\cdot\, \vert s_k)$ conditional on each $s_k$ can be
restricted to the linear span of $\{ \eta(\,\cdot\, \vert s_l) \}_{l}$.

\indent (ii) Conversely, to prevent arbitrage, each $\eta(\,\cdot\, \vert s_k)$ must lie in the linear span of
$\{ \frac{ W^*(\,\cdot\, \vert s_l ) }{\sigma_{\,\cdot\,}^2} \}_{l}$.
\end{proposition}

To make the reduction in Proposition~\ref{cor: finite S 1} explicit, we parameterize portfolios and beliefs on the payoff-relevant span of $\{\eta(\,\cdot\,\mid s_k)\}_{k=1}^K$ by their loadings on these signal-implied payoffs.
Accordingly, write any portfolio in this span as
\[
W(\,\cdot\,)=\sum_{k=1}^K d_k\,\eta(\,\cdot\,\mid s_k),
\qquad \text{with loadings } d=(d_k)\in\mathbb{R}^K .
\]
Likewise, expand the market maker’s signal-contingent ``informed-demand template''
$\widetilde W(\,\cdot\,\mid s_k)$ in the same basis,
\[
\widetilde W(\,\cdot\,\mid s_k)=\sum_{l=1}^K \tilde d^{(k)}_l\,\eta(\,\cdot\,\mid s_l),
\qquad 1\le k\le K ,
\]
and stack the loading vectors as columns to form the belief matrix
\[
\widetilde D \;=\; [\tilde d^{(1)}\ \cdots\ \tilde d^{(K)}]\in\mathbb{R}^{K\times K}.
\]

\begin{definition}
\label{def: information intensity matrix}
The \textbf{information intensity matrix} ${\bf L}\in\mathbb{R}^{K\times K}$ of the trading game is the
positive-semidefinite matrix satisfying
\[
{\bf L}^2
=
\left[
\sum_i
\frac{\eta(x_i\mid s_k)\,\eta(x_i\mid s_l)}{\sigma_i^2}
\right]_{1\le k,l\le K},
\qquad
{\bf L}^T={\bf L}.
\]
\end{definition}

The information intensity matrix ${\bf L}$ is the cross-market analogue of the information intensity parameter $\sigma_v/\sigma_\varepsilon$ from the single-asset Kyle setting~\eqref{eqn: static Kyle solution}.
It aggregates the (noise-adjusted) covariations of the signal-implied payoff profiles $\{\eta(\,\cdot\,\mid s_k)\}_k$ across claims, so a larger ${\bf L}$ (in the natural matrix order) means that cross-market order flow is more diagnostic of which information story is active.

\paragraph{Conversion to Information Units}
Accordingly, ${\bf L}$ provides exactly the right rescaling to convert the raw loadings in $d$ and $\widetilde D$ into \emph{information units}: premultiplying by ${\bf L}$ ``whitens'' the payoff-relevant span so that one unit of loading corresponds to one unit of 
noise-adjusted informational exposure.
We therefore pass to the \emph{information-normalized} loading vector $\hat d$ and belief matrix $\widehat D$ given by
\begin{equation}
\label{eqn: change of basis}
\begin{aligned}
\hat d \;&=\; {\bf L}d,\\
\widehat D \;&=\; {\bf L}\widetilde D \;=\; [\hat d^{(1)}\ \cdots\ \hat d^{(K)}].
\end{aligned}
\end{equation}

\paragraph{Canonical Game}
The information-unit conversion~\eqref{eqn: change of basis} strips away
the idiosyncrasies of the payoff distributions and noise intensities and isolates the strategic core of the
trading game. 
In information units, the payoff-relevant span becomes \emph{whitened} (orthonormal)---each trading direction carries one unit of signal relative to noise. 
The trading game can then be read as trading orthogonal \emph{information contracts}—one per information story—under
unit noise intensity. The next theorem records this equivalence.

\begin{samepage}
\begin{theorem}
\label{thm: canonical game}
\textup{(Canonical Game---Informal Statement)}
\nopagebreak
The original trading game is isomorphic to the \textbf{canonical game} in which:
\begin{itemize}[leftmargin=1.95em, itemsep=0.15em, topsep=0.15em]
\item the traded claims are \textbf{information contracts}---one for each signal $s_k$, paying $1$ under story $k$ and $0$ otherwise;
\item noise trading is i.i.d.\ $\mathcal{N}(0,1)$ across contracts;
\item competitive prices are the market maker's posterior probabilities over the signals.
\end{itemize}

\noindent
The full formal statement is deferred to Appendix Section~\ref{sec: proof of canonical game}.
\end{theorem}
\end{samepage}

\paragraph{Equilibrium in the Canonical Game}
An equilibrium in the canonical game is summarized by a $K\times K$ matrix
$D^*=[\delta^{(1)}\ \cdots\ \delta^{(K)}]$ where the $k$-th column $\delta^{(k)}\in\mathbb{R}^K$ specifies the insider's demand for the information contracts when story $k$ is active.
Given the market maker's belief $D^*$, prices in the canonical game are posterior probabilities over the $K$ stories,
and each $\delta^{(k)}$ is optimal under those very prices.
In other words, $D^*$ is the fixed point that closes the canonical game.

\paragraph{Mapping Back to the Original Game}
Given a canonical-game equilibrium $D^*$, we reverse the information-unit conversion~\eqref{eqn: change of basis}
to express the same equilibrium trades as loadings on the original payoff distributions
$\{\eta(\,\cdot\,\mid s_l)\}_{l=1}^K$.
Define the \textbf{\emph{canonical form}} loadings matrix\footnote{If ${\bf L}$ is singular, interpret ${\bf L}^{-1}$ as the Moore--Penrose pseudoinverse.}
\begin{equation}
\label{eqn: canonical form informed demand matrix}
{\bf L}^{-1}D^* \;=\; [\beta^{(1)}\ \cdots\ \beta^{(K)}]\in\mathbb{R}^{K\times K},
\end{equation}
so the $k$-th column $\beta^{(k)}$ collects the insider’s equilibrium loadings across payoff distributions when signal $s_k$ is realized.
The corresponding informed demand is then
\begin{equation}
\label{eqn: canonical form informed demand}
W^*(\,\cdot\,\mid s_k) \;=\; \sum_{l=1}^K \beta_l^{(k)}\,\eta(\,\cdot\,\mid s_l).
\end{equation}

\subsection{One-Parameter Equilibrium}

We now close the model via the canonical-game reduction, which reveals the extremely tight equilibrium restriction of the general trading game.

The canonical game treats the information stories symmetrically: relabeling stories just relabels markets.  
This symmetry forces a simple long-short equilibrium structure.
Conditional on story $k$, the insider goes long the $k$-th information contract and finances it by shorting the average of the remaining contracts.

Let ${\bf Q}$ denote the $K\times K$ matrix that encodes this \emph{``buy the active story and sell the alternatives''} strategy, column by column:
\begin{equation}
\label{eqn: matrix Q}
{\bf Q}
\;=\;
{\bf I}-\frac{1}{K}\,\bar e\,\bar e^{T},
\qquad \mbox{ where }
\bar e \;=\; (1,\ldots,1)^{T}\in\mathbb{R}^K.
\end{equation}
So the $k$-th column of ${\bf Q}$ puts portfolio weights $1-\tfrac{1}{K}$ on contract $k$ and $-\tfrac{1}{K}$ on each of the remaining contracts.
Accordingly, we parameterize candidate equilibria by a single parameter $\alpha>0$, with corresponding strategy matrices of the form
\begin{equation}
\label{eqn: equilibrium ansatz for transformed Bayesian game}
D^* \;=\; \alpha\,{\bf Q}.
\end{equation}

Imposing insider optimality within this symmetric class reduces the fixed point condition to a single scalar equation in $\alpha$.
Theorem~\ref{thm: equilibrium of original trading game} shows that it admits a solution $\alpha^*>0$, with corresponding equilibrium strategy matrix $D^*=\alpha^*{\bf Q}$. Thus, $\alpha^*$ is the sole endogenous constant characterizing equilibrium.

\begin{samepage}
\begin{theorem}
\label{thm: equilibrium of original trading game}
The canonical game admits an equilibrium $D^*=\alpha^*\,{\bf Q}$ in which the insider trades the \textbf{active} information contract against the
\textbf{average} of the others, with endogenous trading scale $\alpha^*$.
Reversing the information-unit conversion then yields the corresponding equilibrium of the original
trading game with canonical form loadings matrix~\eqref{eqn: canonical form informed demand matrix},
$
{\bf L}^{-1}D^*=\alpha^*\,{\bf L}^{-1}{\bf Q}.
$
\end{theorem}
\end{samepage}

\paragraph{Intuition for $\alpha^*$}
The intuition underlying Theorem~\ref{thm: equilibrium of original trading game} is the same as that highlighted in the toy setting of Section~\ref{sec: main intuition}, now shown to be robust in full Arrow-Debreu generality.
Raising $\alpha$ scales up the long-short portfolio and increases the insider’s exposure to the true story,
but it also makes aggregate order flow more informative, which erodes the insider’s marginal information rent.
Equilibrium $\alpha^*$ obtains at the knife-edge where this tradeoff balances.

The next subsection demonstrates the especially transparent binary signal case (which nests the setting of Section~\ref{sec: main intuition}), where inference collapses to a single log-odds statistic.

\subsection{Binary Signal Case}
\label{sec: e.g. binary signal}

Suppose there are two possible information stories, $S=\{s_1,s_2\}$.
In the canonical game there is one information contract per story, paying $1$ if that story is true and $0$ otherwise.
The natural informed trade is a market-neutral spread: under each signal the insider buys the ``active'' contract and shorts the alternative by the same amount.
This long-short trade corresponds to the strategy matrix
\[
D^*
=
\frac{\alpha}{2}
\begin{bmatrix}
1 & -1\\
-1 & 1
\end{bmatrix},
\qquad \alpha>0,
\]

In this two-story case, the cross-section of information-contract prices $(p_1,p_2)$ is fully summarized by a single posterior log-odds statistic $Z$:
\begin{equation}
\label{eqn: posterior in binary signal case}
(p_1,p_2)\stackrel{d}{=}\left(\frac{e^{Z}}{e^{Z}+1},\,\frac{1}{e^{Z}+1}\right),
\qquad
\mbox{ where }
Z\stackrel{d}{\sim}\mathcal{N}\!\left(\frac{\alpha^2}{2},\,2\alpha^2\right).
\end{equation}
Thus, observing order flow is equivalent to observing the ``evidence score'' $Z$: large $Z$ is strong evidence for $s_1$,
while large negative $Z$ favors $s_2$.
In the toy setting of Section~\ref{sec: main intuition}, the evidence score reduces to the normalized cross-market imbalance
$\Delta(\omega)$ in \eqref{eqn: AD price in toy model}, with $Z = 2\alpha\,\Delta(\omega)$.
Accordingly, in the volatility interpretation of Section~\ref{sec: main intuition}, large $Z$ is strong evidence for high volatility.\footnote{This is a special case of Lemma~\ref{lemma: claim for MM posterior}.}

When the insider does not trade and order flow is pure noise ($\alpha=0$), the evidence score is degenerate at $Z=0$, so prices stay at the prior (no learning).
When the insider scales the spread without bound ($\alpha\to\infty$), $Z$ becomes overwhelmingly informative and prices concentrate on the realized story, so informational rents vanish.
More generally, trading scale $\alpha$ affects payoff only through the evidence score $Z$. Larger $\alpha$ scales up the spread, but it also makes $Z$---and hence order flow---more informative.

Our scalar equilibrium equation is exactly the marginal indifference condition for the trading scale~$\alpha$:
\begin{equation}
\label{eqn: transformed Bayesian game FOC, binary signal}
\Phi(\alpha)\;\equiv\;1-\mathbb{E}[p_1]-\alpha^2\,\mathbb{E}[p_1p_2]\;=\;0,
\end{equation}
Here $1-\mathbb{E}[p_1]$ is the insider's average rent wedge from the active story, while
$p_1p_2=p_1(1-p_1)$ is the market’s residual uncertainty (the posterior variance of the binary signal).
The residual uncertainty is maximized when $Z\approx 0$ (the market is unsure) and it collapses as $|Z|$ grows (the market becomes confident).
Thus, a marginal increase in the trading scale $\alpha$ reveals the most information precisely when the market is unsure,
sharpening inference and compressing marginal information rents. 

The equilibrium scale $\alpha^*$ solves $\Phi(\alpha)=0$, i.e., the marginal gain from increasing exposure is offset by the marginal information leakage. 
We have $\Phi(0)=\tfrac12>0$, while $\Phi(\alpha)<0$ for sufficiently large $\alpha$, and moreover $\Phi(\alpha)\uparrow 0$ as $\alpha\to\infty$.\footnote{The limit $\Phi(\alpha)\uparrow 0$ follows from Fatou's Lemma. 
Existence of $\alpha^*>0$ with $\Phi(\alpha^*)=0$ follows by the Intermediate Value Theorem.}
Hence, there exists $\alpha^*>0$ solving $\Phi(\alpha)=0$, which proves Theorem~\ref{thm: equilibrium of original trading game} in this binary signal case.

\subsection{Comparative Statics of \boldmath $\alpha^*$ \unboldmath}

\begin{figure}[htbp!]

\centering
\stackunder[5pt]{
\scalebox{1}[0.85]{
\begin{tikzpicture}
\input{Endogenous_constant.tex} 
\end{tikzpicture}
}
}{ {\footnotesize $x$-axis: number of signals, $y$-axis: $\alpha^*$}}
\caption{
{\footnotesize
{\bf Comparative Statics of Endogenous Constant \boldmath $\alpha^*$ \unboldmath}
} 
}
\label{fig: endogenous constant}
\end{figure}

We now return to the general model. The canonical-game reduction shows that the equilibrium trading scale
$\alpha^*$ is pinned down solely by the number $K$ of signals and is therefore invariant to the particular signal-implied payoff distributions
$\eta(\,\cdot\,|\,\cdot\,)$ and contract-level noise intensities $(\sigma_i)$. Figure~\ref{fig: endogenous constant} plots the
resulting mapping $K \mapsto \alpha^*(K)$.

As $K$ increases, there are more distinct payoff-distribution ``stories'' the market must disentangle from cross-contract order
flow, so equilibrium informed trading becomes more aggressive.
At the same time, a larger $K$ also increases the extent to which informed demand differs across stories, making order flow more
diagnostic and steepening the informational (price-impact) cost of further scaling. The net effect is that $\alpha^*(K)$ rises
with $K$ but with diminishing increments: $\alpha^*(K)$ is increasing and concave.

Put differently, as more distinct information stories are traded and priced through the cross-section, they support larger informed positions.  
But each additional story contributes less to overall trading intensity, because the cross-sectional pattern of order flow becomes progressively more diagnostic of which story is active.

{\small
\renewcommand{\arraystretch}{1.15}
\setlength{\tabcolsep}{5pt}

\begin{longtable}{@{}>{\raggedright\arraybackslash}p{0.28\textwidth} >{\raggedright\arraybackslash}p{0.68\textwidth}@{}}
\caption{Key Equilibrium Quantities}\label{tab:key_equilibrium_quantities}\\
\toprule
\textbf{Notation} & \textbf{Economic meaning} \\
\midrule
\endfirsthead

\multicolumn{2}{@{}l@{}}{\tablename~\thetable\ (continued)}\\
\toprule
\textbf{Notation} & \textbf{Economic meaning} \\
\midrule
\endhead

\multicolumn{2}{r@{}}{\emph{Continued on next page}}\\
\endfoot

\bottomrule
\endlastfoot

\tabsec{Primitives and indices}
$S=\{s_1,\ldots,s_K\}$ & Set of possible \emph{information stories} (signals) about the payoff distribution (e.g., volatility, skewness, tail risk).\\
$X=\{x_1, x_2, \ldots\}$ & Set of possible future states (e.g., terminal underlying prices); may be infinite.\\
$k,l$ & Indices for information stories/signals.\\
$i,j$ & Indices for Arrow-Debreu states/claims (e.g., option strikes in the options implementation via \eqref{eqn: Breeden-Litzenberger}).\\
$\eta(x_i\mid s_k)$ & Probability of state $x_i$ under story $s_k$ (i.e., the signal-implied distribution over states).\\

\tabsec{Trading and order flow}
$W^*(x_i\mid s_k)$ & Insider’s equilibrium order/position in claim $x_i$ under story $s_k$.\\
$\varepsilon_i$ & Noise order in claim $x_i$ (non-informational order flow).\\
$\omega = (\omega_i)_{i\geq1}$ & Aggregate order flow across claims received by the market maker; $\omega_i=W^*(x_i\mid s_k)+\varepsilon_i$.\\
$\sigma_i$ & Noise-trade intensity in market $i$ (larger $\sigma_i$ means noisier, less informative order flow).\\

\tabsec{Inference, prices, and price impact}
$\pi^*(s_k\mid\omega)$ & Market maker’s posterior probability that story $s_k$ is active, conditional on order flow $\omega$.\\


$P^*(x_i\mid\omega)$ & Equilibrium \emph{state price} (pricing kernel) for claim $x_i$ conditional on $\omega$.\\
$\Lambda=(\Lambda_{i,j})$ & Equilibrium price impact matrix: $\Lambda_{i,j}$ is the response of $P^*(x_i\mid\omega)$ to order flow in claim $x_j$ (own-impact on diagonals, cross-impact off-diagonals).\\

\tabsec{Portfolio construction for informed demand $W^*( \,\cdot\,\mid \,\cdot\,)$ }
{\bf Q} & Signal-tilt matrix: encodes the market-neutral long--short tilt that isolates the realized story (see \eqref{eqn: matrix Q}).\\
{\bf L} & Information intensity matrix: how diagnostic joint cross-contract order flow is about which story is active (see Definition~\ref{def: information intensity matrix}).\\
${\bf L}^{-1}$ & Information adjustment matrix: rotates/scales the raw signal tilt into equilibrium weights, downweighting directions where order flow is more revealing.\\
$\alpha^*$ & Overall position scale, balancing profit vs.\ information revelation (see Section~\ref{sec: e.g. binary signal}).\\
$\beta^{(k)}$ & Story-$s_k$ equilibrium weights (the $k$-th column of $\alpha^*{\bf L}^{-1}{\bf Q}$) used to build $W^*(\,\cdot\,\mid s_k)$ (see \eqref{eqn: informed demand}).\\

\end{longtable}
}

\FloatBarrier

\section{Price Discovery}
\label{sec: discussion}

For ease of reference, Table~\ref{tab:key_equilibrium_quantities} collects the key equilibrium quantities, organized by the model's information channel:
\begin{equation}
\label{eqn: info flow}
\begin{array}{c}
\text{\scriptsize signal}\\[-0.2ex]
s_k
\end{array}
\;\raisebox{-1.6ex}{$\Rightarrow$}\;
\begin{array}{c}
\text{\scriptsize informed demand}\\[-0.2ex]
W^*(\,\cdot\,\mid s_k)
\end{array}
\;\raisebox{-1.6ex}{$\Rightarrow$}\;
\begin{array}{c}
\text{\scriptsize order flow}\\[-0.2ex]
\omega
\end{array}
\;\raisebox{-1.6ex}{$\Rightarrow$}\;
\begin{array}{c}
\text{\scriptsize pricing kernel}\\[-0.2ex]
P^*(\,\cdot\,\mid \omega)
\end{array}
\;\raisebox{-1.6ex}{$\Rightarrow$}\;
\begin{array}{c}
\text{\scriptsize price impact}\\[-0.2ex]
\Lambda
\end{array}
\end{equation}

\noindent
The private signal $s_k$ induces equilibrium informed demand $W^*(\,\cdot\,\mid s_k)$, which is part of aggregate order flow $\omega$.
Conditioning on $\omega$, the market maker sets Arrow-Debreu state prices $P^*(\,\cdot\,\mid \omega)$; the within- and cross-price responses are collected in the price impact matrix $\Lambda$.
The standard formula~\eqref{eqn: Breeden-Litzenberger} translates these Arrow-Debreu quantities to options.

With the equilibrium characterized in Theorem~\ref{thm: equilibrium of original trading game}, we now study its price discovery implications: 
how information-driven demand is transmitted into price impact and the informativeness of prices.

\subsection{Informed Demand}
\label{sec: Generalized Kyle's beta}

The equilibrium informed demand $W^*(\,\cdot\, \vert s_k)$, conditional on a signal $s_k$, describes which states of the world the insider wants to load on, and which states he wants to shed.
Generalizing the intuition of Section~\ref{sec: main intuition}, the insider chooses the direction to trade---the long-short information trade---and then how aggressively to trade in that direction, recognizing that larger orders reveal more information and 
therefore cause higher price impact.

More precisely,
\begin{equation}
\label{eqn: informed demand}
W^*(\,\cdot\,\vert s_k) = \sum_{l = 1}^K \beta^{(k)}_l \eta(\,\cdot\,\vert s_l), \;\; 
\end{equation}
where $\beta^{(k)}$ is the $k$-th column of $\alpha^* {\bf L}^{-1} {\bf Q}$---it collects the equilibrium allocation weights.
Applying the replication formula~\eqref{eqn: Breeden-Litzenberger} then translates this state-contingent AD demand into an options portfolio across strikes.

\paragraph{Informed Demand Portfolio Construction}
The expression~\eqref{eqn: informed demand} can be read as a simple three-step recipe for constructing the insider's portfolio:

\begin{itemize}

\item \textbf{Step 1 \textit{(Raw signal tilt).}} Start from a market-neutral long--short position that isolates the observed signal $s_k$:
go long $1-\frac{1}{K}$ shares of the signal-implied payoff distribution $\eta(\,\cdot\,\vert s_k)$ and short $\frac{1}{K}$ shares of each alternative $\eta(\,\cdot\,\vert s_l)$, $l \neq k$.
\smallskip

\item \textbf{Step 2 \textit{(Liquidity adjustment).}}
Rotate and scale the signal tilt by the matrix ${\bf L}^{-1}$.
${\bf L}$ summarizes how informative order flow is to market makers. Applying ${\bf L}^{-1}$ therefore reduces positions in directions where order flow is more revealing and hence more costly in terms of price impact.
\smallskip

\item \textbf{Step 3 \textit{(Equilibrium aggressiveness).}}
Scale the entire portfolio by the endogenous constant $\alpha^*$, which governs overall trading intensity given the market maker's prior uncertainty over signals (see Figure~\ref{fig: endogenous constant}).

\end{itemize}

Step~1 is the pure \emph{information trade}: it loads on the payoff distribution associated with the insider's signal and nets out the common component across signals.
Step~2 is the \emph{liquidity scaling}: it adjusts the raw bet for how strongly the market maker can infer the signal from cross-market order flow, and therefore how aggressively the insider can trade without moving prices too much. 
Step~3 sets the \emph{overall size} of the trade in equilibrium, increasing when the market maker is more uncertain ex ante about which signal is realized.

The informed demand of Section~\ref{sec: main intuition} is the orthogonal-payoff special case of this general intuition: because 
payoff profiles do not overlap, the liquidity adjustment ${\bf L}^{-1}$ is inactive, so the insider's portfolio problem reduces to choosing the long-short information trade and the equilibrium scale $\alpha^*$.

\paragraph{Single-Asset vs.~AD Informed Demand}

Recall that this same \emph{direction--aggressiveness} decomposition already appears in the single-asset informed demand~\eqref{eqn: static Kyle solution pre} of \citet{kyle1985continuous}. 
There, the insider decides whether to buy or sell based on whether his signal exceeds the prior mean, and then chooses how aggressively to scale that position given the resulting price impact. 
This parallel becomes very transparent when comparing the canonical form (as defined in~\eqref{eqn: canonical form informed demand matrix}) $\alpha^* {\bf L}^{-1} {\bf Q}$  of our Arrow-Debreu equilibrium  to the single-asset trading rule~\eqref{eqn: static Kyle solution pre}:
\[
\text{\emph{Informed Demand}}
=
\left\{
\begin{alignedat}{2}
\makebox[3.3cm][l]{\emph{Single-Asset}:}\quad
& \underbrace{\dfrac{\sigma_{\varepsilon}}{\sigma_v}}_{\text{\emph{information intensity}}}
&\!\times\!&
\underbrace{(s - v_0)}_{\text{\emph{signal effect}}}
\\[1.1ex]
\makebox[3.3cm][l]{\emph{AD Securities}:}\quad
& \underbrace{\alpha^* {\bf L}^{-1}}_{\substack{\text{\emph{information intensity}}\\\text{\& \emph{generalized scaling}}}}
&\!\times\!&
\underbrace{{\bf Q}}_{\substack{\text{\emph{signal effect}}\\\text{\emph{across payoffs}}}}
\end{alignedat}
\right.
\]

\noindent
In both cases, informed demand factors into a \emph{signal effect} term that determines the insider’s long-short direction, and an \emph{information-intensity} term that determines how aggressively that direction can be traded.

\begin{itemize}

\item \textbf{Signal effect ($(s - v_0)$ vs.~${\bf Q}$):}
In the single-asset setting, $(s-v_0)$ says whether the insider wants to be net long or net short.
In the AD setting, ${\bf Q}$ is the payoff-state analogue: it specifies \emph{which states} become relatively more likely under the signal and therefore should be loaded on, and which states should be shorted.

\item \textbf{Information intensity ($\frac{\sigma_{\varepsilon}}{\sigma_v}$ vs.~$\alpha^*{\bf L}^{-1}$):}
In the single-asset setting, $\frac{\sigma_{\varepsilon}}{\sigma_v}$ summarizes how much noise trading obscures the insider’s information; when order flow is less revealing, the insider can scale up.
In the AD setting, ${\bf L}$ summarizes how informative \emph{cross-contract} order flow is about the vector of state payoffs; applying ${\bf L}^{-1}$ therefore scales down exposures in payoff directions 
where the market maker learns quickly (i.e., where trading would generate larger equilibrium price impact), while $\alpha^*$ pins down the overall equilibrium level of aggressiveness.

\end{itemize}

Returning to our motivation, the portfolio characterization of informed demand provides an equilibrium foundation for a broad class of option strategies used in practice. 
To illustrate, we consider a few prominent strategies for trading on the level, volatility, and skewness of an underlying payoff. 
Market participants routinely express such views using options. 
In contrast, existing theoretical treatments of option trading are largely confined to single-option environments, or multi-security frameworks that
either shut down strategic portfolio choice, or else rely on restrictive distributional assumptions; consequently, they cannot endogenously generate the nonlinear, multi-leg option portfolios observed in 
practice. 

In Examples~\ref{example: Kyle with options},~\ref{example: vol straddle}, and~\ref{example: skewness}, the signal is binary ($S=\{s_1,s_2\}$).
We parameterize payoffs within the normal family (discretized), yielding stylized illustrations.
This choice is purely expositional.
The model itself is nonparametric, and other common option strategies can be obtained by adapting the signal and payoff specifications.

\begin{figure}[htbp!]
\centering 

\subfloat[High-mean signal $s_1$]{
\includegraphics[width = 7cm, height = 2.6cm]{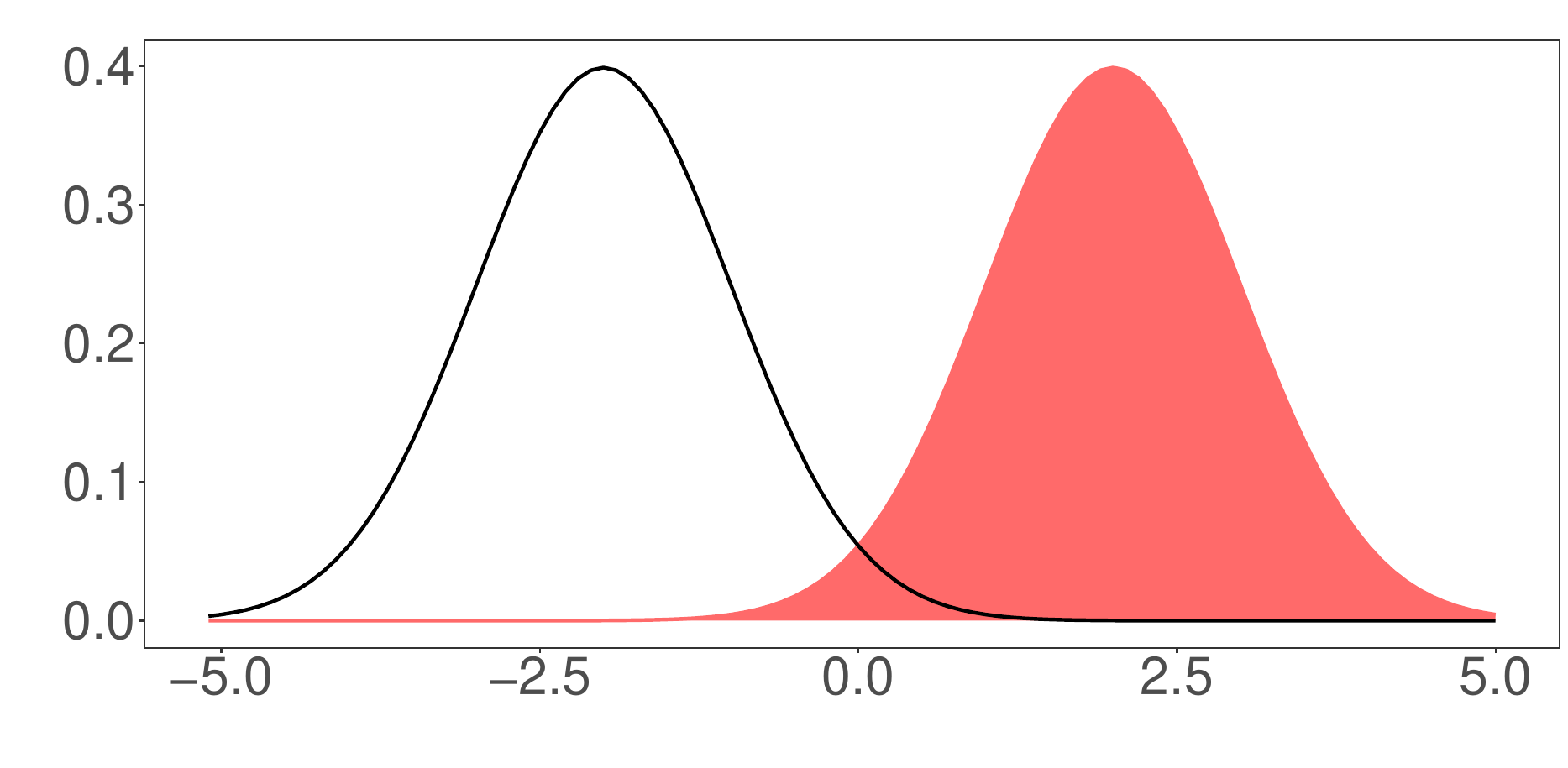} 
\label{fig: Binary_Signal_Distn_s2_observed}
}
\hfil
\subfloat[Informed demand $W^*(\,\cdot\,\vert s_1)$---Bull Spread]{
\includegraphics[width = 7cm, height = 2.6cm]{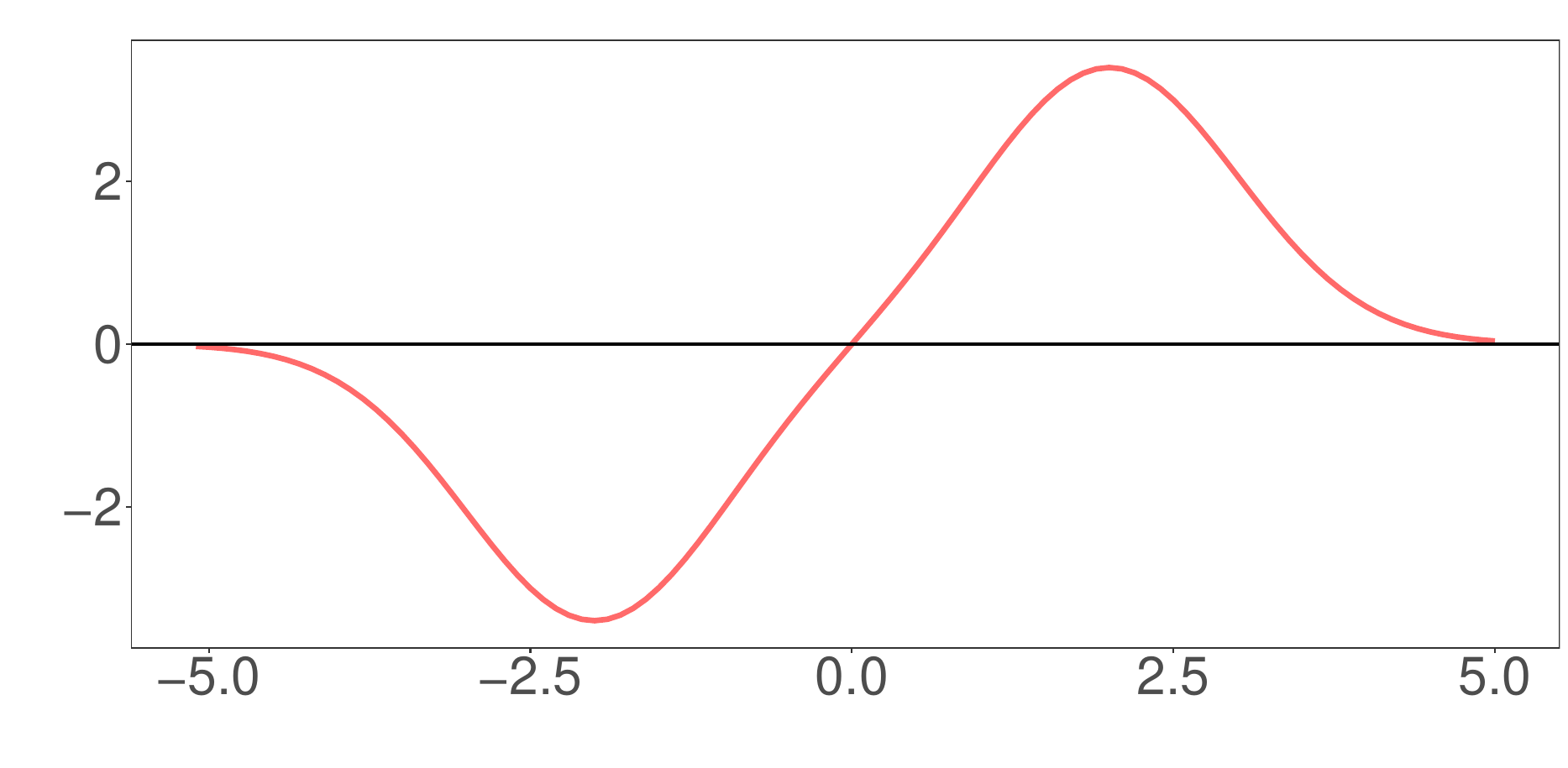} 
\label{fig: Binary_Signal_AD_demand_s2_observed}
}

\par\smallskip
\makebox[7cm][c]{\footnotesize ($x$-axis: state/option strike, $y$-axis: probability)}
\hfil
\makebox[7cm][c]{\footnotesize ($x$-axis: state/option strike, $y$-axis: payoff)}
\par\medskip

\subfloat[Low-mean signal $s_2$]{
\includegraphics[width = 7cm, height = 2.6cm]{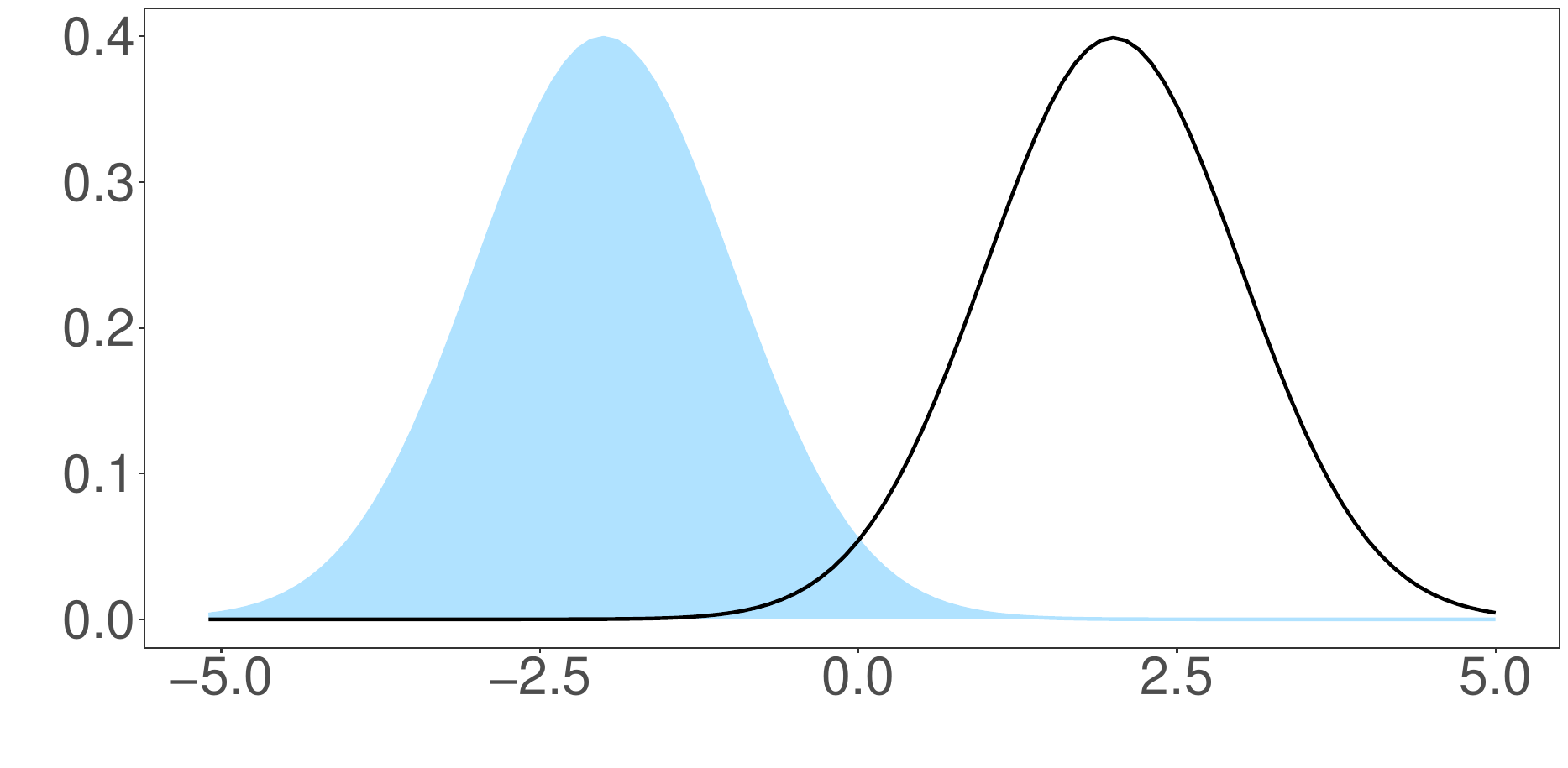} 
\label{fig: Binary_Signal_Distn_s1_observed}
}
\hfil
\subfloat[Informed demand $W^*(\,\cdot\,\vert s_2)$---Bear Spread]{
\includegraphics[width = 7cm, height = 2.6cm]{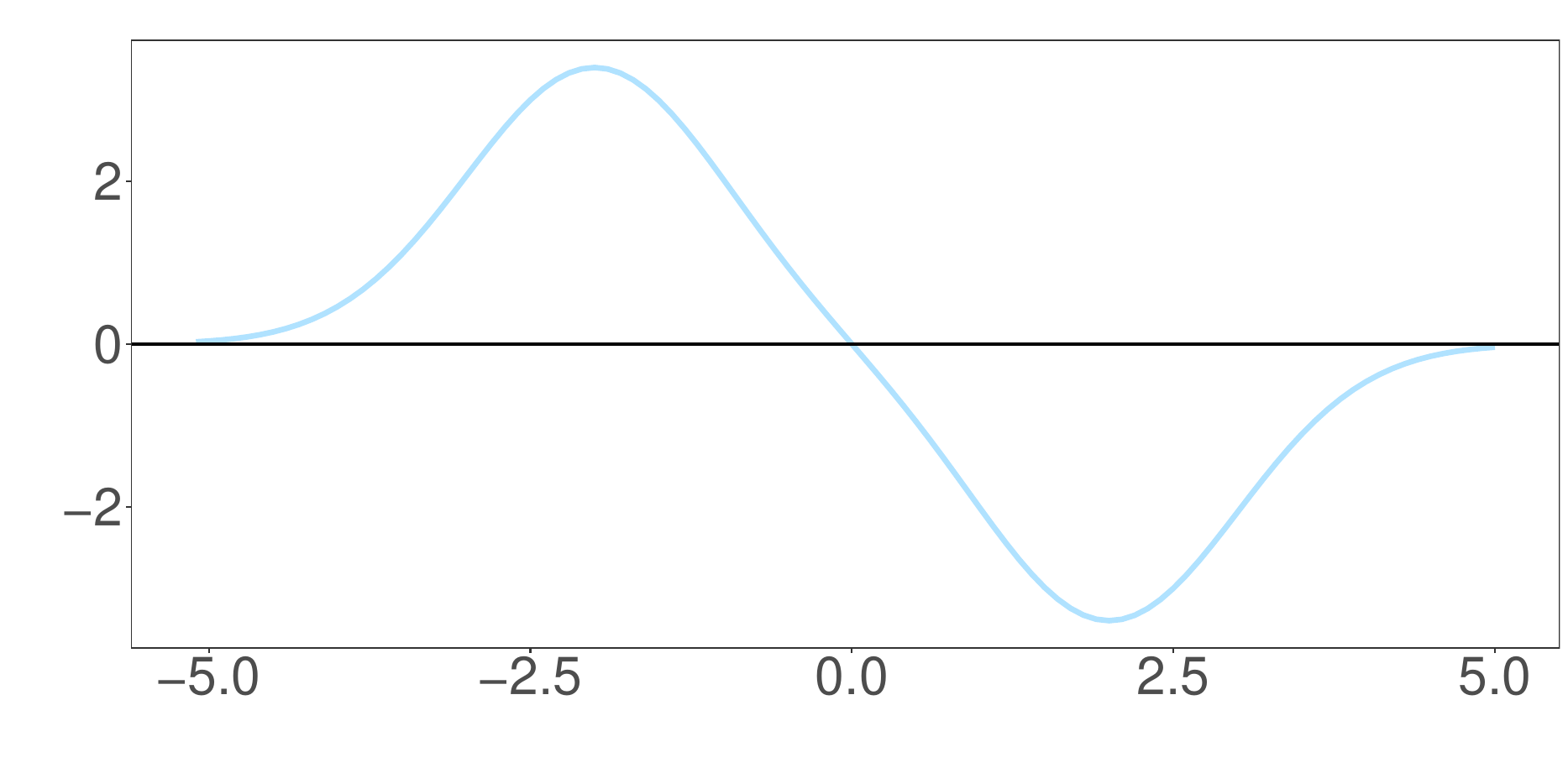} 
\label{fig: Binary_Signal_AD_demand_s1_observed}
}

\caption{
\footnotesize
\textbf{Bull/Bear Vertical Spreads (Example~\ref{example: Kyle with options}).}\\
Left column: the insider's private signal, indicated by the shaded distributions.  
Right column: the corresponding informed demands $W^*$.\\
Top row: high-mean signal $s_1$ and the associated \emph{bull spread}. Bottom row: low-mean signal $s_2$ and the associated \emph{bear spread}.
}
\label{fig: binary AD price discovery for mean}
\end{figure}

Even when private information is only about the mean---the \cite{kyle1985continuous} case---the availability of options reshapes informed demand into a nonlinear, multi-leg portfolio.
Example~\ref{example: Kyle with options} illustrates this.

\Needspace{6\baselineskip}
\begin{samepage}
\begin{example}\textbf{(Trading on the Mean: Bull/Bear Spreads)}
\label{example: Kyle with options}

Suppose the signal selects between two normal payoff distributions with a common variance and means $\mu_1>\mu_2$.
Private information therefore concerns the \emph{level} (expected payoff) of the underlying, as in \cite{kyle1985continuous}, but the insider can trade options.
In terms of the information flow~\eqref{eqn: info flow}, equilibrium maps this mean-shift signal into a standard \textbf{vertical spread} option portfolio.\footnote{See \cite{hull2003options} for standard definitions of vertical spreads.}

\medskip
\noindent\textsc{Figure guide.}
Panels~\ref{fig: Binary_Signal_Distn_s2_observed} and~\ref{fig: Binary_Signal_Distn_s1_observed} (left column) plot the two signal-implied payoff distributions.
Panels~\ref{fig: Binary_Signal_AD_demand_s2_observed} and~\ref{fig: Binary_Signal_AD_demand_s1_observed} (right column) plot the corresponding equilibrium informed demands $W^*(\,\cdot\,\mid s_k)$.
The right-column plots can be read strike-by-strike as holdings in the replicating option portfolio.

\medskip
\noindent\textsc{Portfolio intuition.}
A high-mean signal $s_1$ induces a \textbf{bull spread} (panel~\ref{fig: Binary_Signal_AD_demand_s2_observed}): a long--short vertical spread position that profits from upward revaluation while limiting exposure in far out-of-the-money states.
Conversely, a low-mean signal $s_2$ induces a \textbf{bear spread} (panel~\ref{fig: Binary_Signal_AD_demand_s1_observed}), concentrating downside exposure while economizing on extreme strikes that contribute little to expected profits under the signal-implied distribution.


\end{example}
\end{samepage}

\begin{figure}[h!]

\centering 
\subfloat[High volatility signal $s_1$]{ 
\includegraphics[width = 7cm, height = 2.6cm]{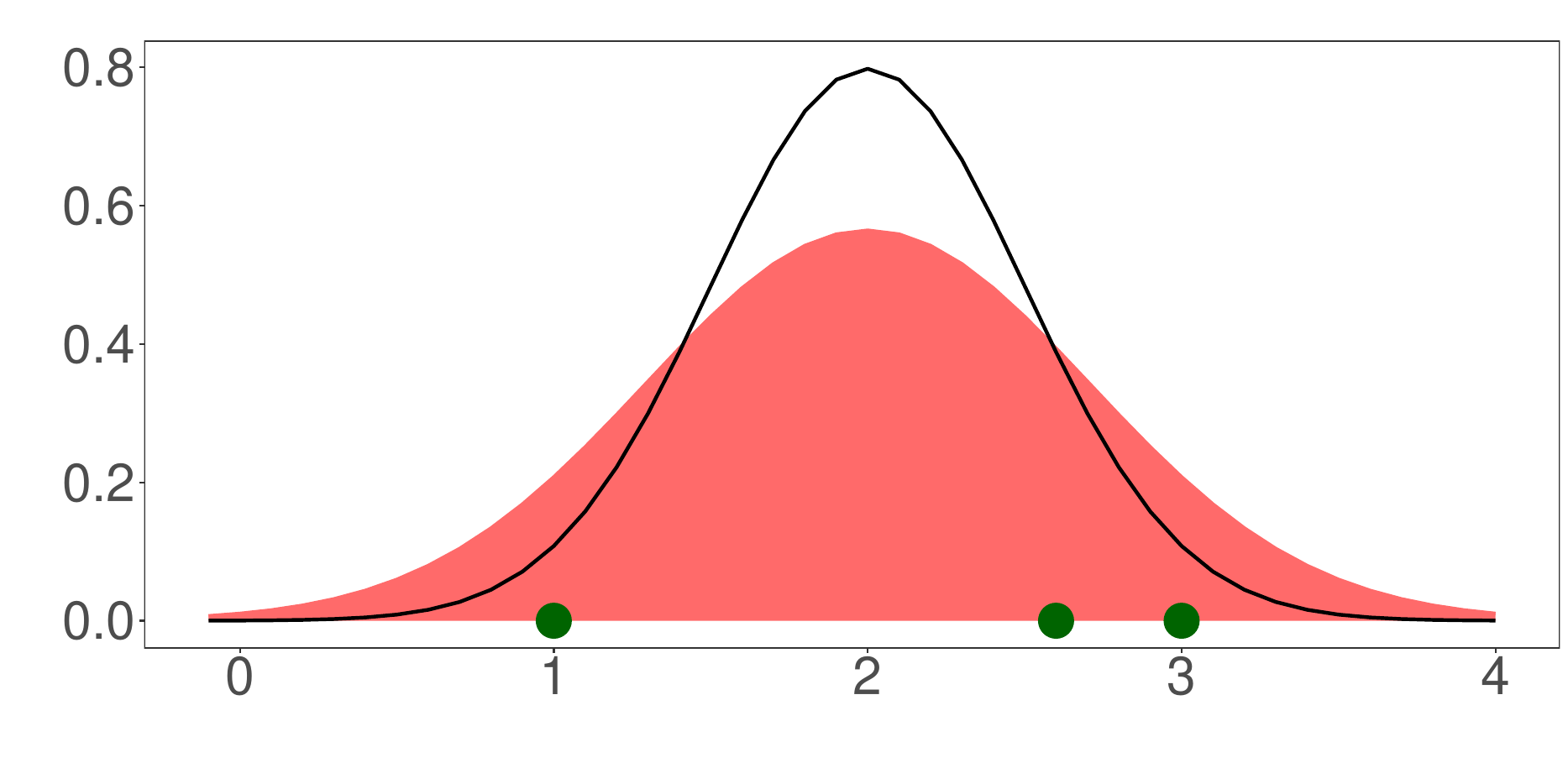} \label{fig: Binary_Signal_Distn_s2_observed, vol}
}
\hfil
\subfloat[Informed demand $W^*(\,\cdot\,\vert s_1)$---Straddle]{ 
\includegraphics[width = 7cm, height = 2.6cm]{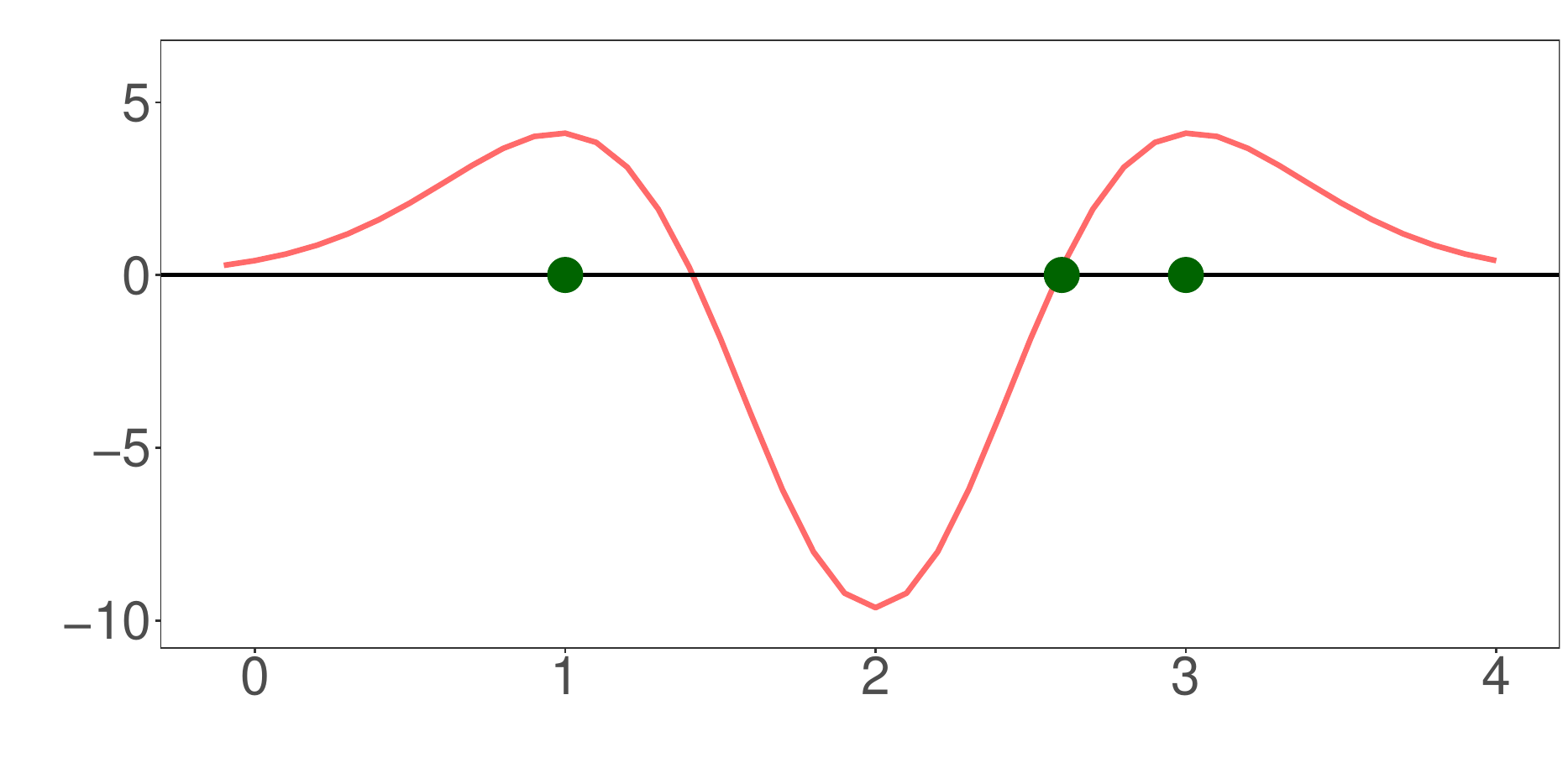} \label{fig: Binary_Signal_AD_demand_s2_observed, vol}
}

\centering 
\subfloat[Low volatility signal $s_2$]{
\includegraphics[width = 7cm, height = 2.6cm]{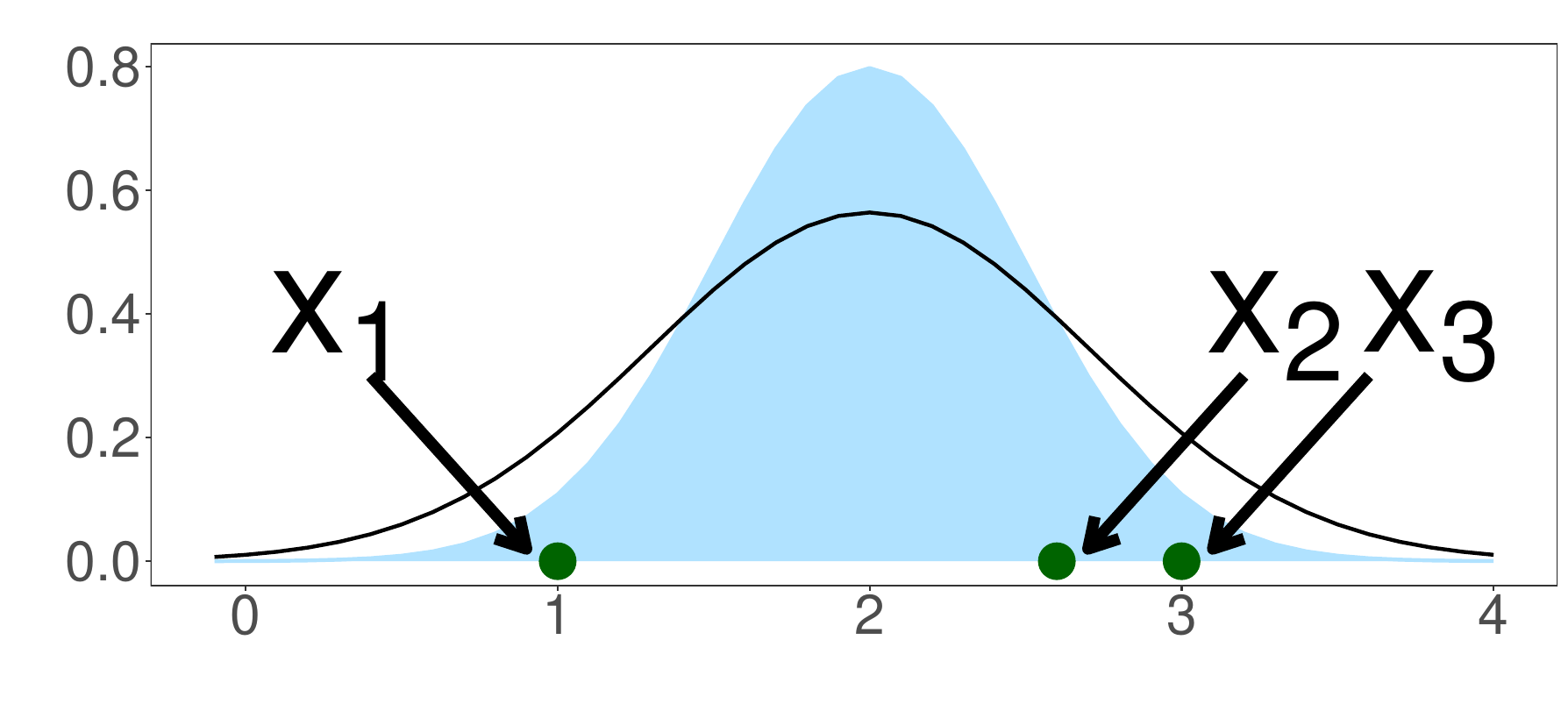} \label{fig: Binary_Signal_Distn_s1_observed, vol}
}
\hfil
\subfloat[Informed demand $W^*(\,\cdot\,\vert s_2)$---Butterfly]{
\includegraphics[width = 7cm, height = 2.6cm]{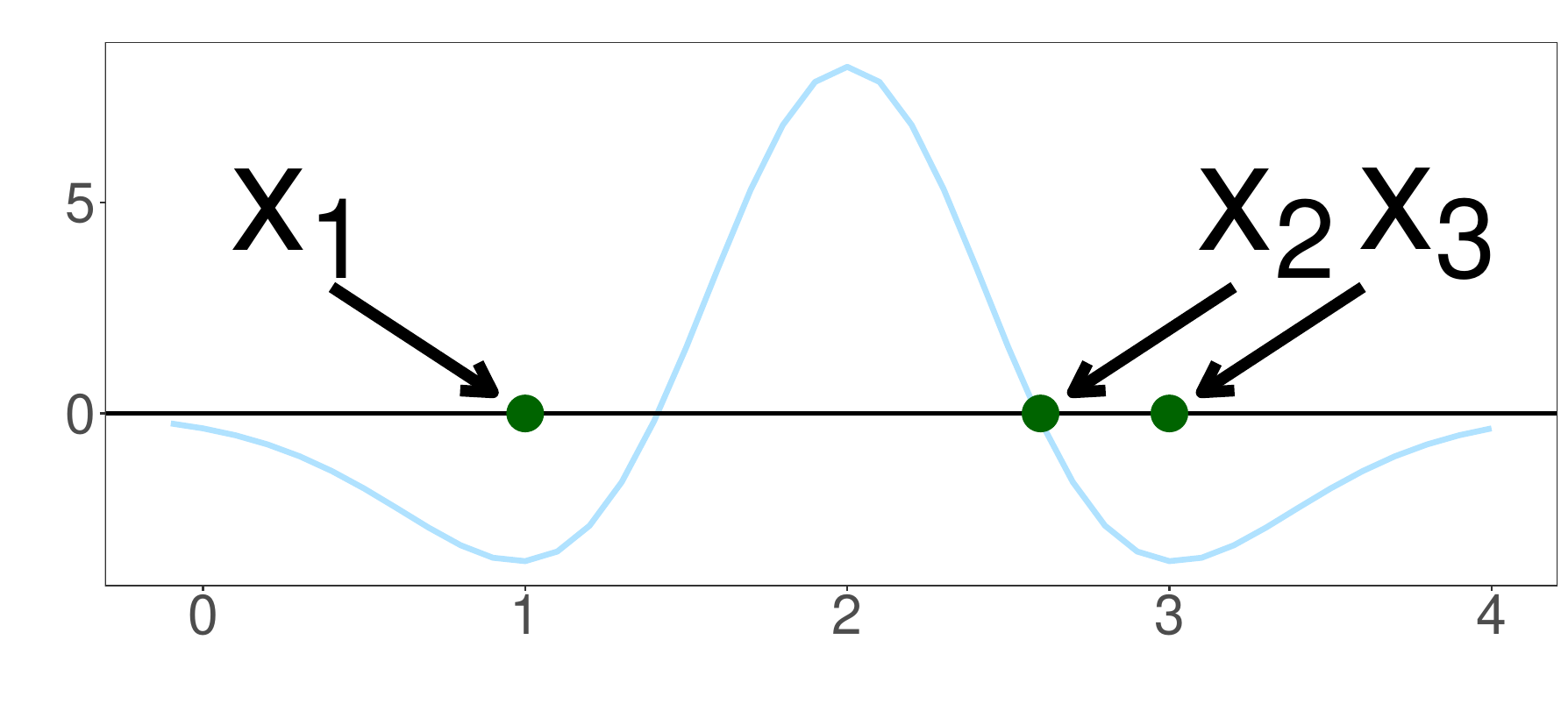} \label{fig: Binary_Signal_AD_demand_s1_observed, vol}
}

\captionsetup{singlelinecheck=off}
\caption{
{\footnotesize
$\;$\\
\textbf{Straddle/Butterfly (Example~\ref{example: vol straddle}).}\\
Left column: the insider's private signal, indicated by the shaded payoff distributions.  
Right column: the corresponding informed demands $W^*$.\\
Top row: high-volatility signal $s_1$ and the associated \emph{straddle}. Bottom row: low-volatility signal $s_2$ and the associated \emph{butterfly}. \vspace{1mm}\\
{\bf Price Impact.} Three states are indicated: $x_1$, $x_2$, and $x_3$.\\
$\bullet \, x_1$ and $x_3$: positive cross-price impact because they have positively correlated (in fact, identical) payoffs.\\
$\bullet \, x_2$: zero price impact on all securities because its payoff is constant (has zero variation) across signals. 
Thus, the informed demand for $x_2$ must be zero. See (b) and (d).
}
}
\label{fig: binary AD price discovery for vol, normal}
\end{figure}

Volatility information concerns shifts in probability mass between the center and the tails, so equilibrium informed demand naturally takes a center--versus--wings form, yielding straddle- and butterfly-type positions.
This is Example~\ref{example: vol straddle}.

\begin{example}\textbf{(Trading on Volatility: Straddle/Butterfly)}
\label{example: vol straddle}
\nopagebreak

Suppose the signal selects between two normal payoff distributions with a common mean and variances $\sigma_1^2>\sigma_2^2$.
Private information therefore concerns the \emph{volatility} of the payoff rather than its level.
In terms of the information flow~\eqref{eqn: info flow}, equilibrium maps this volatility signal into the option portfolios widely used in practice: a \textbf{straddle} when volatility is high, and a \textbf{butterfly} when volatility is low.\footnote{See \cite{hull2003options} for standard definitions of straddles and butterflies.}

\medskip
\noindent\textsc{Figure guide.}
Figure~\ref{fig: binary AD price discovery for vol, normal} summarizes the mapping from signal to portfolio choice.
Panels~\ref{fig: Binary_Signal_Distn_s2_observed, vol} and~\ref{fig: Binary_Signal_Distn_s1_observed, vol} (left column) plot the signal-implied payoff distributions.
Panels~\ref{fig: Binary_Signal_AD_demand_s2_observed, vol} and~\ref{fig: Binary_Signal_AD_demand_s1_observed, vol} (right column) plot the corresponding equilibrium informed demands.
As in the preceding example, the right-column plots can be read strike-by-strike as holdings in the replicating option portfolio.

\medskip
\noindent\textsc{Portfolio intuition.}
Under the high-volatility signal $s_1$, the distribution places relatively more weight on tail outcomes.
Accordingly, the informed demand loads on the wings and avoids the center (panel~\ref{fig: Binary_Signal_AD_demand_s2_observed, vol}), producing a \textbf{straddle}-type position---a wing-loaded long-volatility portfolio.
Conversely, under the low-volatility signal $s_2$ the payoff distribution is more concentrated near the mean.
Panel~\ref{fig: Binary_Signal_AD_demand_s1_observed, vol} shows that informed demand flips sign across strikes, producing a \textbf{butterfly} position: it concentrates exposure near the center and shorts wing payoffs that are unlikely to realize under low volatility.
\end{example}

\begin{figure}[htbp!]

 \centering 
 \subfloat[Right-skewed signal $s_1$]{
 \includegraphics[width = 7cm, height = 2.6cm]{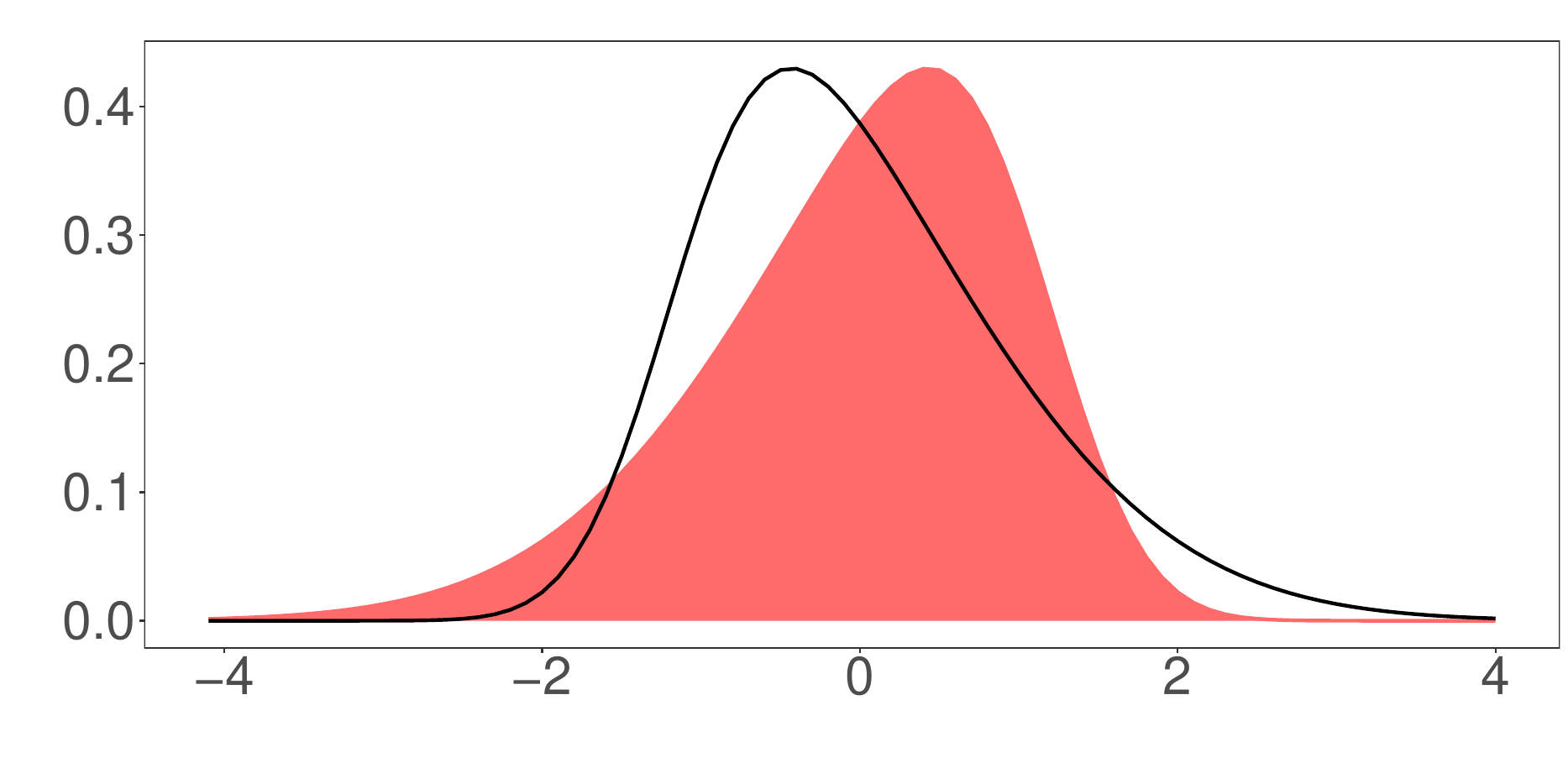} \label{fig: Binary_Signal_Distn_s2_observed, skew}
 }
 \hfil
 \subfloat[Informed demand $W^*(\,\cdot\,\vert s_1)$]{
 \includegraphics[width = 7cm, height = 2.6cm]{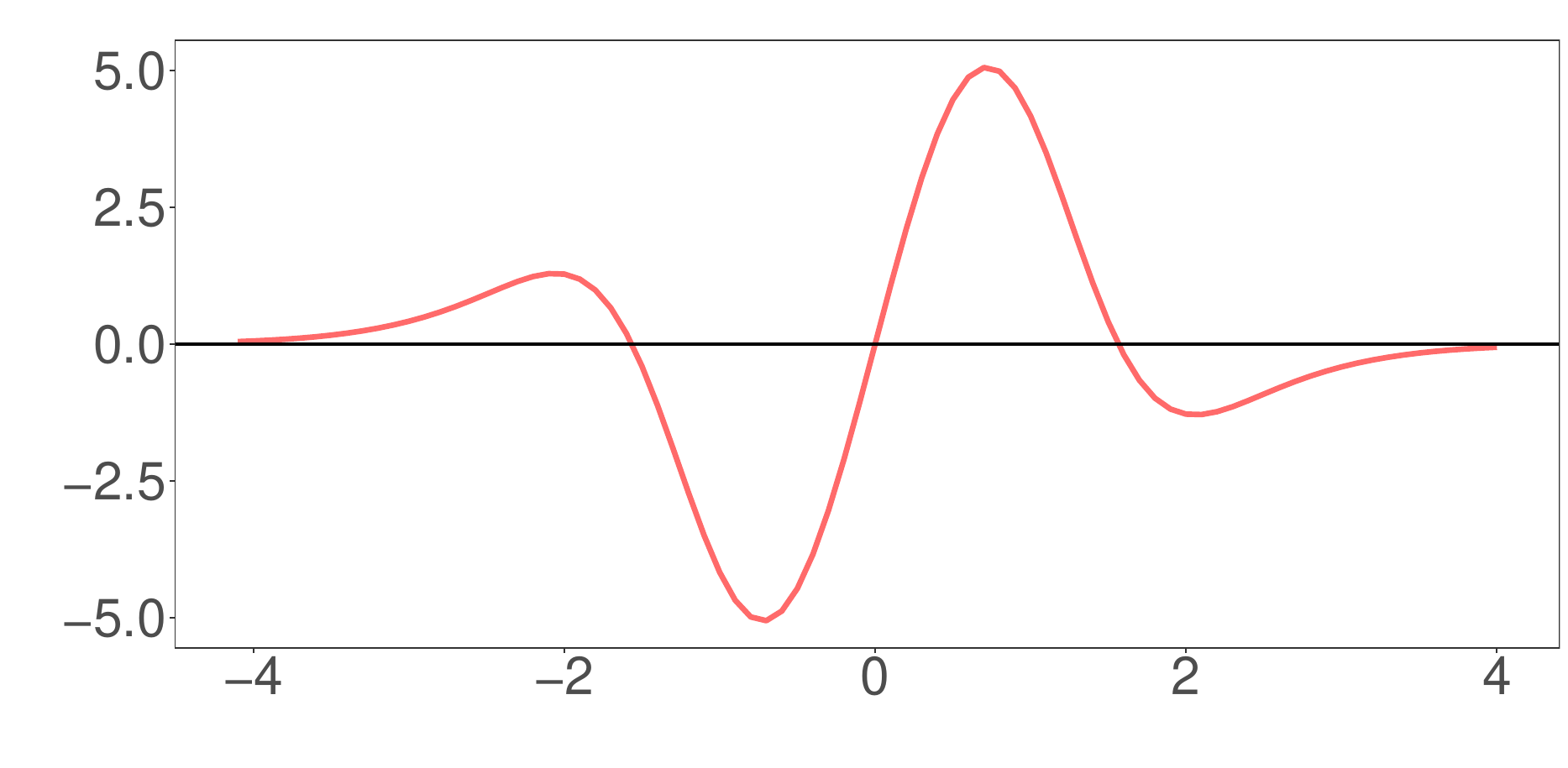} \label{fig: Binary_Signal_AD_demand_s2_observed, skew}
 }

 \centering 
 \subfloat[Left-skewed signal $s_2$]{
 \includegraphics[width = 7cm, height = 2.6cm]{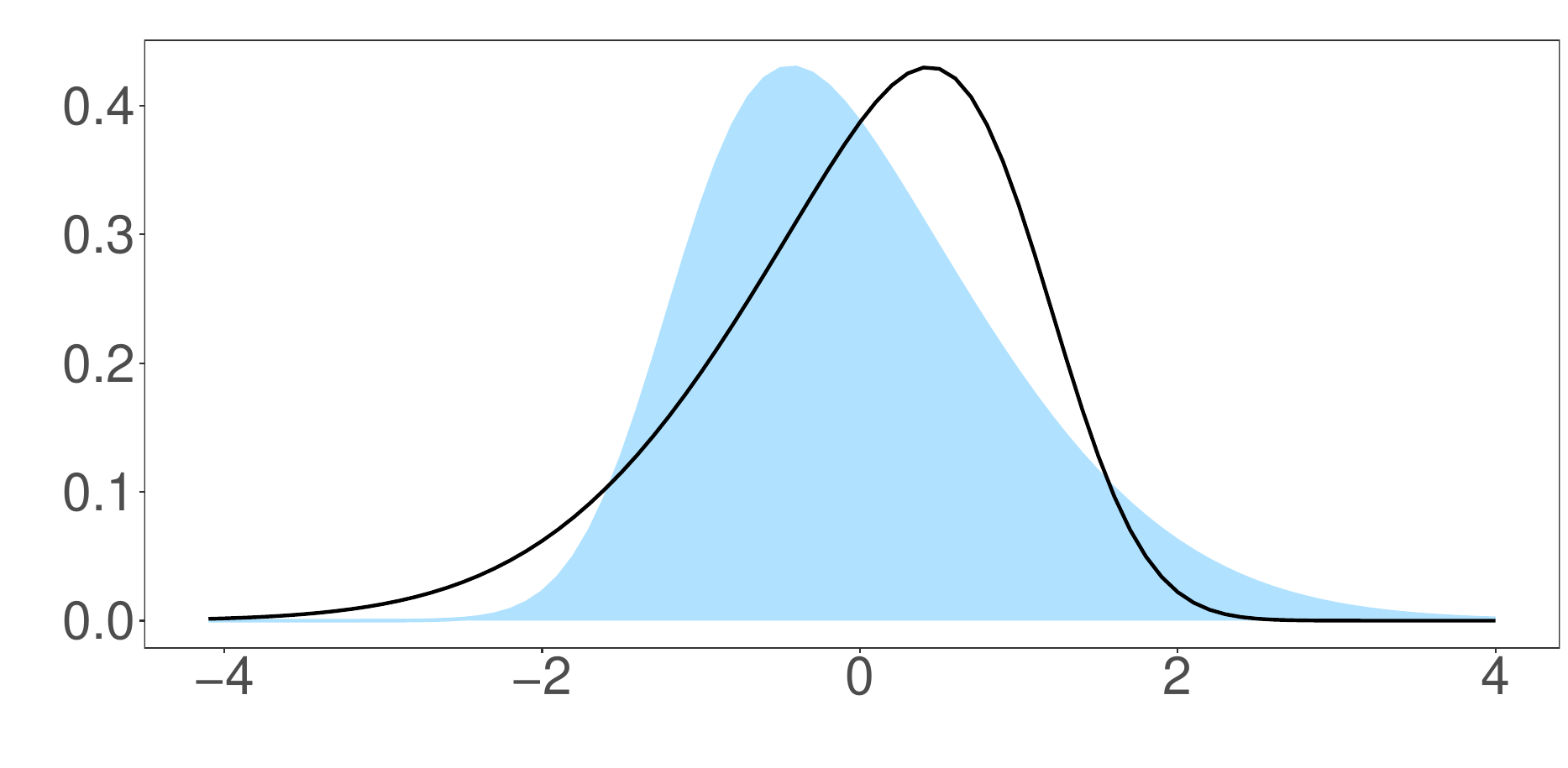} \label{fig: Binary_Signal_Distn_s1_observed, skew}
 }
 \hfil
 \subfloat[Informed demand $W^*(\,\cdot\,\vert s_2)$]{
 \includegraphics[width = 7cm, height = 2.6cm]{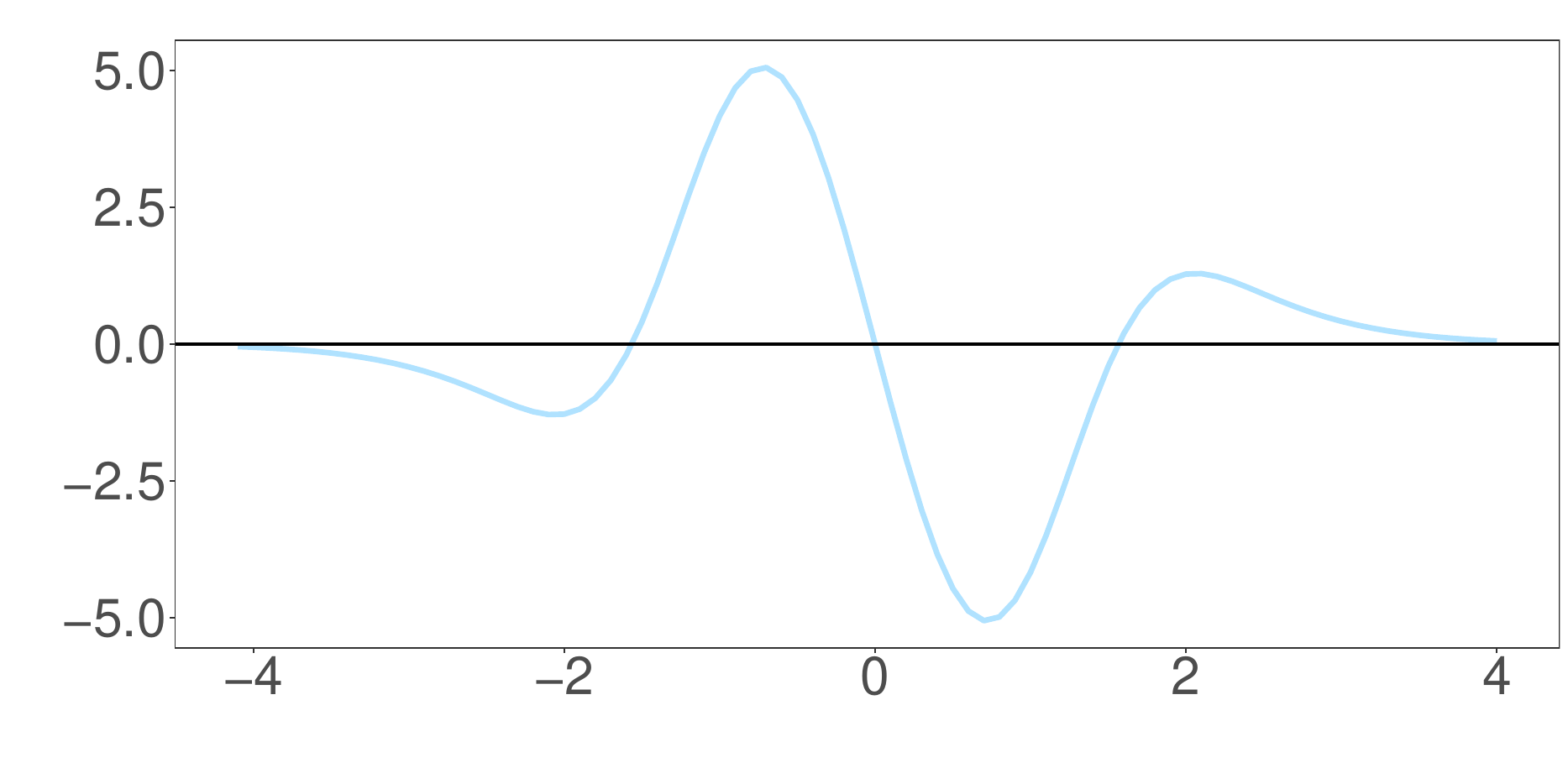} \label{fig: Binary_Signal_AD_demand_s1_observed, skew}
 }

\caption{
\footnotesize
\textbf{Ratio Spreads (Example~\ref{example: skewness}).}\\
Left column: signal-implied payoff distributions (shaded). 
Right column: corresponding informed demands $W^*$.\\
Top row: right-skewed signal. Bottom row: left-skewed signal.
}
\label{fig: binary AD price discovery for skewness}
\end{figure}

Skew is routinely traded in option markets—through ratio spreads and closely related structures.\footnote{See \cite{hull2003options,natenberg2014ovp,mcmillan2012options,vine2011options,overby2007optionsplaybook}.}
Option-implied skewness and skew premia are likewise important in the empirical literature.\footnote{See \cite{DennisMayhew2002,BakshiKapadiaMadan2003,ConradDittmarGhysels2013,BollenWhaley2004,KozhanNeubergerSchneider2013}.}
Example~\ref{example: skewness} shows that a pure skewness signal endogenously recovers this familiar skew trade.

\begin{example}\textbf{(Trading on Skewness: Ratio Spreads)}
\label{example: skewness}
\nopagebreak

Suppose the signal selects between two skew-normal payoff distributions that share the same mean and variance but differ in their skewness parameters, with $\alpha_1>\alpha_2$.
Private information therefore concerns \emph{asymmetry}---how probability mass shifts between left- and right-tail outcomes holding the first two moments fixed.
In equilibrium, this skewness signal maps into a standard skew trade: an option portfolio that tilts exposure toward the tail favored by the signal while offsetting exposure to the opposite tail.

\medskip
\noindent\textsc{Figure guide.}
Figure~\ref{fig: binary AD price discovery for skewness} summarizes the mapping from signal to portfolio choice.
Panels~\ref{fig: Binary_Signal_Distn_s2_observed, skew} and~\ref{fig: Binary_Signal_Distn_s1_observed, skew} (left column) plot the signal-implied payoff distributions.
Panels~\ref{fig: Binary_Signal_AD_demand_s2_observed, skew} and~\ref{fig: Binary_Signal_AD_demand_s1_observed, skew} (right column) plot the corresponding equilibrium informed demands.

\medskip
\noindent\textsc{Portfolio intuition.}
Under the right-skewed signal $s_1$, probability mass shifts toward high-payoff states (panel~\ref{fig: Binary_Signal_Distn_s2_observed, skew}).
Correspondingly, panel~\ref{fig: Binary_Signal_AD_demand_s2_observed, skew} loads on the right tail and takes the opposite side on the left tail, yielding the familiar \textbf{ratio spread} pattern.
Under the left-skewed signal $s_2$, the direction of asymmetry reverses (panel~\ref{fig: Binary_Signal_Distn_s1_observed, skew}), and equilibrium flips the position across strikes (panel~\ref{fig: Binary_Signal_AD_demand_s1_observed, skew}).

\end{example}

Beyond underscoring the practical scope of our model, the above examples also reinforce the point that derivatives are central to price discovery. 
When private information departs from the mean, it corresponds to a cross-strike payoff tilt that cannot be spanned by a single contract; cf.~Proposition~\ref{prop: options}(ii). 
Derivatives matter precisely because they span---and therefore can reveal---such tilts. 
The breadth of observed option strategies points to information that is rarely one-dimensional; mean-only restrictions are mainly methodological.
Our model accommodates that breadth directly.

\subsection{Price Impact}
\label{sec: Generalized Kyle's lambda and Cross Price Impact}

We next characterize the equilibrium price impact matrix $\Lambda$.
Its $(i,j)$-element $\Lambda_{i,j}$ is the equilibrium marginal effect of trades in $x_j$ on the price of $x_i$ (diagonals are within-market impact; off-diagonals are cross-impact).
Table~\ref{tab:key_equilibrium_quantities} recalls the notation used in this subsection.

Corollary~\ref{cor: def of price impact} maps what the market maker learns from order flow into a closed-form expression~\eqref{eqn: price impact y on x} for price impact.
Evaluating \eqref{eqn: price impact y on x} at the equilibrium informed demand $W^*$ yields the \emph{equilibrium expected} cross-price impact between securities $x_i$ and $x_j$:
\begin{equation} 
\label{eqn: lambda x y}
\Lambda_{i, j} = \Big(1- \frac{1}{K}\Big)\cdot \alpha^* \cdot \frac{1}{\sqrt{\sigma_i \sigma_j}} \cdot \mathbb{E}\!\left[ \Cov \Big( \eta(x_i \vert \,\cdot\,), \eta(x_j \vert \,\cdot\,)\,\big|\, \omega \Big) \right].
\end{equation}
Here, $\eta(x_i\mid \,\cdot\,)$ is the \emph{signal-by-signal} payoff profile of claim $x_i$.
Given order flow $\omega$, the market maker forms a posterior \emph{over signals}. Throughout the sequel, we write $\Cov(\cdots,\cdots\mid \omega)$ for the (conditional) covariance under that posterior.
The outer expectation $\mathbb{E}[\,\cdot\,]$ averages over order flow $\omega$ under its equilibrium distribution.
The \emph{realized} (order-flow-conditional) cross-price impact is obtained by dropping the outer expectation from \eqref{eqn: lambda x y}:
\begin{equation}
\label{eqn: realized cross price impact}
\Lambda_{i,j}(\omega)
=
\Big(1-\frac{1}{K}\Big) \cdot \alpha^* \cdot
\frac{1}{\sqrt{\sigma_i\sigma_j}}\cdot
\Cov\!\Big(\eta(x_i\mid\,\cdot\,),\,\eta(x_j\mid\,\cdot\,)\,\big|\,\omega\Big).
\end{equation}

\Needspace{5\baselineskip}
\paragraph{Complementary Factors of Price Impact}
Cross-price impact~\eqref{eqn: lambda x y} is the product of four intuitive factors:

\begin{itemize}

\item \textbf{\textit{Signal-tilt coefficient $\big(1-\frac{1}{K}\big)$}:}
This is the loading on the realized signal in Step~1 of the informed demand~\eqref{eqn: informed demand} portfolio recipe above.
\smallskip

\item \textbf{\textit{Equilibrium aggressiveness $\alpha^*$}:}
This is the overall scale of informed demand~\eqref{eqn: informed demand}---see Step~3.
If the insider trades more aggressively in equilibrium, a given marginal order is more likely to be information-driven, so prices respond one-to-one.
\smallskip

\item \textbf{\textit{Noise adjustment $\frac{1}{\sqrt{\sigma_i\sigma_j}}$:}}
This reflects the standard intuition that more noise trading makes order flow less informative, attenuating how much a trade moves prices across markets; cf.\ footnote~\ref{fn: varying noise intensity} in the toy setting of Section~\ref{sec: main intuition}.
\smallskip

\item \textbf{\textit{Information linkage}:}
The remaining quantity,
\[
\mathbb{E}\!\left[ \Cov \Big( \eta(x_i \vert \,\cdot\,), \eta(x_j \vert \,\cdot\,)\,\big|\, \omega \Big) \right],
\]
is the \emph{(posterior) payoff covariance} between securities $x_i$ and $x_j$---and thus captures whether the market maker’s inference from order flow tends to move their prices in the same direction.

\begin{itemize}[label=\(\circ\), leftmargin=1.6em, itemsep=1pt, topsep=2pt]

\item \textit{Positive covariance:} $x_i$ and $x_j$ are informational complements, i.e., information that points to higher payoffs for $x_j$ also points to higher payoffs for $x_i$.
Accordingly, buy pressure in one security lifts the price in the other---positive cross impact.

\item \textit{Negative covariance:} $x_i$ and $x_j$ are informational substitutes, i.e., information that favors $x_j$ tends to disfavor $x_i$.
Accordingly, buy pressure in one security depresses the price of the other---negative cross impact.

\item \textit{Zero covariance:} information about one security is payoff-irrelevant for the other---therefore cross impact is zero.

\end{itemize}

\end{itemize}

We already saw the information-linkage intuition expressed directly through payoff covariances in the toy model of Section~\ref{sec: main intuition}---when two claims load on the same underlying risk (e.g., the put--call legs of a straddle under a volatility signal), their prices move in tandem. 
Having established that this intuition carries over to the general setting, the next example provides a further concrete illustration.

\Needspace{2\baselineskip}
\begin{example}[Cross Impact: Positive, Negative, and Zero]
\label{example: generalized Kyle lambda}
\nopagebreak

$\;$
\begin{enumerate}[label=(\roman*), itemsep=6pt, topsep=4pt]

\item (\textit{Positive cross-price impact.})
If two securities have identical signal-contingent payoffs, order flow in either security reveals the same payoff-relevant information. 
A buy in one is therefore ``good news'' for the other, so $\Lambda_{i,j}>0$. In Figure~\ref{fig: Binary_Signal_Distn_s1_observed, vol}, the securities labeled $x_1$ and $x_3$ illustrate this case.

\item (\textit{Negative cross-price impact.})
If security $x_i$ pays off only under signal $s_{k_i}$ and $x_j$ only under a different signal $s_{k_j}\neq s_{k_i}$, then their cross-impact is negative: $\Lambda_{i,j}<0$.
Buy pressure in $x_i$ makes the market maker infer that $x_j$ is less likely to pay off (and vice versa).
More precisely, the sign of $\Lambda_{i,j}$ is determined by the (order-flow-conditional) posterior payoff covariance term in \eqref{eqn: realized cross price impact}, which is negative in this case:
\[
\Cov \Big(\eta(x_i\mid \,\cdot\,),\,\eta(x_j\mid\,\cdot\,)\,\big|\,\omega\Big)
=
-\eta(x_i\mid s_{k_i})\,\pi^*_1(s_{k_i}\mid\omega)\cdot
\eta(x_j\mid s_{k_j})\,\pi^*_1(s_{k_j}\mid\omega)
\;<\;0.
\]

\item (\textit{Zero cross-price impact.})
If $x_j$’s payoff is \emph{signal-invariant}—$\eta(x_j\vert \,\cdot \,)$ is constant—then trades in $x_j$ are not informative about the signal.
As a result, order flow in $x_j$ does not shift the market maker’s beliefs and therefore has zero price impact: $\Lambda_{i,j}=0$ for all $i$.
Figure~\ref{fig: Binary_Signal_Distn_s1_observed, vol} illustrates this case for the security labeled $x_2$.

\end{enumerate}
\end{example}

In equilibrium, a security generates zero cross-market price impact exactly when its informed demand is zero.
If order flow in $x_j$ never moves prices, then any non-zero informed trade for $x_j$ would create an arbitrage opportunity (Theorem~\ref{thm: informed trader 2}).
Conversely, when informed demand is zero, observed order flow is pure noise and therefore has no price impact.
Figures~\ref{fig: Binary_Signal_AD_demand_s2_observed, vol} and~\ref{fig: Binary_Signal_AD_demand_s1_observed, vol} illustrate this equivalence for the zero-impact security $x_2$ by showing $W^*(x_2\vert\,\cdot\,)=0$.

As an immediate corollary, cross-market price impact between two state-contingent claims---in particular, derivatives---is determined by the payoff-weighted aggregation of the price impact kernel~$\Lambda$.

\begin{corollary}
\label{cor: price impact between derivatives}
\textnormal{(Price Impact Between Derivatives)}
\nopagebreak
$\;$

Let $\varphi_1, \varphi_2 \colon X \rightarrow \mathbb{R}$ be state-contingent claims (e.g., options).
Then the cross-price impact between $\varphi_1$ and $\varphi_2$ is the payoff-weighted bilinear form
\begin{equation}
\label{eqn: derivatives cross price impact}
\sum_{i,j} \varphi_1(x_i)\,\varphi_2(x_j)\,\Lambda_{i,j}.
\end{equation}
\end{corollary}

Specializing Corollary~\ref{cor: price impact between derivatives} to the volatility environment in Example~\ref{example: vol straddle} yields a sharp comparative static---within-straddle cross impact is higher when volatility is higher: 

\begin{proposition}
\label{prop: cross price impact between call and put in a straddle and vol}
Under the specification of Example~\ref{example: vol straddle}, the put-call cross-price impact within a straddle is higher under the high-volatility signal than under the low-volatility signal.
\end{proposition}

In a long-volatility straddle, high put-call cross-price impact is the empirical footprint of volatility being learned from order flow---buy pressure in one leg raises the other price because both prices load on the same volatility-relevant states.
More generally, our framework allows for \emph{arbitrary} signals (beyond volatility or higher moments), so the same cross-market learning intuition delivers a broad set of testable predictions; see Section~\ref{sec: discussion subsection}.

\subsection{Information Efficiency of Prices}
\label{sec: AD price discovery}

Price impact captures the marginal price response to order flow.
Information efficiency is the complementary \emph{level} question---how much private information is reflected in prices.
In the single-asset setting of \cite{kyle1985continuous}, the equilibrium price moves halfway toward the insider's value.
We ask the same question in our richer Arrow-Debreu setting---with many traded claims and an arbitrary signal, what does the full cross-section of prices reveal?

Empirically, multi-market price discovery is typically studied through reduced-form decompositions of cointegrated price series with a common efficient component.\footnote{See \cite{Hasbrouck1995} and \cite{gonzalo1995estimation}.}
These approaches deliver a \emph{relative} ``who leads'' attribution across arbitrage-linked markets rather than an absolute, model-based measure of overall informativeness.
Derivative applications require first collapsing option prices to a common implied series.\footnote{See, for example, \cite{chakravarty2004informed}.}

Our contribution is complementary and structural: conditional on aggregate order flow $\omega$, equilibrium prices embed a posterior $\pi_1^*(\,\cdot\,\mid\omega)$ over signals that summarizes the 
informativeness of the \emph{entire} cross-section of prices (see Table~\ref{tab:key_equilibrium_quantities}).
We begin by characterizing this posterior in closed form.

\begin{proposition}
\label{prop: MM posterior finite S} 
Conditional on the realized signal $s_k$, let $q^{(k)}(\omega)$ denote the market maker’s equilibrium posterior belief over the
competing payoff-distribution ``stories'' (signals) after observing aggregate order flow $\omega$,
\[
q^{(k)}(\omega)
=\big(\pi_1^*(s_1\mid\omega),\ldots,\pi_1^*(s_K\mid\omega)\big)\in\Delta(S).
\]
Across realizations of $\omega$, $q^{(k)}(\omega)$ is logistic-normal:
\begin{equation}
\label{eqn: canonical posterior}
q^{(k)}(\omega)
\;\propto\;
\big(e^{Z_1},\,\ldots,\,e^{Z_{k-1}},\,e^{(\alpha^*)^2+Z_k},\,e^{Z_{k+1}},\,\ldots,\,e^{Z_K}\big)^T,
\qquad
\sum_{\ell=1}^K q^{(k)}_\ell(\omega)=1,
\end{equation}
where $Z \stackrel{d}{\sim} \mathcal{N}\!\big(0,(\alpha^*)^2\,{\bf Q}\big)$ is a Gaussian \textbf{evidence score} vector
that summarizes how strongly order flow favors each story.

\bigskip
\end{proposition}


Proposition~\ref{prop: MM posterior finite S} characterizes the belief-updating step in the information flow~\eqref{eqn: info flow}. 
The logistic-normal posterior~\eqref{eqn: canonical posterior} is especially transparent---it is summarized by the Gaussian evidence score vector $Z$.
Intuitively, $Z_\ell$ aggregates the components of order flow that ``look like'' signal $s_\ell$, while the covariance
$(\alpha^*)^2{\bf Q}$ governs which signals are easy to tell apart from order flow.
The realized signal $s_k$ receives an additional log-odds boost $(\alpha^*)^2$, reflecting that informed order flow
systematically tilts $\omega$ toward the true information type.

Given these posterior weights, the equilibrium state prices are simply the
posterior mixture of the signal-implied payoff distributions $\eta(\,\cdot\,\mid s_\ell)$.

\begin{corollary}[Equilibrium State Prices]
\label{cor: equil AD prices}
Suppose the realized signal is $s_k$. Conditional on aggregate order flow $\omega$, equilibrium state prices are the posterior-weighted average of the possible payoff distributions
\[
P^*(\cdot\mid\omega)
=\sum_{\ell=1}^K q_\ell^{(k)}(\omega)\,\eta(\cdot\mid s_\ell),
\]
where $q^{(k)}(\omega)$ is the posterior weight vector~\eqref{eqn: canonical posterior} characterized in Proposition~\ref{prop: MM posterior finite S}.
Averaging over equilibrium order flow yields the expected state prices 
\begin{equation}
\label{eqn: equil expected AD price}
\overline{P}^*(\cdot)\;\equiv\;\mathbb{E}\!\left[P^*(\cdot\mid\omega)\right]
=\sum_{\ell=1}^K \mathbb{E}\!\left[q_\ell^{(k)}(\omega)\right]\eta(\cdot\mid s_\ell).
\end{equation}
\end{corollary}

\paragraph{Information Efficiency Measure} 
The posterior-average representation in Corollary~\ref{cor: equil AD prices} immediately yields a canonical measure of informativeness for the full price cross-section:
\begin{equation}
\label{eqn: info efficiency measure}
\mathcal{I}\;\equiv\;\mathbb{E}\!\left[q^{(k)}_k(\omega)\right],
\end{equation}
i.e., the average weight placed on the true payoff distribution $\eta(\,\cdot\,\mid s_k)$ in the expected state prices~\eqref{eqn: equil expected AD price}.
With the uniform prior, $\mathcal{I}=1/K$ corresponds to no learning from order flow, while $\mathcal{I}\to 1$ corresponds to near-full revelation.

\begin{corollary}
\label{cor: invariance of info efficiency}
$\;$

(i) Overall information efficiency of state prices, measured by $\mathcal{I}$ in \eqref{eqn: info efficiency measure}, depends only on the number of possible signals $K=|S|$. 

(ii) In particular, it is invariant to the possible payoff distributions $\eta(\,\cdot\,\vert\,\cdot\,)$ and the contract-level noise-trade intensities $(\sigma_i)$.
\end{corollary}

Thus, the sole primitive governing \emph{aggregate} price informativeness is the dimension $K$ of latent private information---the amount of information incorporated into the \emph{pricing kernel} depends only on the number of distinct information types the market must distinguish, 
not on how those information types map into state payoffs, nor on contract-level noise.
Aggregate information efficiency depends only on the number of distinct payoff-distribution stories, not on their specific details.

The next proposition characterizes the comparative statics of information efficiency as the latent informational dimension expands.

\begin{figure}[htbp!]

\centering 
\stackunder[5pt]{
\scalebox{1}[0.85]{
\begin{tikzpicture}
\input{info_efficiency.tex} \label{fig: expected post prob plot}
\end{tikzpicture}
}
}{ {\footnotesize $x$-axis: number of signals, $y$-axis: information efficiency index {$\mathcal{I}$}}}
\caption{
{\footnotesize
{\bf Comparative Statics of Information Efficiency}
}
} 
\label{fig: expected post prob}
\end{figure}

\begin{proposition}
\label{prop: info efficiency}
As the signal space expands, so that the number of distinct payoff-distribution ``stories'' $K$ increases, aggregate price informativeness $\mathcal{I}$ declines, but with diminishing marginal losses. Equivalently, $\mathcal{I}(K)$ is decreasing and convex.
\end{proposition}

Figure~\ref{fig: expected post prob} visualizes Proposition~\ref{prop: info efficiency}---as $K$ increases, aggregate price informativeness $\mathcal{I}(K)$ declines, and the decline flattens (the curve is decreasing and convex).
This is intuitive and aligns with the earlier comparative statics for $\alpha^*$ (Figure~\ref{fig: endogenous constant}), which governs the overall scale of informed trading and, in turn, how sharply order flow separates across information types.
The market maker tries to infer which of $K$ competing information types/stories is driving order flow.
When $K$ is larger, the same order-flow realization admits more plausible stories, so posterior beliefs are more diffuse and the cross-section of prices is less informative about the true type.
The marginal reduction is diminishing because trading adjusts endogenously---higher-dimensional prior uncertainty increases the returns to informed trading. 
The resulting higher informed trading intensity (higher $\alpha^*$) makes order flow more diagnostic of the underlying information type, 
thereby partially offsetting the informativeness loss from having more competing stories.

\medskip
\paragraph{Kyle's Variance-Ratio Measure}
In \cite{kyle1985continuous}, private information is one-dimensional (a single asset's value), so the prior-posterior variance ratio of that asset price naturally quantifies how much uncertainty is resolved by trading.
In our setting, private information is inherently multi-dimensional: it concerns which payoff distribution (e.g., a higher-moment or tail-risk tilt) governs state prices, and it is learned from joint order flow across many claims.
Security-by-security variance ratios are therefore neither aggregative nor invariant---they depend on the particular set of traded claims used to span payoffs and do not capture the cross-claim information spillovers.
The posterior-weight measure \eqref{eqn: info efficiency measure} directly answers the level question posed here---how much weight the equilibrium pricing kernel places on the true payoff distribution.

\medskip

Corollary~\ref{cor: invariance of info efficiency}(ii) is an invariance principle for price discovery.
Changing $\eta(\,\cdot\, \vert \,\cdot\,)$ reshuffles which claims the informed trader loads on and where price impact shows up across markets, while changing $(\sigma_i)$ 
reshuffles which markets are liquid or informative.
But with full spanning, equilibrium trading endogenously rotates and scales across claims, so the price system incorporates information to the same extent.

This invariance is the price-discovery analogue of the textbook complete-markets invariance in risk sharing.
With complete markets, risk sharing depends only on \emph{aggregate} risk---not on how that risk is distributed across agents.
Likewise, once traded claims span payoffs, \emph{aggregate} price informativeness depends only on the \emph{aggregate} dimension of private uncertainty---not on which informed types have which specific knowledge about the payoff distribution.

Empirically, this invariance is also natural. Liquidity and trading activity often migrate across derivative markets---e.g., volume and depth shift across strikes and maturities, and between listed claims and synthetic substitutes. 
Such movements will predictably reallocate where price impact shows up and which contracts appear more informative.  
However, they need not change the informativeness of the entire cross-section of prices taken together---as measured, for example, by the out-of-sample forecasting power of objects implied by the full option surface for subsequent realized outcomes.
Accordingly, when this aggregate informativeness changes under such cross-contract reallocations, it points to frictions that impede the rotation of informed trading across contracts, such as trading or margin constraints, 
segmentation, limited participation, etc.

\section{Empirical Implications}
\label{sec: discussion subsection}

While our framework applies to general contingent-claim markets, we examine its immediate empirical implications through the lens of equity options.
Option markets are a natural venue for trading broad information about underlying payoffs, with rich price and order-flow data that have enabled extensive empirical work.

A large empirical literature shows that option prices and order flow embed incremental information about the distribution of the underlying payoff and, in many settings, contribute materially to 
price discovery. 
This evidence spans both return-directional information and higher-moment information.\footnote{For instance, option-implied variance forecasts subsequent 
realized variance, and the variance risk premium predicts expected returns (see \cite{BakshiKapadiaMadan2003,CarrWu2009,BollerslevTauchenZhou2009}); option-implied skewness has also been shown to predict cross-sectional returns (see \cite{ConradDittmarGhysels2013}).}
But in the absence of a formal equilibrium account of how information is transmitted across option contracts and strategies, empirical work has necessarily been somewhat ad hoc---different studies rely on different predictors and 
proxies, and they adopt different contract-specific sorting 
characteristics.\footnote{Examples include \cite{Hasbrouck1995}, which considers different price-discovery measures; \cite{BollenWhaley2004}, which uses net buying pressure to study movements in the implied-volatility surface; \cite{Muravyev2016}, which uses option order flow to measure informativeness; and \cite{KaeckVanKervelSeeger2022}, which studies cross-contract impact in index options.}

Our model provides an identifying mapping from option-market observables to latent informational dimensions, yielding equilibrium restrictions that can be taken directly to the data. 
Specifically, it identifies (i) the option portfolios that informed traders use to express a given informational dimension via the informed-demand~\eqref{eqn: informed demand}, and (ii) the equilibrium pattern of within- and 
cross-market price impacts via Corollary~\ref{cor: price impact between derivatives}. With this identification in hand, \emph{cross-price impact} becomes an estimand with a structural interpretation: it captures both \emph{what is being traded on} (which dimension of the payoff distribution) and 
\emph{how that information propagates across option markets}. This, in turn, imposes cross-equation restrictions on how impact loadings vary across strikes and strategies, and it generates predictions for how these impacts co-move with subsequent realizations of the 
targeted features of the return distribution. Some implications sharpen familiar empirical intuition, while others are less obvious ex ante.

\paragraph{Multivariate Price Impact}
Empirical work on adverse selection typically estimates \emph{within}-market price impact using univariate regressions.\footnote{
A common specification is the univariate regression
\begin{equation}
\Delta p_{t} \;=\; \lambda \,\omega_{t} \;+\; u_{t},
\end{equation}
where $\omega_{t}$ is a signed net order imbalance and $\lambda$ captures own-market impact.
See \cite{glosten1988estimating,lin1995trade,huang1997components,goyenko2009liquidity,hendershott2011does,makarov2020trading}.}
Our equilibrium framework shows that the relevant price-impact estimand is inherently multivariate.
Because informed traders optimally express information through \emph{portfolios} of options, equilibrium price discovery is governed by a system of contemporaneous cross-contract price impacts. 
These effects are summarized theoretically by the impact matrix $(\Lambda_{i,j})$ characterized in Corollary~\ref{cor: price impact between derivatives}, and empirically by its estimated counterpart $\widehat{\Lambda}$.
Contract-by-contract impact regressions can suffer from simultaneity bias when cross-market adjustments are excluded and correlated with own-contract order flow, potentially misattributing cross-market price responses to within-market liquidity or
informed trading.\footnote{The price impact matrix $\Lambda$ can be estimated by the multivariate  regression,
\begin{equation}
\label{eqn: Lambda estimation specification}
\Delta p_t \;=\; \Lambda\,\omega_t \;+\; u_t,
\end{equation}
where $\Delta p_t$ stacks changes in option prices across a given set of contracts and $\omega_t$ stacks the corresponding signed net order imbalances.
The diagonal entries of $\Lambda$ summarize within-contract impact, while off-diagonals $\Lambda_{i,j}$ measure \emph{cross-price impacts}---the contemporaneous response of contract $i$ to order flow in contract $j$.}

\paragraph{Higher-Moment Adverse Selection}
The cross-impact matrix $\Lambda$  expands the notion of adverse selection beyond mean-shift information to general features of the payoff distribution—such as volatility, asymmetry, and tail risk. 
Accordingly, its empirical counterpart $\widehat{\Lambda}$ is not merely a more complete measure of price impact; it is a model-implied adverse-selection statistic that identifies \emph{which} feature of the payoff distribution the market is learning about.

This connects to a large empirical literature that recovers volatility and higher moments from option prices, either through parametric mappings (implied volatility) or through model-free constructions of risk-neutral moments.\footnote{
See \cite{britten2000option,Jiang2005,DennisMayhew2002,BakshiMadan2000,BollenWhaley2004,XingZhangZhao2010,CremersWeinbaum2010,BollerslevTodorov2011,BollerslevTodorovXu2015}.}
While these reduced-form approaches can recover the risk-neutral moments embedded in option prices, inferring \emph{physical} moments requires equilibrium identifying restrictions.

In characterizing the equilibrium informational channel, we map distinct informational dimensions to distinct option portfolios, and those portfolios load into the option strike structure through systematic cross-impact patterns.
This mapping is not restricted to higher-moment information: it applies to any feature spanned by option payoffs, yielding feature-specific predictions for sign and concentration patterns in $\widehat{\Lambda}$ across strikes and strategies.

\paragraph{Predictability of Higher Moments}
We now elaborate on these restrictions in the empirically salient case of higher-moment predictability, starting with volatility.
Intuitively, volatility-related information should load symmetrically on volatility-spanning positions.
More precisely, higher cross-price impact between the call--put pair in a straddle indicates a greater degree of adverse selection about volatility across option markets.
In equilibrium, this intensifies the extent to which current option prices incorporate volatility-related private information, implying that periods of high within-straddle cross impact should be followed by upward revisions in realized volatility.
Proposition~\ref{prop: cross price impact between call and put in a straddle and vol} therefore translates directly to the following empirical hypothesis:

\begin{hypothesis}
\label{hypo: vol}
Higher cross-price impact across the legs of a straddle~(resp.~butterfly) predicts an increase~(resp.~decrease) in the volatility of the underlying return.\footnote{Hypothesis~\ref{hypo: vol} can be tested as follows. First estimate $\widehat{\Lambda}$ from the multivariate impact system \eqref{eqn: Lambda estimation specification}. 
Next construct a straddle-based cross-impact measure. For a straddle at strike $K$ and maturity $T$, a natural choice is the symmetrized within-straddle cross impact (and analogously for a butterfly-based measure),
\[
\widehat{CI}^{\text{strad}}_{t}(K,T) =  \widehat{\Lambda}_{C(K,T),P(K,T)}+\widehat{\Lambda}_{P(K,T),C(K,T)}.
\] 
The hypothesis can then be tested via a predictive regression of realized volatility over horizon $h$ on the lagged cross-impact measure,
\[
RV_{t,t+h}=\alpha+\beta\,\widehat{CI}^{\text{strad}}_{t}(K,T)+\Gamma'X_t+\varepsilon_{t+h},
\]
with $\beta>0$ for straddles and $\beta<0$ for butterflies as implied by Proposition~\ref{prop: cross price impact between call and put in a straddle and vol}.}
\medskip
\end{hypothesis}

The volatility case is a concrete illustration of a broader implication for higher-moment predictability.
Combining the informed-demand characterization~\eqref{eqn: informed demand} with the cross-impact formula in Corollary~\ref{cor: price impact between derivatives}, the theory delivers a portfolio-specific forecasting restriction: when informed traders trade a 
moment-spanning option portfolio, the relevant adverse-selection statistic is the cross-price impact between the portfolio’s legs.
In Example~\ref{example: skewness}, for instance, higher cross-price impact between the long--short option positions used to trade right-skewness predicts an increase in the skewness of the underlying return.
This logic is consistent with evidence that option-implied skewness can be informative about return asymmetries (e.g., \citet{ConradDittmarGhysels2013}), but existing empirical work typically proceeds through reduced-form moment proxies rather than systematic  
predictions about \emph{where} information enters and \emph{how} it propagates across option markets.
We have the following general empirical hypothesis:

\begin{hypothesis}
\label{hypo: beyond vol}
Cross-price impact between the option legs prescribed by the informed-demand formula~\eqref{eqn: informed demand} predicts the corresponding targeted higher moment(s) of the underlying return.\footnote{Hypothesis~\ref{hypo: beyond vol} can be tested analogously. 
After estimating $\widehat{\Lambda}$ from the multivariate impact system~\eqref{eqn: Lambda estimation specification}, form the moment-targeting option portfolio prescribed by the informed-demand formula~\eqref{eqn: informed demand}~(with legs indexed by $\mathcal{L}^{(m)}$) and 
compute the within-portfolio cross-impact statistic,
\[
\widehat{CI}^{(m)}_{t} = \!\!\sum_{\substack{a,b\in \mathcal{L}^{(m)}\\ a\neq b}}\! w^{(m)}_{a}\Big(\widehat{\Lambda}_{a,b}+\widehat{\Lambda}_{b,a}\Big)w^{(m)}_{b},
\]
where $w^{(m)}$ are the theory-implied leg weights. The hypothesis can then be tested via a predictive regression of the corresponding realized moment over horizon $h$ on the lagged cross-impact statistic,
\[
RM^{(m)}_{t,t+h}=\alpha+\beta\,\widehat{CI}^{(m)}_{t}+\Gamma'X_t+\varepsilon_{t+h},
\]
with $\beta$ signed according to whether the prescribed portfolio is long or short the targeted moment.}
\end{hypothesis}

Under Hypothesis~\ref{hypo: beyond vol}, the cross-price impact between the theory-prescribed option legs is not merely a liquidity object; it is a structural measure of higher-moment adverse selection. 
In particular, when cross-impact is elevated in the direction implied by the informed-demand~\eqref{eqn: informed demand}, it signals a temporary mispricing of the targeted moment that subsequently corrects as price discovery unfolds. 
The corresponding prescribed strategy should then earn positive profits as that information is incorporated into option prices. 
Conversely, rejection of Hypothesis~\ref{hypo: beyond vol} for certain moments would indicate that option markets are already informationally efficient along those dimensions.

A natural empirical question is how strongly this mechanism operates in practice. This, in turn, determines which features of the underlying payoff---and at what frequencies---are efficiently incorporated by option markets, thereby providing a structural lens for empirical work that relies on implied moments and option-based predictors.

\Needspace{2.5\baselineskip}
\begin{question}
\label{question: higher moment options price discovery}
Which moments and over which horizons allow the prescribed strategies in Hypothesis~\ref{hypo: beyond vol} to earn positive profits?
Equivalently, to what extent does higher-moment price discovery occur across option markets?
\end{question}

\paragraph{Cross-Section of Option Returns}

The cross-section of option returns remains a puzzle (see \cite{bali2013does}, \cite{cao2013cross}, \cite{christoffersen2018illiquidity}, and \cite{zhan2022option}). 
From the perspective of our framework, the mechanism in Hypothesis~\ref{hypo: beyond vol} translates into asset-pricing implications: 
if cross-price impact identifies which higher-moment information is being incorporated and how quickly, then heterogeneity in cross-impact patterns should translate into systematic differences in option expected returns. 
Intuitively, contracts (or prescribed option portfolios) whose cross impacts indicate stronger adverse selection are precisely those for which prices embed more adjustment and/or greater exposure to information shocks; either channel can generate a systematic component in average returns that a factor should capture.
This leads to the following conjecture:

\begin{conjecture}
\label{conjec: systematic factors}
\nopagebreak
Systematic factors for the cross-section of option returns can be constructed from the prescribed option portfolios that target higher moments and earn positive returns.\footnote{Conjecture~\ref{conjec: systematic factors} can be taken to the data as follows. 
First estimate $\widehat{\Lambda}$ from the multivariate impact system \eqref{eqn: Lambda estimation specification}. 
Next, for each date $t$ (and within fixed maturity/moneyness buckets), form the theory-prescribed moment-targeting option portfolio 
from the informed-demand formula~\eqref{eqn: informed demand} for each underlying $i$, with legs indexed by $\mathcal{L}^{(m)}_{i,t}$ and 
theory-implied weights $w^{(m)}_{i,t}$. 
Construct the corresponding within-portfolio cross-impact sorting characteristic
\[
\widehat{CI}^{(m)}_{i,t}\equiv \!\!\sum_{\substack{a,b\in \mathcal{L}^{(m)}_{i,t}\\ a\neq b}}\! w^{(m)}_{a,i,t}\Big(\widehat{\Lambda}_{a,b}+\widehat{\Lambda}_{b,a}\Big)w^{(m)}_{b,i,t},
\]
and compute the subsequent portfolio return $r^{(m)}_{i,t+1}$.
Sort the cross-section of prescribed portfolios into quantiles based on $\widehat{CI}^{(m)}_{i,t}$ and form the factor-mimicking portfolio as a high-minus-low spread,
\[
F^{(m)}_{t+1}\equiv \overline{r}^{(m)}_{H,t+1}-\overline{r}^{(m)}_{L,t+1},
\]
where $H$ and $L$ denote the top and bottom sorting bins. The conjecture predicts $\mathbb{E}[F^{(m)}_{t+1}]>0$ for informational dimensions $m$ along which the prescribed strategies earn positive returns. 
Finally, the pricing content of $F^{(m)}$ can be evaluated using standard option-asset-pricing tests, e.g.,
Fama-MacBeth cross-sectional regressions of average returns on estimated factor loadings, with the conjecture implying that exposure to $F^{(m)}$ should help explain cross-sectional variation in expected option returns and reduce pricing errors.}
\end{conjecture}

Conjecture~\ref{conjec: systematic factors} thus offers candidate option-return factors rooted in information aggregation across contracts. 
These candidate factors are distinct from (though complementary to) the contract- or stock-level characteristics considered in the existing literature.

\begin{figure}[h!]
\centering
\subfloat[$x$-axis: log-moneyness, $y$-axis: implied vol]{
\scalebox{1}[0.85]{
\begin{tikzpicture}
\input{insider_smile.tex} \label{fig: one smile}
\end{tikzpicture}
}
}
\hfil
\subfloat[$x$-axis: log-moneyness, $y$-axis: implied vol]{
\scalebox{1}[0.85]{
\begin{tikzpicture}
\input{insider_smile_multiple.tex} \label{fig: multiple smiles}
\end{tikzpicture}
}
}

\caption{
{\footnotesize
{\bf Insider-Induced Volatility Smile}\\
(a) displays the insider-induced implied volatility smile across option strikes under Example~\ref{example: vol straddle}.\\ 
(b) displays the insider-induced volatility smiles for different values of true volatility: $0.25$ (solid line, same as panel~(a)), $0.3$ (dashed line), and $0.35$ (dotted line).\vspace{1mm}\\
{\bf Computation.} For an order flow realization $\omega$, we use the pricing kernel $P^*(\, \cdot \, \vert \omega ; s_1)$ to compute the equilibrium option prices across strikes $K$. 
Based on these option prices, we compute the implied volatility $\sigma_{\scriptscriptstyle IV}(\omega, K)$ for each strike $K$.
We repeat this for $1,000$ realizations of $\omega$ and plot the average implied-volatility curve over log-moneyness. 
} 
}
\label{fig: insider smile}
\end{figure}

\paragraph{Insider-Induced Volatility Smile}
\label{sec: insider smile}

The volatility smile is among the most robust empirical regularities in option markets, yet much of the existing literature treats it as an \emph{input} to be fitted by reduced-form specifications, effectively taking as primitive an object that is induced by equilibrium option prices.
Our framework instead derives the smile from strategic trading and cross-market information transmission, providing the first explanation of the smile as a natural equilibrium \emph{outcome}.

Example~\ref{example: vol straddle} provides a natural setting to study informed volatility trading and its implications for the implied-volatility curve.
When an insider trades on high volatility, the resulting equilibrium implied volatilities reproduce the familiar smirk/skew toward lower strikes (see Figure~\ref{fig: one smile}).
As volatility information is expressed through trading, joint option supply schedules aggregate that information across strikes through cross-price impact, endogenously shaping the implied-volatility curve into a smile.

The mechanism is straightforward. For a given volatility state, demand shifts toward option positions that load on large price moves, raising option prices and implied volatilities in the put and call wings.
Cross-strike price impact then propagates this pressure across the curve: because payoffs at deep in-the-money and deep out-of-the-money strikes are positively correlated (as illustrated by strikes $x_1$ and $x_3$ in Figure~\ref{fig: binary AD price discovery for vol, normal}),
order flow in one wing raises prices in the other wing as well.
Moreover, these cross-wing spillovers intensify farther from at-the-money, where payoff covariance is higher.
The result is the characteristic convexity of the implied-volatility curve---a smile that emerges from the aggregation of volatility information across strikes.

Across volatility states, the same mechanism implies a clear comparative static: higher volatility leads to a more convex smile (see Figure~\ref{fig: multiple smiles}).
The reason is that higher volatility increases the covariance between payoffs at deep in-the-money and deep out-of-the-money strikes, thereby amplifying cross-strike price impact.
As a result, smile curvature rises with the intensity of volatility information being aggregated across strikes.
Variation in smile curvature is thus the observable footprint of that aggregation.\footnote{This comparative static is again testable. Fix a maturity $T$. For each day $t$, estimate $\widehat{\Lambda}_t(T)$ from \eqref{eqn: Lambda estimation specification} and construct a cross-wing impact measure $\widehat{CI}^{\mathrm{wing}}_t(T)$, 
for example, as the average cross-impact between the left and right wings. The model predicts that smile curvature at maturity $T$ is increasing in $\widehat{CI}^{\mathrm{wing}}_t(T)$ and is higher in high-volatility states.}

\section{Conclusion}
\label{sec: conclusion}

Building on \cite{arrow1954existence} and \cite{kyle1985continuous}, we develop a general model of price discovery across contingent-claim markets.
In the same spirit as \cite{kyle1985continuous}---which rationalizes intuitive informed-trading behavior in the single-asset case---our Arrow--Debreu generalization elevates that equilibrium rationale to settings in which many contingent claims trade jointly.
The result is a constructive and tractable equilibrium account of observed market practices and empirical regularities across derivative markets, along with new, testable predictions.

The tractability and generality of our model invite natural follow-on questions.
How does risk aversion reshape informed portfolios and the pricing kernel, and how does asymmetric information distort risk sharing across contingent claims?
In a dynamic setting, how do information spillovers propagate across strikes and maturities?
Beyond equity options, the same approach can be applied to futures and forwards, credit derivatives, and interest-rate derivatives, where order-flow-driven prices map into term-structure, default-intensity, and correlation.
We hope to pursue these and related questions in future work.

\newpage

\appendix

\setcounter{section}{0}
\setcounter{theorem}{0}

\makeatletter

\renewcommand{\thetheorem}{\thesection.\arabic{theorem}}
\renewcommand{\theequation}{\thesection.\arabic{equation}}

\section{Appendix}

\setcounter{equation}{0}

\subsection{Model Assumptions}
\label{sec: assumptions}

The formal assumptions of the general model are as follows.

\begin{assumption}
\label{assumption: MM's posterior}
$\;$

\nopagebreak
(i) Possible signals lie in the probability space $(S, \pi_0)$ where $S = \{s_1, \cdots, s_K\}$, and the probability measure $\pi_0$ is the market maker's prior. 

(ii) Possible realizations of order flow across states lie in the measurable space $(\Omega, \mathcal{F})$, where $\Omega = \mathbb{R}^{\infty}$ (the set of countable sequences) and 
$\mathcal{F}$ is the Borel $\sigma$-field generated by the coordinate functions 
$$
\omega = (\omega_i) \mapsto \omega_j, \; \Omega \rightarrow \mathbb{R}, \;\; \mbox{$j = 1, 2, \cdots$}.
$$

(iii) The probability measure $\mathbb{P}_{\! \scriptscriptstyle 0}$ on $(\Omega, \mathcal{F}, \mathbb{P}_{\! \scriptscriptstyle 0})$ specifies the canonical process $\omega  \mapsto \omega_i$, $i = 1, 2, \cdots$, as
the stochastic sequence 
$$
\varepsilon_i \! \stackrel{i.i.d.}{\sim} \! \mathcal{N}(0, \sigma_i^2) \;\; \mbox{$i = 1, 2, \cdots$}.
$$ 
In other words, $\mathbb{P}_{\! \scriptscriptstyle 0}$ specifies the probability law of noise-only order flow $\omega$ when insider demand is zero across markets.

(iv) The insider's AD portfolio $W \colon X \rightarrow \mathbb{R}$ satisfies 
$$
\exp \left( \frac12 \sum\limits_{i = 1}^{\infty} \left( \frac{W(x_i)}{\sigma_i} \right)^2 \right) < \infty.
$$

\end{assumption}

\subsection{Proof of Theorem~\ref{thm: Market Maker's Posterior}}

We recall the standard (discrete version of) Girsanov's Theorem \cite[Chapter V, Section 3]{shiryaev1999essentials}: Under Assumption~\ref{assumption: MM's posterior}, let $W \colon X \rightarrow \mathbb{R}$ be an insider portfolio 
and let $\mathbb{P}_{\!{\scriptscriptstyle W}}$ be the probability measure on $(\Omega, \mathcal{F})$ defined by the Radon-Nikodym density
\begin{equation}
\label{eqn: Radon-Nikodym density, omega by omega}
\frac{\mathrm{d} \mathbb{P}_{\scriptscriptstyle W}}{ \mathrm{d} \mathbb{P}_{\! \scriptscriptstyle 0}} = \exp \left( \sum\limits_{i=1}^{\infty} \frac{ W (x_i) \omega_i }{\sigma_i^2}  - \frac12 \sum\limits_{i=1}^{\infty} \frac{ W(x_i)^2 }{\sigma_i^2} \right).
\end{equation}
Then the canonical process $\omega  \mapsto \omega_i$ on $(\Omega, \mathcal{F}, \mathbb{P}_{\!{\scriptscriptstyle W}})$  specifies the stochastic sequence
$$
W(x_i) + \varepsilon_i, \; i = 1, 2, \cdots.
$$
That is, $\mathbb{P}_{\! \scriptscriptstyle W}$ specifies the distribution of order flow $\omega$ when insider demand is $W$.

Therefore, the probability measure on $\Omega \times S$ given by 
$$
\left( \exp \left( \sum\limits_{i=1}^{\infty}  \frac{ \widetilde{W}(x_i \vert s_k) \omega_i }{\sigma_i^2}  - \frac12 \sum\limits_{i=1}^{\infty} \frac{ \widetilde{W}(x_i \vert s_k)^2 }{\sigma_i^2} \right) \cdot \mathbb{P}_{\! \scriptscriptstyle 0} \right) \otimes \pi_0 
$$
correctly specifies the intended joint probability law of $(\omega, s_k)$ according to the market maker's belief $\widetilde{W}(\,\cdot\, \vert \,\cdot\,)$.

Since
$$
\exp \left( \sum\limits_{i=1}^{\infty}  \frac{ \widetilde{W}(x_i \vert s_k)  \omega_i }{\sigma_i^2}  - \frac12 \sum\limits_{i=1}^{\infty} \frac{ \widetilde{W}(x_i \vert s_k)^2 }{\sigma_i^2} \right)
$$
is clearly jointly measurable in $(\omega, s_k)$, it follows from the Fubini-Tonelli Theorem that
$\{ \pi_1(\,\cdot\, \vert \omega \kern 0.045em ; \widetilde{W}) \}_{\omega \in \Omega}$ is the $\omega$-disintegration of
the family $\{ \mathbb{P}_{\scriptscriptstyle \widetilde{W}(\,\cdot\, \vert s_k)} \}_{s_k \in S}$.
In other words, the posterior $\pi_1(\,\cdot\, \vert \omega \kern 0.045em ; \widetilde{W})$ of~\eqref{eqn: candidate expression for posterior} specifies the market maker's posterior probability measure on $S$ conditional on $\omega$, with the normalizing random 
variable
$$
C(\omega) = \sum\limits_{k=1}^K  \exp \left( \sum\limits_{i=1}^{\infty} \frac{ \widetilde{W}(x_i \vert s_k) \omega_i}{\sigma_i^2}  - \frac12 \sum\limits_{i=1}^{\infty} \frac{ \widetilde{W}(x_i \vert s_k)^2}{\sigma_i^2} \right) \cdot \pi_0(s_k).
$$
This proves the theorem.

\subsection{Proof of Theorem~\ref{thm: informed trader 1}}
\label{sec: proof of FOC}

The marginal utility functional $\partial \! \kern 0.08em J(\,\cdot\,; W) \colon \mathbb{R}^{\infty} \rightarrow \mathbb{R}$ is the G\^{a}teaux-derivative of the expected utility functional $J(\, \cdot \, \vert s_k; \widetilde{W}) \colon \mathbb{R}^{\infty} \rightarrow \R$,
as defined in Equation~\eqref{eqn: informed trader's problem}, at $W$.
(Both $J(\, \cdot \, \vert s_k; \widetilde{W})$ and its G\^{a}teaux-derivative $\partial \! \kern 0.08em J(\,\cdot\,; W)$ are conditional on $s_k$ and  parameterized by $\widetilde{W}$.)

$J(W \vert s_k; \widetilde{W}) = J_p(W) - J_c(W)$ is the difference between the payoff functional
$$
J_p(W) = \sum\limits_i W(x_i) \eta(x_i \vert s_k) 
$$ 
and the cost functional
$$
J_c(W) = \sum_i W(x_i) \overline{P}(x_i, W; \widetilde{W}).
$$
We need to show its G\^{a}teaux-derivative decomposes into
$$
\underbrace{ \partial \! \kern 0.08em J(\,\cdot\, ; W) }_\text{marginal utility} = \underbrace{ \partial \! \kern 0.08em J_{p}(\,\cdot\, ; W) }_\text{marginal payoff} - \underbrace{ \big(\, \partial \! \kern 0.08em J_{\scriptscriptstyle \! AD}(\,\cdot\, ; W) + \partial \! \kern 0.08em J_{\scriptscriptstyle \! K}(\,\cdot\, ; W) \,\big) }_\text{marginal cost}
$$

Because the G\^{a}teaux derivative of the payoff functional $J_p(\,\cdot\,)$ 
is trivially identified with $\eta(\,\cdot\, \vert s_k)$, the characterization of the marginal payoff functional $\partial \! \kern 0.08em J_{p}(\,\cdot\, ; W)$ in Equation~\eqref{eqn: marginal payoff} is immediate. 

Now, consider the cost functional $J_c(\,\cdot\,)$.
To make the dependence of expected AD price $\overline{P}(x_i, W; \widetilde{W})$ on $W$ more explicit, we write out
\begin{align}
\nopagebreak
\overline{P}(x_i, W; \widetilde{W}) 
&= \mathbb{E}^{\mathbb{P}_{\scriptscriptstyle W}} [ \sum_k \eta(x_i \vert s_k) \pi_1(s_k \mid \omega; \widetilde{W}) ] \nonumber \\
&= \mathbb{E}^{\mathbb{P}_{\scriptscriptstyle W}} [ \sum_k \eta(x_i \vert s_k) \, \frac{ {\rm e}^{ \mathcal{I}(\omega, s_k ; \widetilde{W} )} }{C(\omega)} \, \pi_0(s_k)] \nonumber \\
&= \mathbb{E}^{\mathbb{P}_{\scriptscriptstyle 0}} [ \sum_k \eta(x_i \vert s_k) \, \frac{ {\rm e}^{\mathcal{I}(\omega, s_k ; \widetilde{W} ) + \sum\limits_j \frac{\widetilde{W}(x_j \vert s_k) W(x_j)}{\sigma_j^2 } }  }{C'(\omega)} \, \pi_0(s_k)], \label{eqn: pushed forward AD bar}
\end{align}
where 
\begin{equation}
\label{eqn: caligraphic I}
\mathcal{I}(\omega, s_k ; \widetilde{W}) = \sum_j \frac{ \widetilde{W}(x_j \vert s_k) \omega_j}{\sigma_j^2}  - \frac12 \sum_j \frac{ \widetilde{W}(x_j \vert s_k)^2 }{\sigma_j^2} 
\end{equation}
and 
\begin{equation}
\label{eqn: C prime}
C'(\omega) = \sum_l {\rm e}^{\mathcal{I}(\omega, s_l ; \widetilde{W} ) + \sum_j \frac{\widetilde{W}(x_j \vert s_l) W(x_j)}{\sigma_j^2 } } \pi_0(s_l).
\end{equation}
The equality~\eqref{eqn: pushed forward AD bar} holds because the law of the canonical process $\omega \mapsto \omega_j$ under $\mathbb{P}_{\scriptscriptstyle W}$ is the same as the
law of $\omega \mapsto W(x_j) + \omega_j$ under $\mathbb{P}_{\! \scriptscriptstyle 0}$.

Let $v \colon X \rightarrow \R$ be a marginal portfolio and define $f(\varepsilon) = J_c(W + \varepsilon v)$.
The marginal cost functional (i.e., G\^{a}teaux derivative) $\partial \! \kern 0.08em J_c(W)$ of $J_c(W)$ evaluated at $v$ can be computed by invoking the Dominated Convergence Theorem and differentiating under the summation signs:
\begin{align}
\label{eqn: two integrals}
\partial \! \kern 0.08em J_c(W) &= f'(0) \nonumber \\
&= \sum_i v(x_i) \overline{P}(x_i, W; \widetilde{W}) + \sum_i W(x_i) g(x_i)
\end{align}
\noindent for some $g \colon X \rightarrow \mathbb{R}$. 
The first sum in \eqref{eqn: two integrals} verifies the AD term $\partial \! \kern 0.08em J_{\scriptscriptstyle \! AD}(v; W)$ of Equation~\eqref{eqn: AD term}. 
It remains to show the price impact term $\partial \! \kern 0.08em J_{\scriptscriptstyle \! K}(v; W)$ of Equation~\eqref{eqn: Kyle term in FOC} is the second sum in \eqref{eqn: two integrals}.

The function $g$ is of the form $g(x_i) = \mathbb{E}^{\mathbb{P}_{\! \scriptscriptstyle 0}}[ \psi(\omega; x_i, \widetilde{W})]$, $i = 1, 2, \cdots$. 
The random variable $\psi(\,\cdot\,; x_i, \widetilde{W})$ defined on $\Omega$ is given by
\begin{equation}
\label{eqn: psi}
\psi(\omega; x_i, \widetilde{W}) =
\sum_k \left(  \eta(x_i \vert s_k) \frac{ C'(\omega) l(s, \omega)  \sum\limits_j \frac{\widetilde{W}(x_j \vert s_k)}{\sigma_j^2 } v(x_j)  - l(s, \omega) \sum\limits_l \left( l(s_l, \omega) \sum\limits_j \frac{\widetilde{W}(x_j \vert s_l)}{\sigma_j^2 } v(x_j)  \pi_0(s_l) \right)}{ C'(\omega)^2 } \pi_0(s_k) \right)
\end{equation}
where
$$
C'(\omega) = \sum_k \left( {\rm e}^{ \sum\limits_j \frac{ \widetilde{W}(x_j \vert s_k) W(x_j)}{\sigma_j^2} + \cdots } \pi_0(s_k) \right) = \sum_k l(s_k, \omega) \pi_0(s_k)
$$
is the normalization constant of the market maker's posterior under the probability measure $\mathbb{P}_{\! \scriptscriptstyle 0}$, and 
\begin{equation}
\label{eqn: dots}
l(s_k, \omega)= {\rm e}^{\sum\limits_j \frac{ \widetilde{W}(x_j \vert s_k) W(x_j)}{\sigma_j^2} + \cdots }
\end{equation}
is the likelihood of $\omega$ under $\mathbb{P}_{\! \scriptscriptstyle 0}$. For clarity of notation, in \eqref{eqn: dots} we have put ``$\cdots$'' for terms not relevant for this calculation.

Now, to interpret $g(x_i) = \mathbb{E}^{\mathbb{P}_{\! \scriptscriptstyle 0}}[ \psi(\omega; x_i, \widetilde{W})]$, observe that 
$$
\frac{ l(s_k, \omega) \pi_0( s_k )}{C'(\omega) }, \; k = 1, \cdots, K
$$
is the market maker's posterior conditional on $\omega$.
It is then clear from Equation~\eqref{eqn: psi} that
$\psi(\omega; x_i, \widetilde{W})$ is the difference between the posterior expectation of the product of
$\eta(x_i \vert \,\cdot\,)$ and $\Pi_{insider}(v, \,\cdot\, ; \widetilde{W}) = \sum\limits_j \frac{\widetilde{W}(x_j \vert \,\cdot\,) v(x_j)}{\sigma_j^2 } $ (as in Definition~\ref{def: overlap measures})
and the product of their posterior expectations. In other words, $\psi(\omega; x, \widetilde{W})$ is equal to
the posterior covariance 
$$
\psi(\omega; x, \widetilde{W}) = \Cov \! \left( \eta(x, \,\cdot\, ), \Pi_{insider}(v, \,\cdot\, ; \widetilde{W}) \big| \omega  \right). 
$$
This shows that the second term in $\partial J_c(W)$ is precisely the price impact term $\partial \! \kern 0.08em J_{\scriptscriptstyle \! K}(v; W)$ of Equation~\eqref{eqn: Kyle term in FOC}. This proves the theorem.

\subsection{Proof of Theorem~\ref{thm: informed trader 2} }
\label{sec: proof of no arb}

For $c \colon S \rightarrow \R$, define the affine subspace of portfolios 
$
\mathcal{V}_c(\widetilde{W}) \!\equiv\! \{ W \colon \Pi_{insider}(W, \,\cdot\, ; \widetilde{W}) = c\}.
$
The zero price impact portfolios are those in the subspace $\mathcal{V}_0(\widetilde{W})$ corresponding to $c = 0$.

We observe that, for a given $c$, the market maker's (expected) AD prices $\overline{P}(\,\cdot\,, W; \widetilde{W})$ do not change with 
respect to $W \! \in \! \mathcal{V}_c(\widetilde{W})$.\footnote{This can be seen explicitly from Equation~\eqref{eqn: pushed forward AD bar}.} 
That is, conditional on  $\Pi_{insider}(W, \,\cdot\, ; \widetilde{W})$, a portfolio $W$ does not change the (expected) AD prices.
In particular, this is true for the zero price impact subspace $\mathcal{V}_0(\widetilde{W})$. Also, $\mathcal{V}_0(\widetilde{W})$ is invariant under scaling.
It follows that, for the insider's problem \eqref{eqn: informed trader's problem} to be well-posed, a portfolio $W \in \mathcal{V}_0(\widetilde{W})$ must give him zero expected utility. 
Otherwise, he can obtain unbounded utility by scaling up his portfolio indefinitely without incurring price impact.
For example, suppose $J(W \vert s_k; \widetilde{W}) > 0$ for some $W \in \mathcal{V}_0(\widetilde{W})$.
Then the insider's expected utility $J(\alpha W \vert s_k; \widetilde{W}) \rightarrow \infty$ as $\alpha \rightarrow \infty$ because the scaled portfolio $\alpha W \in \mathcal{V}_0(\widetilde{W})$ causes no price impact as $\alpha \rightarrow \infty$. 
Similarly, $J(W \vert s_k; \widetilde{W}) < 0$ for some $W \in \mathcal{V}_0(\widetilde{W})$ would allow arbitrage.

More generally, for any two portfolios $W_1, W_2 \in \mathcal{V}_c(\widetilde{W})$, no-arbitrage requires that $J(W_1 \vert s_k; \widetilde{W}) = J(W_2 \vert s_k; \widetilde{W})$.
Otherwise, the long-short portfolio $W_1 - W_2 \in \mathcal{V}_0(\widetilde{W})$ would be an arbitrage opportunity.
In other words, to preclude arbitrage, the insider's expected utility functional $J(\,\cdot\, \vert s_k; \widetilde{W})$ must be constant on the closed affine subspace $\mathcal{V}_c(\widetilde{W})$ for each $c \in C(S, \R)$.
This proves the theorem.

\subsection{Theorem~\ref{thm: canonical game}}
\label{sec: proof of canonical game}

\subsubsection{Formal Statement of Theorem}


\begin{canonicalgame}
The Bayesian trading game between the insider and market maker is isomorphic to the {\bf {\em canonical game}}, defined as follows:

\begin{itemize}

\item  There are markets for information contracts $k = 1, \cdots, K$.

\item Conditional on signal $s_k$, contract $k$ has payoff of $1$ while all others have payoffs of $0$.  

\item The insider observes the signal, while the market maker has a uniform prior on signals.

\item The insider submits orders $\hat{d} \in \mathbb{R}^{K}$ for the information contracts, and the market maker receives order flow  $\hat{\omega} = \hat{d} + \widehat{N}$, where $\widehat{N}_k$ are i.i.d.~$\mathcal{N}(0,1)$ noise trades for $k = 1, \cdots, K$. 

\item The market maker has a belief $\widehat{D}$ as specified in \eqref{eqn: change of basis}. 
Upon receiving $\hat{\omega}$, he updates the posterior probability of each signal $s_k$ (which is also the contract $k$ price) to 
\begin{equation}
\label{eqn: pi_1 finite S new}
\hat{\pi}_1(k \vert \hat{\omega} \kern 0.045em ; \widehat{D}) \propto  e^{ ( \hat{d}^{(k)} )^{T} \hat{\omega} - \frac12 (\hat{d}^{(k)} )^{T} \hat{d}^{(k)} }.
\end{equation}

\item Conditional on $s_k$ and given market maker belief $\widehat{D}$, the insider maximizes expected profit: 
\begin{equation}
\label{eqn: further transformed insider's problem for finite S case}
\max_{\hat{d} \, \in \, \mathbb{R}^K} \,  \hat{d}_k - \hat{\pi}_1(\hat{d} \kern 0.045em ; \widehat{D})^T\hat{d}  \equiv \max_{\hat{d} \, \in \, \mathbb{R}^K} J(\hat{d} \vert k \kern 0.045em ; \widehat{D})
\end{equation}
where $\hat{\pi}_1(\hat{d} \kern 0.045em; \widehat{D})$ is the expected value of the prices \eqref{eqn: pi_1 finite S new} over the distribution of $\hat{\omega}$.
\end{itemize}
\medskip
\end{canonicalgame}


\subsubsection{Proof of Theorem}

Under the general specification, the spanning conditions of Proposition~\ref{cor: finite S 1} imply the equilibrium restriction that the noise trading intensity is a constant $\sigma \! > \! 0$ across markets. 
Under additional assumptions (e.g., when $\eta(\,\cdot\,\vert s_k)$'s have disjoint supports as in Section~\ref{sec: main intuition}), this spanning condition can hold with a varying noise intensity.
In such cases, the proof here goes through verbatim with $\sigma_i$, $i = 1, 2, \cdots$, in place of $\sigma$.  

For order flow $\omega \in \Omega$, let $B(\omega) = ( \sum\limits_i \frac{ \eta(x_i \vert s_k) \omega_i}{\sigma^2} )_{k =1, \cdots, K} \in \mathbb{R}^K$.
Then, in terms of $d$ and $\widetilde{D}$ (the latter is defined in \eqref{eqn: change of basis}), the general market maker posterior over signals obtained in Theorem~\ref{thm: Market Maker's Posterior}(i) can be written explicitly as the probability mass function
\begin{equation}
\label{eqn: pi_1 finite S}
\pi_1(k \vert \omega, d ; \widetilde{D}) = \frac{ e^{ ( \tilde{d}^{(k)} )^{T} {\bf L}^2 d + ( \tilde{d}^{(k)} )^{T} B(\omega) - \frac12 (\tilde{d}^{(k)} )^{T} {\bf L^2} \tilde{d}^{(k)} } } { \sum\limits_{l=1}^K e^{ ( \tilde{d}^{(l)} )^{T} {\bf L^2} d + ( \tilde{d}^{(l)} )^{T} B(\omega) - \frac12 (\tilde{d}^{(l)} )^{T} {\bf L}^2 \tilde{d}^{(l)} } }, \;\; k = 1, \cdots, K
\end{equation}
on $S$. 
For the expectation of $\big( \pi_1(k \vert \omega, d ; \widetilde{D}) \big)_{k = 1, \cdots, K} \in \mathbb{R}^K$, we write it as 
\begin{equation}
\label{eqn: AD bar finite dimensional}
\bar{\pi}_1(d; \widetilde{D}) = \left( \mathbb{E}^{\mathbb{P}_{\scriptscriptstyle 0}}[ \pi_1(k \vert \omega, d ; \widetilde{D}) ]\right)_{k = 1, \cdots, K} \in \mathbb{R}^K,
\end{equation}
where $\mathbb{E}^{\mathbb{P}_{\scriptscriptstyle 0}}[\,\cdot\,]$ is taken with respect to distribution over possible order flows $\omega$ if the insider chooses $d$ and the market maker holds belief $\widetilde{D}$.
The insider's portfolio choice problem~\eqref{eqn: informed trader's problem} conditional on observing $s_k$ now takes the simple form
\begin{equation}
\label{eqn: transformed insider's problem for finite S case}
\max_{d \in \mathbb{R}^K} e_k^T {\bf L}^2 d - \bar{\pi}_1(d; \widetilde{D})^T {\bf L}^2 d \equiv \max_{d \in \mathbb{R}^K} J(d \vert k; \widetilde{D}).
\end{equation}
This reduces the Bayesian trading game between the insider and the market maker to one where
the market maker's posterior belief is specified by \eqref{eqn: pi_1 finite S}, and the insider's problem is \eqref{eqn: transformed insider's problem for finite S case}.

The canonical transformation of \eqref{eqn: change of basis}
$$
\hat{d} = {\bf L} d
$$
replaces ${\bf L}^2$ and $d$ in the market maker's posterior $\pi_1(k \vert \omega, d ; \widetilde{D})$ of \eqref{eqn: pi_1 finite S} by ${\bf I}$ and $\hat{d}$, respectively.
Under this transformation, the $k$-th component of the random vector $\widehat{N} = {\bf L}^{-1} B(\omega) $ can be re-written as $\widehat{N}_k = \sum\limits_i \xi^{(k)}_i \tilde{\varepsilon}_i$ 
where $\sum\limits_i \xi^{(k)}_i \xi^{(l)}_i = \delta_{kl}$ and $(\tilde{\varepsilon}_i)$ is an i.i.d.~standard normal sequence. Therefore, by (the discrete version of) It\^{o} isometry,
\begin{equation}
\label{eqn: Ito isometry result}
\widehat{N}_k \stackrel{i.i.d.}{\sim} \mathcal{N}(0,1), \;\; k = 1, \cdots, K.
\end{equation}
This gives the market maker's posterior \eqref{eqn: pi_1 finite S new} in the canonical game.

Similarly, the canonical transformation replaces ${\bf L}^2$ and $d$ in the insider's objective function $J(d \vert k ; \widetilde{D})$ of \eqref{eqn: transformed insider's problem for finite S case} by ${\bf I}$ and $\hat{d}$, respectively.
This verifies the insider's problem \eqref{eqn: further transformed insider's problem for finite S case} in the canonical game and proves the theorem.

\subsubsection{Equilibrium in Canonical Game}
In the canonical game, an equilibrium is specified by a $K \times K$ strategy matrix $D^* = [\delta^{(1)} \; \cdots \; \delta^{(K)}]$ such that, under the same market maker belief $D^*$,
each strategy $\delta^{(k)}\in \mathbb{R}^K$ solves the insider's problem \eqref{eqn: further transformed insider's problem for finite S case} conditional on $s_k$, i.e.,
\begin{equation}
\label{eqn: transformed equil def for finite S case}
\delta^{(k)} \in \argmax_{\hat{d} \, \in \, \mathbb{R}^K} J(\hat{d} \vert k \kern 0.045em ; D^*), \; \mbox{for each} \; k = 1, \cdots, K.
\end{equation}

\subsection{Proof of Theorem~\ref{thm: equilibrium of original trading game}}
\label{sec: proof of existence}

\paragraph{Insider FOC in Canonical Game}
Let $p \in [0,1]^I$ denote the random vector \eqref{eqn: pi_1 finite S new} of the market maker's posterior in the canonical game (with the dependence on $\hat{\omega}$ and $\widehat{D}$ understood) and
$\mbox{diag}(p) \in \mathbb{R}^{K \times K}$ denote the matrix with $p$ along the diagonal and zeros elsewhere.
Then, conditional on observing $s_k$ and given market maker belief $\widehat{D}$, the first-order condition for the insider's problem \eqref{eqn: further transformed insider's problem for finite S case}
in the canonical game is
\begin{equation}
\label{eqn: transformed Bayesian game FOC}
e_k 
- 
\underbrace{
\mathbb{E}^{(\hat{d}; \widehat{D})}[p]
\vphantom{\left(
\mathbb{E}^{(\hat{d}; \widehat{D})}
\!
\left[
\mbox{diag}(p)
-
p
p^T
\right]
\right)}
}_\text{AD term}
- 
\underbrace{ 
\widehat{D} 
\cdot
\left(
\mathbb{E}^{(\hat{d}; \widehat{D})}
\!
\left[
\mbox{diag}(p)
-
p
p^T
\right]
\right)
\cdot 
\hat{d}
}_\text{price impact term}
= 0
\end{equation}
where $e_k$ denotes the $k$-th standard basis vector, and the expectation $\mathbb{E}^{(\hat{d}; \widehat{D})}[\,\cdot\,]$ is taken over the distribution of $\hat{\omega}$ induced by the insider's choice $\hat{d}$ 
when the market maker holds belief $\widehat{D}$.
The terms in the first-order condition \eqref{eqn: transformed Bayesian game FOC} are the isomorphic counterparts to those in Theorem~\ref{thm: informed trader 1}.

\begin{lemma} 
\label{lemma: convex price impact}

Suppose $\alpha' > 0$ solves the equilibrium equation $\Phi(\alpha) = 0$, and suppose the market maker holds belief $\alpha' {\bf Q}$ in the canonical game.
Then, for the insider's problem \eqref{eqn: further transformed insider's problem for finite S case} conditional on each $s_k$ in the canonical game,
the first-order condition is sufficient for optimality. 
Therefore, since the insider's strategy $\alpha' {\bf Q}$ satisfies the collated first-order conditions $\Phi(\alpha') {\bf Q} = 0$, $\alpha' {\bf Q}$ is an equilibrium of the canonical game as defined in \eqref{eqn: transformed equil def for finite S case}. 
\end{lemma}

\begin{proof}
Given the market maker's belief $\widehat{D} = \alpha'{\bf Q}$ in the canonical game, the Hessian matrix $H$ of the insider's objective function \eqref{eqn: further transformed insider's problem for finite S case} conditional on a signal $s_{k}$ can be computed by differentiating 
directly his first-order condition \eqref{eqn: transformed Bayesian game FOC}, which gives 
$$
- \alpha' {\bf Q}
\underbrace{ 
\mathbb{E}
\left[
\begin{bmatrix}
p_{1} & -p_1 p_2 & \cdots \\
-p_1 p_2 & p_2 & \cdots \\
\vdots & \vdots & \ddots
\end{bmatrix}
\right]
}_\text{$H$}
= - \alpha' {\bf Q} H.
$$
By the Cauchy-Schwarz inequality,
$$
\mathbb{E}[p_l p_m]^2 \leq \mathbb{E}[p_l^2]\mathbb{E}[p_m^2] \leq \mathbb{E}[p_l]\mathbb{E}[p_m], \;\; 1 \leq l, m \leq K.
$$
Therefore, $H$ is positive semidefinite. 
Since ${\bf Q}H = H{\bf Q}$, ${\bf Q}H$ is also positive semidefinite. Therefore, the Hessian $- \alpha' {\bf Q}H$ is negative semidefinite. 
It follows that the insider's objective function is concave conditional on any given signal. This proves the lemma.
\end{proof}

\begin{lemma}
\label{lemma: claim for MM posterior} 

Let
$$
{\bf p} = [p_1 \, \cdots \, p_K]^T
$$
denote the random vector \eqref{eqn: pi_1 finite S new} of the market maker's (normalized) posterior in the canonical game conditional on the insider observing signal $s_k$ (with the dependence on $\hat{\omega} = \hat{d} +\widehat{N}$, and $\widehat{D}$ understood).
${\bf p}$ is a random probability measure on the signal space $S = \{ s_1, \cdots, s_K \}$.
Let $\alpha > 0$ and $Z \stackrel{d}{\sim} \mathcal{N}(0, \alpha^2{\bf Q})$ be a random vector that is multivariate normal with mean $0$ and covariance 
matrix $\alpha^2{\bf Q}$.

Then, under the equilibrium postulate $\alpha{\bf Q}$, the probability law of {\bf p} is given by\footnote{``$\sum\limits_{l} \cdots$'' is a random normalization factor so that $\sum\limits_k p_k = 1$.}
$$
{\bf p} \stackrel{d}{\sim} \left( \frac{e^{Z_1}}{\sum\limits_{l} \cdots}, \cdots, \frac{ e^{\alpha^2 + Z_k} }{\sum\limits_{l} \cdots}, \cdots \frac{e^{Z_K}}{\sum\limits_{l} \cdots} \right)^T.
$$
\end{lemma}

\begin{proof}
Substituting the equilibrium postulate $\alpha {\bf Q}$ into the market maker's posterior~\eqref{eqn: pi_1 finite S} conditional on order flow $\omega$, we have
\begin{equation}
\label{eqn: aa eqn 1}
\pi_1(s_k \vert \omega, \beta^{(k)}; \alpha{\bf Q}) = \frac{ e^{ \alpha^2 (1 - \frac{1}{K}) \, + \, \alpha e_k^T {\bf Q} {\bf L}^{-1} B(\omega) } } { \sum\limits_{l} \cdots },
\end{equation}
and, for $m \neq k$,
\begin{equation}
\label{eqn: aa eqn 2}
\pi_1(s_m \vert \omega, \beta^{(k)}; \alpha{\bf Q}) = \frac{ e^{ \alpha^2 (-\frac{1}{K}) \, + \, \alpha e_m^T {\bf Q} {\bf L}^{-1} B(\omega)  } } { \sum\limits_{l} \cdots },
\end{equation}
where $\beta^{(k)}$ is the $k$-th column of $\alpha{\bf Q}$, i.e., the insider strategy postulated by $\alpha{\bf Q}$  conditional on observing signal $s_k$, and the denominator $\sum\limits_{l} \cdots$ is the normalization factor. 

The common factor $e^{ \alpha^2 (-\frac{1}{K})}$ in \eqref{eqn: aa eqn 1} and \eqref{eqn: aa eqn 2} cancels after normalization.
It remains to consider the random vector ${\bf L}^{-1}  B(\omega)$.
By the same It\^o isometry argument as that for \eqref{eqn: Ito isometry result},
${\bf L}^{-1} B(\omega) \in \mathbb{R}^K$ has the standard multivariate normal distribution.
Therefore,
$\alpha {\bf Q} {\bf L}^{-1} B(\omega) \stackrel{d}{\sim} \mathcal{N}(0, \alpha^2 {\bf Q})$.
This proves the lemma.
\end{proof}

\paragraph{Proof of Theorem}
$\;$
 
{\em \underline{The Equilibrium Equation.}}

Substituting the equilibrium postulate $\widehat{D} = \alpha {\bf Q}$ of \eqref{eqn: equilibrium ansatz for transformed Bayesian game}
into the insider's first-order condition~\eqref{eqn: transformed Bayesian game FOC} for the canonical game conditional on (say) $s_1$ gives
\begin{equation}
\label{eqn: thm proof appendix}
\mathbb{E}
\left[
e_1 -
\begin{bmatrix}
p_1 \\
\vdots \\
p_K 
\end{bmatrix}
- 
\alpha^2
{\bf Q}
\begin{bmatrix}
p_{1} & & \\
& \ddots & \\
& & p_{K}
\end{bmatrix}
{\bf Q}
e_1
+
\alpha^2
{\bf Q}
\begin{bmatrix}
p_1 \\
\vdots \\
p_K 
\end{bmatrix}
\begin{bmatrix}
p_1 & \cdots & p_K 
\end{bmatrix}
{\bf Q}
e_1
\right]
= 0,
\end{equation}
where $\mathbb{E}[\, \cdot \, ]$ is taken with respect to the probability law of the random probability measure
$
{\bf p} = 
\begin{bmatrix}
p_1 \\
\vdots \\
p_K 
\end{bmatrix}
$
under the postulated equilibrium.

The first-order condition \eqref{eqn: thm proof appendix} conditional on $s_1$ simplifies to  
\begin{equation}
\label{eqn: eqn in proof}
\mathbb{E} 
\left[
e_1 -
\begin{bmatrix}
p_1 \\
p_2 \\
\vdots \\
p_K 
\end{bmatrix}
-
\alpha^2
\big(
\begin{bmatrix}
p_1 \\
0 \\
\vdots \\
0 
\end{bmatrix}
-
p_1
\begin{bmatrix}
p_1 \\
p_2 \\
\vdots \\
p_K 
\end{bmatrix}
\big)
\right] 
=
\mathbb{E} 
\left[
(1 - \alpha^2 p_1)
\begin{bmatrix}
1- p_1 \\
-p_2 \\
\vdots \\
-p_K 
\end{bmatrix} 
\right] = 0.
\end{equation}

The probability law of ${\bf p}$ is as characterized in Lemma~\ref{lemma: claim for MM posterior} above, with $k = 1$.
Under this probability law, the moments in Equation~\eqref{eqn: eqn in proof} are the moments of a logistic-normal 
distribution.

We now show that, by substituting for the appropriate relationships among the corresponding moments, \eqref{eqn: eqn in proof} can be written as
\begin{equation}
\label{eqn: equil eqn in proof}
\Phi(\alpha) {\bf Q} e_1 = 0,
\end{equation}
for some $\Phi \colon [0, \infty) \rightarrow \mathbb{R}$.
To show this,
it suffices to show
\begin{equation}
\label{eqn: eqn in proof 2.5}
\mathbb{E}[ (1-\alpha^2 p_1) (1-p_1)] = (K-1) \mathbb{E} [ (1-\alpha^2 p_1) p_l ], \;\; l = 2, \cdots, K.
\end{equation}
\eqref{eqn: eqn in proof 2.5} reduces to 
\begin{equation}
\label{eqn: eqn in proof 2}
\mathbb{E} [ (1-\alpha^2 p_1) (1-p_1) ] = (K-1) \mathbb{E}[ (1-\alpha^2 p_1) p_2]
\end{equation}
because, under the probability law of ${\bf p}$ conditional on $s_1$ characterized in Lemma~\ref{lemma: claim for MM posterior},
$\mathbb{E}[ p_1 p_l ] = \mathbb{E} \left[ p_1 p_m \right]$ for all $l, m \neq 1$.
In turn, \eqref{eqn: eqn in proof 2} holds because
$$
\mathbb{E} [1 -p_1 ] = (K-1) \mathbb{E} [p_2 ], \,\mbox{and} \,\,\,
\mathbb{E} [ p_1 (1-p_1) ] = \mathbb{E} [ p_1 (p_2 + \cdots + p_K) ] = (K-1) \mathbb{E} [ p_1 p_2 ]
$$
under the same logistic-normal probability law.

Therefore, we can take $\Phi(\alpha)$ in \eqref{eqn: equil eqn in proof} to be (up to a scalar multiple) the left-hand side of \eqref{eqn: eqn in proof 2.5}, i.e.,
\begin{equation}
\label{eqn: scalar equil eqn in proof}
\Phi(\alpha) = \mathbb{E} [ 1 - p_1 - \alpha^2 p_1 + \alpha p_1^2 ]
\end{equation}
where 
$$
p_1 = \frac{ e^{\alpha^2 + Z_1} }{ e^{\alpha^2 + Z_1} + \sum\limits_{l \neq 1} e^{Z_l} }, \;\; Z = (Z_l)_{1 \leq l \leq K} \stackrel{d}{\sim} \mathcal{N}(0, \alpha^2{\bf Q}).
$$

By symmetry, the first-order condition for $k \neq 1$ is identical to \eqref{eqn: equil eqn in proof} after permuting the indices $1$ and $k$.
Collating these $K$ symmetric first-order conditions, 
$$
\Phi(\alpha) {\bf Q} e_k = 0, \; 1 \leq k \leq K,
$$ side-by-side 
gives the matrix equation $\Phi(\alpha) {\bf Q} = 0$.
By Lemma~\ref{lemma: convex price impact}, an equilibrium is given by a solution $\alpha^* > 0$ to the scalar equilibrium equation $\Phi(\alpha) = 0$.

{\em \underline{Existence of Equilibrium.}}

First, we have $\Phi(0) = \mathbb{E} [ 1 - p_1 ] > 0$. Second, $1 - p_1 - \alpha^2 p_1 + \alpha p_1^2 \rightarrow 0$ and $1 - p_1 - \alpha^2 p_1 + \alpha p_1^2 < 0$ eventually as $\alpha \rightarrow \infty$, with probability one.
By Fatou's Lemma, we have that $\Phi(\alpha) \rightarrow 0$ from below as $\alpha \rightarrow \infty$.
Therefore, by the Intermediate Value Theorem, there exists $\alpha^* > 0$ such that $\Phi(\alpha^*) = 0$. This proves the theorem.

\subsection{Proof of Proposition~\ref{prop: MM posterior finite S}}

This is a special case of Lemma~\ref{lemma: claim for MM posterior}, with $\alpha = \alpha^*$.

\bibliographystyle{chicago}
\bibliography{Draft_36_AER}


\end{document}

%% file: Endogenous_constant.tex
\begin{tikzpicture}[x=1pt,y=1pt]
\definecolor{fillColor}{RGB}{255,255,255}
\path[use as bounding box,fill=fillColor,fill opacity=0.00] (0,0) rectangle (199.47, 93.95);
\begin{scope}
\path[clip] ( 18.12, 21.72) rectangle (198.15, 92.63);
\definecolor{drawColor}{RGB}{0,0,0}

\path[draw=drawColor,line width= 1.2pt,line join=round,line cap=round] ( 24.79, 24.35) --
	( 28.19, 65.99) --
	( 31.59, 69.71) --
	( 34.99, 72.06) --
	( 38.39, 73.92) --
	( 41.80, 75.30) --
	( 45.20, 76.48) --
	( 48.60, 77.48) --
	( 52.00, 78.37) --
	( 55.40, 79.11) --
	( 58.81, 79.79) --
	( 62.21, 80.45) --
	( 65.61, 81.04) --
	( 69.01, 81.54) --
	( 72.41, 81.99) --
	( 75.82, 82.48) --
	( 79.22, 82.92) --
	( 82.62, 83.27) --
	( 86.02, 83.61) --
	( 89.42, 84.00) --
	( 92.82, 84.33) --
	( 96.23, 84.67) --
	( 99.63, 84.97) --
	(103.03, 85.21) --
	(106.43, 85.51) --
	(109.83, 85.75) --
	(113.24, 86.05) --
	(116.64, 86.25) --
	(120.04, 86.57) --
	(123.44, 86.66) --
	(126.84, 86.94) --
	(130.24, 87.19) --
	(133.65, 87.41) --
	(137.05, 87.57) --
	(140.45, 87.73) --
	(143.85, 87.90) --
	(147.25, 88.16) --
	(150.66, 88.32) --
	(154.06, 88.51) --
	(157.46, 88.60) --
	(160.86, 88.77) --
	(164.26, 88.99) --
	(167.66, 89.08) --
	(171.07, 89.27) --
	(174.47, 89.38) --
	(177.87, 89.49) --
	(181.27, 89.74) --
	(184.67, 89.84) --
	(188.08, 89.98) --
	(191.48, 90.00);
\end{scope}
\begin{scope}
\path[clip] (  0.00,  0.00) rectangle (199.47, 93.95);
\definecolor{drawColor}{RGB}{0,0,0}

\path[draw=drawColor,line width= 0.4pt,line join=round,line cap=round] ( 18.12, 21.72) --
	(198.15, 21.72) --
	(198.15, 92.63) --
	( 18.12, 92.63) --
	( 18.12, 21.72);
\end{scope}
\begin{scope}
\path[clip] (  0.00,  0.00) rectangle (199.47, 93.95);
\definecolor{drawColor}{RGB}{0,0,0}

\path[draw=drawColor,line width= 0.4pt,line join=round,line cap=round] ( 23.09, 21.72) -- (193.18, 21.72);

\path[draw=drawColor,line width= 0.4pt,line join=round,line cap=round] ( 23.09, 21.72) -- ( 23.09, 18.17);

\path[draw=drawColor,line width= 0.4pt,line join=round,line cap=round] ( 57.11, 21.72) -- ( 57.11, 18.17);

\path[draw=drawColor,line width= 0.4pt,line join=round,line cap=round] ( 91.12, 21.72) -- ( 91.12, 18.17);

\path[draw=drawColor,line width= 0.4pt,line join=round,line cap=round] (125.14, 21.72) -- (125.14, 18.17);

\path[draw=drawColor,line width= 0.4pt,line join=round,line cap=round] (159.16, 21.72) -- (159.16, 18.17);

\path[draw=drawColor,line width= 0.4pt,line join=round,line cap=round] (193.18, 21.72) -- (193.18, 18.17);

\node[text=drawColor,anchor=base,inner sep=0pt, outer sep=0pt, scale=  0.70] at ( 23.09,  8.52) {0};

\node[text=drawColor,anchor=base,inner sep=0pt, outer sep=0pt, scale=  0.70] at ( 57.11,  8.52) {20};

\node[text=drawColor,anchor=base,inner sep=0pt, outer sep=0pt, scale=  0.70] at ( 91.12,  8.52) {40};

\node[text=drawColor,anchor=base,inner sep=0pt, outer sep=0pt, scale=  0.70] at (125.14,  8.52) {60};

\node[text=drawColor,anchor=base,inner sep=0pt, outer sep=0pt, scale=  0.70] at (159.16,  8.52) {80};

\node[text=drawColor,anchor=base,inner sep=0pt, outer sep=0pt, scale=  0.70] at (193.18,  8.52) {100};

\path[draw=drawColor,line width= 0.4pt,line join=round,line cap=round] ( 18.12, 24.35) -- ( 18.12, 92.63);

\path[draw=drawColor,line width= 0.4pt,line join=round,line cap=round] ( 18.12, 24.35) -- ( 14.57, 24.35);

\path[draw=drawColor,line width= 0.4pt,line join=round,line cap=round] ( 18.12, 38.01) -- ( 14.57, 38.01);

\path[draw=drawColor,line width= 0.4pt,line join=round,line cap=round] ( 18.12, 51.67) -- ( 14.57, 51.67);

\path[draw=drawColor,line width= 0.4pt,line join=round,line cap=round] ( 18.12, 65.34) -- ( 14.57, 65.34);

\path[draw=drawColor,line width= 0.4pt,line join=round,line cap=round] ( 18.12, 79.00) -- ( 14.57, 79.00);

\node[text=drawColor,anchor=base east,inner sep=0pt, outer sep=0pt, scale=  0.70] at ( 14.52, 21.94) {0};

\node[text=drawColor,anchor=base east,inner sep=0pt, outer sep=0pt, scale=  0.70] at ( 14.52, 35.60) {0.5};

\node[text=drawColor,anchor=base east,inner sep=0pt, outer sep=0pt, scale=  0.70] at ( 14.52, 49.26) {1};

\node[text=drawColor,anchor=base east,inner sep=0pt, outer sep=0pt, scale=  0.70] at ( 14.52, 62.93) {1.5};

\node[text=drawColor,anchor=base east,inner sep=0pt, outer sep=0pt, scale=  0.70] at ( 14.52, 76.59) {2};
\end{scope}
\end{tikzpicture}

%% file: info_efficiency.tex
\begin{tikzpicture}[x=1pt,y=1pt]
\definecolor{fillColor}{RGB}{255,255,255}
\path[use as bounding box,fill=fillColor,fill opacity=0.00] (0,0) rectangle (199.47, 93.95);
\begin{scope}
\path[clip] ( 18.12, 21.72) rectangle (198.15, 92.63);
\definecolor{drawColor}{RGB}{0,0,0}

\path[draw=drawColor,line width= 1.2pt,line join=round,line cap=round] ( 24.79, 90.00) --
	( 28.19, 59.18) --
	( 31.59, 50.29) --
	( 34.99, 45.40) --
	( 38.39, 42.32) --
	( 41.80, 40.06) --
	( 45.20, 38.31) --
	( 48.60, 37.11) --
	( 52.00, 36.02) --
	( 55.40, 34.97) --
	( 58.81, 34.00) --
	( 62.21, 33.28) --
	( 65.61, 32.70) --
	( 69.01, 32.05) --
	( 72.41, 31.63) --
	( 75.82, 31.08) --
	( 79.22, 30.61) --
	( 82.62, 30.17) --
	( 86.02, 29.78) --
	( 89.42, 29.52) --
	( 92.82, 29.40) --
	( 96.23, 28.95) --
	( 99.63, 28.64) --
	(103.03, 28.45) --
	(106.43, 28.07) --
	(109.83, 27.76) --
	(113.24, 27.88) --
	(116.64, 27.53) --
	(120.04, 27.30) --
	(123.44, 26.94) --
	(126.84, 26.86) --
	(130.24, 26.68) --
	(133.65, 26.63) --
	(137.05, 26.29) --
	(140.45, 26.10) --
	(143.85, 26.06) --
	(147.25, 26.05) --
	(150.66, 25.81) --
	(154.06, 25.73) --
	(157.46, 25.59) --
	(160.86, 25.40) --
	(164.26, 25.24) --
	(167.66, 25.28) --
	(171.07, 25.04) --
	(174.47, 24.94) --
	(177.87, 24.79) --
	(181.27, 24.73) --
	(184.67, 24.80) --
	(188.08, 24.63) --
	(191.48, 24.35);
\end{scope}
\begin{scope}
\path[clip] (  0.00,  0.00) rectangle (199.47, 93.95);
\definecolor{drawColor}{RGB}{0,0,0}

\path[draw=drawColor,line width= 0.4pt,line join=round,line cap=round] ( 18.12, 21.72) --
	(198.15, 21.72) --
	(198.15, 92.63) --
	( 18.12, 92.63) --
	( 18.12, 21.72);
\end{scope}
\begin{scope}
\path[clip] (  0.00,  0.00) rectangle (199.47, 93.95);
\definecolor{drawColor}{RGB}{0,0,0}

\path[draw=drawColor,line width= 0.4pt,line join=round,line cap=round] ( 24.79, 21.72) -- (193.18, 21.72);

\path[draw=drawColor,line width= 0.4pt,line join=round,line cap=round] ( 24.79, 21.72) -- ( 24.79, 18.17);

\path[draw=drawColor,line width= 0.4pt,line join=round,line cap=round] ( 57.11, 21.72) -- ( 57.11, 18.17);

\path[draw=drawColor,line width= 0.4pt,line join=round,line cap=round] ( 91.12, 21.72) -- ( 91.12, 18.17);

\path[draw=drawColor,line width= 0.4pt,line join=round,line cap=round] (125.14, 21.72) -- (125.14, 18.17);

\path[draw=drawColor,line width= 0.4pt,line join=round,line cap=round] (159.16, 21.72) -- (159.16, 18.17);

\path[draw=drawColor,line width= 0.4pt,line join=round,line cap=round] (193.18, 21.72) -- (193.18, 18.17);

\node[text=drawColor,anchor=base,inner sep=0pt, outer sep=0pt, scale=  0.70] at ( 24.79,  8.52) {1};

\node[text=drawColor,anchor=base,inner sep=0pt, outer sep=0pt, scale=  0.70] at ( 57.11,  8.52) {20};

\node[text=drawColor,anchor=base,inner sep=0pt, outer sep=0pt, scale=  0.70] at ( 91.12,  8.52) {40};

\node[text=drawColor,anchor=base,inner sep=0pt, outer sep=0pt, scale=  0.70] at (125.14,  8.52) {60};

\node[text=drawColor,anchor=base,inner sep=0pt, outer sep=0pt, scale=  0.70] at (159.16,  8.52) {80};

\node[text=drawColor,anchor=base,inner sep=0pt, outer sep=0pt, scale=  0.70] at (193.18,  8.52) {100};

\path[draw=drawColor,line width= 0.4pt,line join=round,line cap=round] ( 18.12, 21.72) -- ( 18.12, 90.00);

\path[draw=drawColor,line width= 0.4pt,line join=round,line cap=round] ( 18.12, 32.25) -- ( 14.57, 32.25);

\path[draw=drawColor,line width= 0.4pt,line join=round,line cap=round] ( 18.12, 51.50) -- ( 14.57, 51.50);

\path[draw=drawColor,line width= 0.4pt,line join=round,line cap=round] ( 18.12, 70.75) -- ( 14.57, 70.75);

\path[draw=drawColor,line width= 0.4pt,line join=round,line cap=round] ( 18.12, 90.00) -- ( 14.57, 90.00);

\node[text=drawColor,anchor=base east,inner sep=0pt, outer sep=0pt, scale=  0.70] at ( 14.52, 29.84) {0.4};

\node[text=drawColor,anchor=base east,inner sep=0pt, outer sep=0pt, scale=  0.70] at ( 14.52, 49.09) {0.6};

\node[text=drawColor,anchor=base east,inner sep=0pt, outer sep=0pt, scale=  0.70] at ( 14.52, 68.34) {0.8};

\node[text=drawColor,anchor=base east,inner sep=0pt, outer sep=0pt, scale=  0.70] at ( 14.52, 87.59) {1};
\end{scope}
\end{tikzpicture}

%% file: insider_smile.tex
\begin{tikzpicture}[x=1pt,y=1pt]
\definecolor{fillColor}{RGB}{255,255,255}
\path[use as bounding box,fill=fillColor,fill opacity=0.00] (0,0) rectangle (199.47, 93.95);
\begin{scope}
\path[clip] ( 18.12, 21.72) rectangle (198.15, 92.63);
\definecolor{drawColor}{RGB}{0,0,0}

\path[draw=drawColor,line width= 1.2pt,line join=round,line cap=round] ( 24.79, 78.32) --
	( 25.41, 77.95) --
	( 26.03, 77.55) --
	( 26.65, 77.13) --
	( 27.27, 76.69) --
	( 27.89, 76.23) --
	( 28.51, 75.76) --
	( 29.13, 75.28) --
	( 29.74, 74.78) --
	( 30.36, 74.28) --
	( 30.98, 73.76) --
	( 31.60, 73.24) --
	( 32.22, 72.71) --
	( 32.84, 72.17) --
	( 33.46, 71.63) --
	( 34.08, 71.09) --
	( 34.70, 70.54) --
	( 35.32, 69.99) --
	( 35.94, 69.43) --
	( 36.56, 68.87) --
	( 37.18, 68.31) --
	( 37.80, 67.75) --
	( 38.42, 67.19) --
	( 39.04, 66.63) --
	( 39.66, 66.06) --
	( 40.28, 65.50) --
	( 40.90, 64.93) --
	( 41.52, 64.37) --
	( 42.14, 63.80) --
	( 42.76, 63.24) --
	( 43.38, 62.68) --
	( 44.00, 62.11) --
	( 44.62, 61.55) --
	( 45.24, 60.99) --
	( 45.86, 60.44) --
	( 46.48, 59.88) --
	( 47.10, 59.33) --
	( 47.72, 58.78) --
	( 48.33, 58.23) --
	( 48.95, 57.69) --
	( 49.57, 57.15) --
	( 50.19, 56.61) --
	( 50.81, 56.07) --
	( 51.43, 55.54) --
	( 52.05, 55.02) --
	( 52.67, 54.49) --
	( 53.29, 53.98) --
	( 53.91, 53.46) --
	( 54.53, 52.96) --
	( 55.15, 52.45) --
	( 55.77, 51.96) --
	( 56.39, 51.47) --
	( 57.01, 50.98) --
	( 57.63, 50.51) --
	( 58.25, 50.04) --
	( 58.87, 49.57) --
	( 59.49, 49.12) --
	( 60.11, 48.67) --
	( 60.73, 48.23) --
	( 61.35, 47.80) --
	( 61.97, 47.38) --
	( 62.59, 46.96) --
	( 63.21, 46.55) --
	( 63.83, 46.16) --
	( 64.45, 45.77) --
	( 65.07, 45.39) --
	( 65.69, 45.02) --
	( 66.31, 44.67) --
	( 66.92, 44.32) --
	( 67.54, 43.98) --
	( 68.16, 43.65) --
	( 68.78, 43.33) --
	( 69.40, 43.01) --
	( 70.02, 42.71) --
	( 70.64, 42.42) --
	( 71.26, 42.14) --
	( 71.88, 41.86) --
	( 72.50, 41.60) --
	( 73.12, 41.34) --
	( 73.74, 41.09) --
	( 74.36, 40.85) --
	( 74.98, 40.62) --
	( 75.60, 40.40) --
	( 76.22, 40.18) --
	( 76.84, 39.97) --
	( 77.46, 39.77) --
	( 78.08, 39.57) --
	( 78.70, 39.38) --
	( 79.32, 39.20) --
	( 79.94, 39.02) --
	( 80.56, 38.85) --
	( 81.18, 38.68) --
	( 81.80, 38.52) --
	( 82.42, 38.36) --
	( 83.04, 38.21) --
	( 83.66, 38.06) --
	( 84.28, 37.91) --
	( 84.90, 37.77) --
	( 85.51, 37.64) --
	( 86.13, 37.50) --
	( 86.75, 37.37) --
	( 87.37, 37.25) --
	( 87.99, 37.12) --
	( 88.61, 37.00) --
	( 89.23, 36.89) --
	( 89.85, 36.77) --
	( 90.47, 36.66) --
	( 91.09, 36.55) --
	( 91.71, 36.44) --
	( 92.33, 36.34) --
	( 92.95, 36.24) --
	( 93.57, 36.13) --
	( 94.19, 36.04) --
	( 94.81, 35.94) --
	( 95.43, 35.85) --
	( 96.05, 35.75) --
	( 96.67, 35.66) --
	( 97.29, 35.58) --
	( 97.91, 35.49) --
	( 98.53, 35.40) --
	( 99.15, 35.32) --
	( 99.77, 35.24) --
	(100.39, 35.16) --
	(101.01, 35.08) --
	(101.63, 35.01) --
	(102.25, 34.93) --
	(102.87, 34.86) --
	(103.49, 34.79) --
	(104.10, 34.72) --
	(104.72, 34.66) --
	(105.34, 34.59) --
	(105.96, 34.53) --
	(106.58, 34.47) --
	(107.20, 34.41) --
	(107.82, 34.18) --
	(108.44, 34.04) --
	(109.06, 32.62) --
	(109.68, 32.60) --
	(110.30, 32.55) --
	(110.92, 32.49) --
	(111.54, 32.44) --
	(112.16, 32.39) --
	(112.78, 32.34) --
	(113.40, 32.28) --
	(114.02, 32.23) --
	(114.64, 32.18) --
	(115.26, 32.12) --
	(115.88, 32.07) --
	(116.50, 32.02) --
	(117.12, 31.97) --
	(117.74, 31.91) --
	(118.36, 31.86) --
	(118.98, 31.81) --
	(119.60, 31.76) --
	(120.22, 31.71) --
	(120.84, 31.66) --
	(121.46, 31.61) --
	(122.08, 31.56) --
	(122.69, 31.52) --
	(123.31, 31.47) --
	(123.93, 31.42) --
	(124.55, 31.38) --
	(125.17, 31.33) --
	(125.79, 31.29) --
	(126.41, 31.25) --
	(127.03, 31.21) --
	(127.65, 31.17) --
	(128.27, 31.13) --
	(128.89, 31.09) --
	(129.51, 31.06) --
	(130.13, 31.02) --
	(130.75, 30.99) --
	(131.37, 30.96) --
	(131.99, 30.93) --
	(132.61, 30.90) --
	(133.23, 30.88) --
	(133.85, 30.85) --
	(134.47, 30.83) --
	(135.09, 30.81) --
	(135.71, 30.80) --
	(136.33, 30.78) --
	(136.95, 30.77) --
	(137.57, 30.77) --
	(138.19, 30.76) --
	(138.81, 30.76) --
	(139.43, 30.76) --
	(140.05, 30.77) --
	(140.67, 30.77) --
	(141.28, 30.79) --
	(141.90, 30.80) --
	(142.52, 30.82) --
	(143.14, 30.85) --
	(143.76, 30.87) --
	(144.38, 30.91) --
	(145.00, 30.94) --
	(145.62, 30.98) --
	(146.24, 31.03) --
	(146.86, 31.08) --
	(147.48, 31.14) --
	(148.10, 31.20) --
	(148.72, 31.26) --
	(149.34, 31.33) --
	(149.96, 31.40) --
	(150.58, 31.48) --
	(151.20, 31.57) --
	(151.82, 31.65) --
	(152.44, 31.74) --
	(153.06, 31.84) --
	(153.68, 31.94) --
	(154.30, 32.04) --
	(154.92, 32.15) --
	(155.54, 32.26) --
	(156.16, 32.37) --
	(156.78, 32.49) --
	(157.40, 32.61) --
	(158.02, 32.73) --
	(158.64, 32.85) --
	(159.25, 32.98) --
	(159.87, 33.11) --
	(160.49, 33.23) --
	(161.11, 33.36) --
	(161.73, 33.49) --
	(162.35, 33.63) --
	(162.97, 33.76) --
	(163.59, 33.89) --
	(164.21, 34.02) --
	(164.83, 34.15) --
	(165.45, 34.28) --
	(166.07, 34.41) --
	(166.69, 34.54) --
	(167.31, 34.67) --
	(167.93, 34.80) --
	(168.55, 34.92) --
	(169.17, 35.05) --
	(169.79, 35.17) --
	(170.41, 35.29) --
	(171.03, 35.41) --
	(171.65, 35.52) --
	(172.27, 35.63) --
	(172.89, 35.74) --
	(173.51, 35.85) --
	(174.13, 35.96) --
	(174.75, 36.06) --
	(175.37, 36.15) --
	(175.99, 36.25) --
	(176.61, 36.34) --
	(177.23, 36.43) --
	(177.84, 36.51) --
	(178.46, 36.59) --
	(179.08, 36.66) --
	(179.70, 36.73) --
	(180.32, 36.80) --
	(180.94, 36.86) --
	(181.56, 36.92) --
	(182.18, 36.97) --
	(182.80, 37.02) --
	(183.42, 37.06) --
	(184.04, 37.09) --
	(184.66, 37.12) --
	(185.28, 37.14) --
	(185.90, 37.16) --
	(186.52, 37.17) --
	(187.14, 37.17) --
	(187.76, 37.17) --
	(188.38, 37.15) --
	(189.00, 37.13) --
	(189.62, 37.10) --
	(190.24, 37.06) --
	(190.86, 37.01) --
	(191.48, 36.95);
\end{scope}
\begin{scope}
\path[clip] (  0.00,  0.00) rectangle (199.47, 93.95);
\definecolor{drawColor}{RGB}{0,0,0}

\path[draw=drawColor,line width= 0.4pt,line join=round,line cap=round] ( 18.12, 21.72) --
	(198.15, 21.72) --
	(198.15, 92.63) --
	( 18.12, 92.63) --
	( 18.12, 21.72);
\end{scope}
\begin{scope}
\path[clip] (  0.00,  0.00) rectangle (199.47, 93.95);
\definecolor{drawColor}{RGB}{0,0,0}

\path[draw=drawColor,line width= 0.4pt,line join=round,line cap=round] ( 24.17, 21.72) -- (198.15, 21.72);

\path[draw=drawColor,line width= 0.4pt,line join=round,line cap=round] ( 24.17, 21.72) -- ( 24.17, 18.17);

\path[draw=drawColor,line width= 0.4pt,line join=round,line cap=round] ( 55.15, 21.72) -- ( 55.15, 18.17);

\path[draw=drawColor,line width= 0.4pt,line join=round,line cap=round] ( 86.13, 21.72) -- ( 86.13, 18.17);

\path[draw=drawColor,line width= 0.4pt,line join=round,line cap=round] (117.12, 21.72) -- (117.12, 18.17);

\path[draw=drawColor,line width= 0.4pt,line join=round,line cap=round] (148.10, 21.72) -- (148.10, 18.17);

\path[draw=drawColor,line width= 0.4pt,line join=round,line cap=round] (179.08, 21.72) -- (179.08, 18.17);

\node[text=drawColor,anchor=base,inner sep=0pt, outer sep=0pt, scale=  0.70] at ( 24.17,  8.52) {0.01};

\node[text=drawColor,anchor=base,inner sep=0pt, outer sep=0pt, scale=  0.70] at ( 55.15,  8.52) {0.34};

\node[text=drawColor,anchor=base,inner sep=0pt, outer sep=0pt, scale=  0.70] at ( 86.13,  8.52) {0.67};

\node[text=drawColor,anchor=base,inner sep=0pt, outer sep=0pt, scale=  0.70] at (117.12,  8.52) {1.00};

\node[text=drawColor,anchor=base,inner sep=0pt, outer sep=0pt, scale=  0.70] at (148.10,  8.52) {1.33};

\node[text=drawColor,anchor=base,inner sep=0pt, outer sep=0pt, scale=  0.70] at (179.08,  8.52) {1.66};

\path[draw=drawColor,line width= 0.4pt,line join=round,line cap=round] ( 18.12, 24.35) -- ( 18.12, 90.00);

\path[draw=drawColor,line width= 0.4pt,line join=round,line cap=round] ( 18.12, 24.35) -- ( 14.57, 24.35);

\path[draw=drawColor,line width= 0.4pt,line join=round,line cap=round] ( 18.12, 35.29) -- ( 14.57, 35.29);

\path[draw=drawColor,line width= 0.4pt,line join=round,line cap=round] ( 18.12, 46.23) -- ( 14.57, 46.23);

\path[draw=drawColor,line width= 0.4pt,line join=round,line cap=round] ( 18.12, 57.18) -- ( 14.57, 57.18);

\path[draw=drawColor,line width= 0.4pt,line join=round,line cap=round] ( 18.12, 68.12) -- ( 14.57, 68.12);

\path[draw=drawColor,line width= 0.4pt,line join=round,line cap=round] ( 18.12, 79.06) -- ( 14.57, 79.06);

\path[draw=drawColor,line width= 0.4pt,line join=round,line cap=round] ( 18.12, 90.00) -- ( 14.57, 90.00);

\node[text=drawColor,anchor=base east,inner sep=0pt, outer sep=0pt, scale=  0.70] at ( 14.52, 21.94) {0.2};

\node[text=drawColor,anchor=base east,inner sep=0pt, outer sep=0pt, scale=  0.70] at ( 14.52, 32.88) {0.3};

\node[text=drawColor,anchor=base east,inner sep=0pt, outer sep=0pt, scale=  0.70] at ( 14.52, 43.82) {0.4};

\node[text=drawColor,anchor=base east,inner sep=0pt, outer sep=0pt, scale=  0.70] at ( 14.52, 54.76) {0.5};

\node[text=drawColor,anchor=base east,inner sep=0pt, outer sep=0pt, scale=  0.70] at ( 14.52, 65.71) {0.6};

\node[text=drawColor,anchor=base east,inner sep=0pt, outer sep=0pt, scale=  0.70] at ( 14.52, 76.65) {0.7};

\node[text=drawColor,anchor=base east,inner sep=0pt, outer sep=0pt, scale=  0.70] at ( 14.52, 87.59) {0.8};
\end{scope}
\end{tikzpicture}

%% file: insider_smile_multiple.tex
\begin{tikzpicture}[x=1pt,y=1pt]
\definecolor{fillColor}{RGB}{255,255,255}
\path[use as bounding box,fill=fillColor,fill opacity=0.00] (0,0) rectangle (199.47, 93.95);
\begin{scope}
\path[clip] ( 18.12, 21.72) rectangle (198.15, 92.63);
\definecolor{drawColor}{RGB}{0,0,0}

\path[draw=drawColor,line width= 1.2pt,line join=round,line cap=round] ( 24.79, 78.32) --
	( 25.41, 77.95) --
	( 26.03, 77.55) --
	( 26.65, 77.13) --
	( 27.27, 76.69) --
	( 27.89, 76.23) --
	( 28.51, 75.76) --
	( 29.13, 75.28) --
	( 29.74, 74.78) --
	( 30.36, 74.28) --
	( 30.98, 73.76) --
	( 31.60, 73.24) --
	( 32.22, 72.71) --
	( 32.84, 72.17) --
	( 33.46, 71.63) --
	( 34.08, 71.09) --
	( 34.70, 70.54) --
	( 35.32, 69.99) --
	( 35.94, 69.43) --
	( 36.56, 68.87) --
	( 37.18, 68.31) --
	( 37.80, 67.75) --
	( 38.42, 67.19) --
	( 39.04, 66.63) --
	( 39.66, 66.06) --
	( 40.28, 65.50) --
	( 40.90, 64.93) --
	( 41.52, 64.37) --
	( 42.14, 63.80) --
	( 42.76, 63.24) --
	( 43.38, 62.68) --
	( 44.00, 62.11) --
	( 44.62, 61.55) --
	( 45.24, 60.99) --
	( 45.86, 60.44) --
	( 46.48, 59.88) --
	( 47.10, 59.33) --
	( 47.72, 58.78) --
	( 48.33, 58.23) --
	( 48.95, 57.69) --
	( 49.57, 57.15) --
	( 50.19, 56.61) --
	( 50.81, 56.07) --
	( 51.43, 55.54) --
	( 52.05, 55.02) --
	( 52.67, 54.49) --
	( 53.29, 53.98) --
	( 53.91, 53.46) --
	( 54.53, 52.96) --
	( 55.15, 52.45) --
	( 55.77, 51.96) --
	( 56.39, 51.47) --
	( 57.01, 50.98) --
	( 57.63, 50.51) --
	( 58.25, 50.04) --
	( 58.87, 49.57) --
	( 59.49, 49.12) --
	( 60.11, 48.67) --
	( 60.73, 48.23) --
	( 61.35, 47.80) --
	( 61.97, 47.38) --
	( 62.59, 46.96) --
	( 63.21, 46.55) --
	( 63.83, 46.16) --
	( 64.45, 45.77) --
	( 65.07, 45.39) --
	( 65.69, 45.02) --
	( 66.31, 44.67) --
	( 66.92, 44.32) --
	( 67.54, 43.98) --
	( 68.16, 43.65) --
	( 68.78, 43.33) --
	( 69.40, 43.01) --
	( 70.02, 42.71) --
	( 70.64, 42.42) --
	( 71.26, 42.14) --
	( 71.88, 41.86) --
	( 72.50, 41.60) --
	( 73.12, 41.34) --
	( 73.74, 41.09) --
	( 74.36, 40.85) --
	( 74.98, 40.62) --
	( 75.60, 40.40) --
	( 76.22, 40.18) --
	( 76.84, 39.97) --
	( 77.46, 39.77) --
	( 78.08, 39.57) --
	( 78.70, 39.38) --
	( 79.32, 39.20) --
	( 79.94, 39.02) --
	( 80.56, 38.85) --
	( 81.18, 38.68) --
	( 81.80, 38.52) --
	( 82.42, 38.36) --
	( 83.04, 38.21) --
	( 83.66, 38.06) --
	( 84.28, 37.91) --
	( 84.90, 37.77) --
	( 85.51, 37.64) --
	( 86.13, 37.50) --
	( 86.75, 37.37) --
	( 87.37, 37.25) --
	( 87.99, 37.12) --
	( 88.61, 37.00) --
	( 89.23, 36.89) --
	( 89.85, 36.77) --
	( 90.47, 36.66) --
	( 91.09, 36.55) --
	( 91.71, 36.44) --
	( 92.33, 36.34) --
	( 92.95, 36.24) --
	( 93.57, 36.13) --
	( 94.19, 36.04) --
	( 94.81, 35.94) --
	( 95.43, 35.85) --
	( 96.05, 35.75) --
	( 96.67, 35.66) --
	( 97.29, 35.58) --
	( 97.91, 35.49) --
	( 98.53, 35.40) --
	( 99.15, 35.32) --
	( 99.77, 35.24) --
	(100.39, 35.16) --
	(101.01, 35.08) --
	(101.63, 35.01) --
	(102.25, 34.93) --
	(102.87, 34.86) --
	(103.49, 34.79) --
	(104.10, 34.72) --
	(104.72, 34.66) --
	(105.34, 34.59) --
	(105.96, 34.53) --
	(106.58, 34.47) --
	(107.20, 34.41) --
	(107.82, 34.18) --
	(108.44, 34.04) --
	(109.06, 32.62) --
	(109.68, 32.60) --
	(110.30, 32.55) --
	(110.92, 32.49) --
	(111.54, 32.44) --
	(112.16, 32.39) --
	(112.78, 32.34) --
	(113.40, 32.28) --
	(114.02, 32.23) --
	(114.64, 32.18) --
	(115.26, 32.12) --
	(115.88, 32.07) --
	(116.50, 32.02) --
	(117.12, 31.97) --
	(117.74, 31.91) --
	(118.36, 31.86) --
	(118.98, 31.81) --
	(119.60, 31.76) --
	(120.22, 31.71) --
	(120.84, 31.66) --
	(121.46, 31.61) --
	(122.08, 31.56) --
	(122.69, 31.52) --
	(123.31, 31.47) --
	(123.93, 31.42) --
	(124.55, 31.38) --
	(125.17, 31.33) --
	(125.79, 31.29) --
	(126.41, 31.25) --
	(127.03, 31.21) --
	(127.65, 31.17) --
	(128.27, 31.13) --
	(128.89, 31.09) --
	(129.51, 31.06) --
	(130.13, 31.02) --
	(130.75, 30.99) --
	(131.37, 30.96) --
	(131.99, 30.93) --
	(132.61, 30.90) --
	(133.23, 30.88) --
	(133.85, 30.85) --
	(134.47, 30.83) --
	(135.09, 30.81) --
	(135.71, 30.80) --
	(136.33, 30.78) --
	(136.95, 30.77) --
	(137.57, 30.77) --
	(138.19, 30.76) --
	(138.81, 30.76) --
	(139.43, 30.76) --
	(140.05, 30.77) --
	(140.67, 30.77) --
	(141.28, 30.79) --
	(141.90, 30.80) --
	(142.52, 30.82) --
	(143.14, 30.85) --
	(143.76, 30.87) --
	(144.38, 30.91) --
	(145.00, 30.94) --
	(145.62, 30.98) --
	(146.24, 31.03) --
	(146.86, 31.08) --
	(147.48, 31.14) --
	(148.10, 31.20) --
	(148.72, 31.26) --
	(149.34, 31.33) --
	(149.96, 31.40) --
	(150.58, 31.48) --
	(151.20, 31.57) --
	(151.82, 31.65) --
	(152.44, 31.74) --
	(153.06, 31.84) --
	(153.68, 31.94) --
	(154.30, 32.04) --
	(154.92, 32.15) --
	(155.54, 32.26) --
	(156.16, 32.37) --
	(156.78, 32.49) --
	(157.40, 32.61) --
	(158.02, 32.73) --
	(158.64, 32.85) --
	(159.25, 32.98) --
	(159.87, 33.11) --
	(160.49, 33.23) --
	(161.11, 33.36) --
	(161.73, 33.49) --
	(162.35, 33.63) --
	(162.97, 33.76) --
	(163.59, 33.89) --
	(164.21, 34.02) --
	(164.83, 34.15) --
	(165.45, 34.28) --
	(166.07, 34.41) --
	(166.69, 34.54) --
	(167.31, 34.67) --
	(167.93, 34.80) --
	(168.55, 34.92) --
	(169.17, 35.05) --
	(169.79, 35.17) --
	(170.41, 35.29) --
	(171.03, 35.41) --
	(171.65, 35.52) --
	(172.27, 35.63) --
	(172.89, 35.74) --
	(173.51, 35.85) --
	(174.13, 35.96) --
	(174.75, 36.06) --
	(175.37, 36.15) --
	(175.99, 36.25) --
	(176.61, 36.34) --
	(177.23, 36.43) --
	(177.84, 36.51) --
	(178.46, 36.59) --
	(179.08, 36.66) --
	(179.70, 36.73) --
	(180.32, 36.80) --
	(180.94, 36.86) --
	(181.56, 36.92) --
	(182.18, 36.97) --
	(182.80, 37.02) --
	(183.42, 37.06) --
	(184.04, 37.09) --
	(184.66, 37.12) --
	(185.28, 37.14) --
	(185.90, 37.16) --
	(186.52, 37.17) --
	(187.14, 37.17) --
	(187.76, 37.17) --
	(188.38, 37.15) --
	(189.00, 37.13) --
	(189.62, 37.10) --
	(190.24, 37.06) --
	(190.86, 37.01) --
	(191.48, 36.95);
\end{scope}
\begin{scope}
\path[clip] (  0.00,  0.00) rectangle (199.47, 93.95);
\definecolor{drawColor}{RGB}{0,0,0}

\path[draw=drawColor,line width= 0.4pt,line join=round,line cap=round] ( 18.12, 21.72) --
	(198.15, 21.72) --
	(198.15, 92.63) --
	( 18.12, 92.63) --
	( 18.12, 21.72);
\end{scope}
\begin{scope}
\path[clip] (  0.00,  0.00) rectangle (199.47, 93.95);
\definecolor{drawColor}{RGB}{0,0,0}

\path[draw=drawColor,line width= 0.4pt,line join=round,line cap=round] ( 24.17, 21.72) -- (198.15, 21.72);

\path[draw=drawColor,line width= 0.4pt,line join=round,line cap=round] ( 24.17, 21.72) -- ( 24.17, 18.17);

\path[draw=drawColor,line width= 0.4pt,line join=round,line cap=round] ( 55.15, 21.72) -- ( 55.15, 18.17);

\path[draw=drawColor,line width= 0.4pt,line join=round,line cap=round] ( 86.13, 21.72) -- ( 86.13, 18.17);

\path[draw=drawColor,line width= 0.4pt,line join=round,line cap=round] (117.12, 21.72) -- (117.12, 18.17);

\path[draw=drawColor,line width= 0.4pt,line join=round,line cap=round] (148.10, 21.72) -- (148.10, 18.17);

\path[draw=drawColor,line width= 0.4pt,line join=round,line cap=round] (179.08, 21.72) -- (179.08, 18.17);

\node[text=drawColor,anchor=base,inner sep=0pt, outer sep=0pt, scale=  0.70] at ( 24.17,  8.52) {0.01};

\node[text=drawColor,anchor=base,inner sep=0pt, outer sep=0pt, scale=  0.70] at ( 55.15,  8.52) {0.34};

\node[text=drawColor,anchor=base,inner sep=0pt, outer sep=0pt, scale=  0.70] at ( 86.13,  8.52) {0.67};

\node[text=drawColor,anchor=base,inner sep=0pt, outer sep=0pt, scale=  0.70] at (117.12,  8.52) {1.00};

\node[text=drawColor,anchor=base,inner sep=0pt, outer sep=0pt, scale=  0.70] at (148.10,  8.52) {1.33};

\node[text=drawColor,anchor=base,inner sep=0pt, outer sep=0pt, scale=  0.70] at (179.08,  8.52) {1.66};

\path[draw=drawColor,line width= 0.4pt,line join=round,line cap=round] ( 18.12, 24.35) -- ( 18.12, 90.00);

\path[draw=drawColor,line width= 0.4pt,line join=round,line cap=round] ( 18.12, 24.35) -- ( 14.57, 24.35);

\path[draw=drawColor,line width= 0.4pt,line join=round,line cap=round] ( 18.12, 35.29) -- ( 14.57, 35.29);

\path[draw=drawColor,line width= 0.4pt,line join=round,line cap=round] ( 18.12, 46.23) -- ( 14.57, 46.23);

\path[draw=drawColor,line width= 0.4pt,line join=round,line cap=round] ( 18.12, 57.18) -- ( 14.57, 57.18);

\path[draw=drawColor,line width= 0.4pt,line join=round,line cap=round] ( 18.12, 68.12) -- ( 14.57, 68.12);

\path[draw=drawColor,line width= 0.4pt,line join=round,line cap=round] ( 18.12, 79.06) -- ( 14.57, 79.06);

\path[draw=drawColor,line width= 0.4pt,line join=round,line cap=round] ( 18.12, 90.00) -- ( 14.57, 90.00);

\node[text=drawColor,anchor=base east,inner sep=0pt, outer sep=0pt, scale=  0.70] at ( 14.52, 21.94) {0.2};

\node[text=drawColor,anchor=base east,inner sep=0pt, outer sep=0pt, scale=  0.70] at ( 14.52, 32.88) {0.3};

\node[text=drawColor,anchor=base east,inner sep=0pt, outer sep=0pt, scale=  0.70] at ( 14.52, 43.82) {0.4};

\node[text=drawColor,anchor=base east,inner sep=0pt, outer sep=0pt, scale=  0.70] at ( 14.52, 54.76) {0.5};

\node[text=drawColor,anchor=base east,inner sep=0pt, outer sep=0pt, scale=  0.70] at ( 14.52, 65.71) {0.6};

\node[text=drawColor,anchor=base east,inner sep=0pt, outer sep=0pt, scale=  0.70] at ( 14.52, 76.65) {0.7};

\node[text=drawColor,anchor=base east,inner sep=0pt, outer sep=0pt, scale=  0.70] at ( 14.52, 87.59) {0.8};
\end{scope}
\begin{scope}
\path[clip] ( 18.12, 21.72) rectangle (198.15, 92.63);
\definecolor{drawColor}{RGB}{0,0,0}

\path[draw=drawColor,line width= 1.2pt,dash pattern=on 4pt off 4pt ,line join=round,line cap=round] ( 24.79, 83.75) --
	( 25.41, 83.38) --
	( 26.03, 82.98) --
	( 26.65, 82.56) --
	( 27.27, 82.11) --
	( 27.89, 81.64) --
	( 28.51, 81.16) --
	( 29.13, 80.66) --
	( 29.74, 80.14) --
	( 30.36, 79.62) --
	( 30.98, 79.08) --
	( 31.60, 78.54) --
	( 32.22, 77.98) --
	( 32.84, 77.42) --
	( 33.46, 76.85) --
	( 34.08, 76.28) --
	( 34.70, 75.70) --
	( 35.32, 75.12) --
	( 35.94, 74.53) --
	( 36.56, 73.94) --
	( 37.18, 73.35) --
	( 37.80, 72.75) --
	( 38.42, 72.15) --
	( 39.04, 71.55) --
	( 39.66, 70.95) --
	( 40.28, 70.34) --
	( 40.90, 69.74) --
	( 41.52, 69.13) --
	( 42.14, 68.53) --
	( 42.76, 67.92) --
	( 43.38, 67.31) --
	( 44.00, 66.71) --
	( 44.62, 66.10) --
	( 45.24, 65.49) --
	( 45.86, 64.89) --
	( 46.48, 64.28) --
	( 47.10, 63.68) --
	( 47.72, 63.08) --
	( 48.33, 62.47) --
	( 48.95, 61.87) --
	( 49.57, 61.27) --
	( 50.19, 60.68) --
	( 50.81, 60.08) --
	( 51.43, 59.49) --
	( 52.05, 58.90) --
	( 52.67, 58.31) --
	( 53.29, 57.73) --
	( 53.91, 57.15) --
	( 54.53, 56.57) --
	( 55.15, 55.99) --
	( 55.77, 55.42) --
	( 56.39, 54.86) --
	( 57.01, 54.30) --
	( 57.63, 53.74) --
	( 58.25, 53.19) --
	( 58.87, 52.65) --
	( 59.49, 52.11) --
	( 60.11, 51.58) --
	( 60.73, 51.06) --
	( 61.35, 50.54) --
	( 61.97, 50.04) --
	( 62.59, 49.54) --
	( 63.21, 49.05) --
	( 63.83, 48.57) --
	( 64.45, 48.10) --
	( 65.07, 47.64) --
	( 65.69, 47.19) --
	( 66.31, 46.75) --
	( 66.92, 46.33) --
	( 67.54, 45.91) --
	( 68.16, 45.51) --
	( 68.78, 45.11) --
	( 69.40, 44.73) --
	( 70.02, 44.37) --
	( 70.64, 44.01) --
	( 71.26, 43.66) --
	( 71.88, 43.33) --
	( 72.50, 43.01) --
	( 73.12, 42.70) --
	( 73.74, 42.40) --
	( 74.36, 42.11) --
	( 74.98, 41.84) --
	( 75.60, 41.57) --
	( 76.22, 41.31) --
	( 76.84, 41.06) --
	( 77.46, 40.82) --
	( 78.08, 40.59) --
	( 78.70, 40.37) --
	( 79.32, 40.15) --
	( 79.94, 39.95) --
	( 80.56, 39.75) --
	( 81.18, 39.55) --
	( 81.80, 39.36) --
	( 82.42, 39.18) --
	( 83.04, 39.01) --
	( 83.66, 38.84) --
	( 84.28, 38.67) --
	( 84.90, 38.51) --
	( 85.51, 38.36) --
	( 86.13, 38.21) --
	( 86.75, 38.06) --
	( 87.37, 37.92) --
	( 87.99, 37.78) --
	( 88.61, 37.64) --
	( 89.23, 37.51) --
	( 89.85, 37.38) --
	( 90.47, 37.25) --
	( 91.09, 37.13) --
	( 91.71, 37.01) --
	( 92.33, 36.89) --
	( 92.95, 36.78) --
	( 93.57, 36.67) --
	( 94.19, 36.56) --
	( 94.81, 36.45) --
	( 95.43, 36.34) --
	( 96.05, 36.24) --
	( 96.67, 36.14) --
	( 97.29, 36.04) --
	( 97.91, 35.94) --
	( 98.53, 35.85) --
	( 99.15, 35.75) --
	( 99.77, 35.66) --
	(100.39, 35.57) --
	(101.01, 35.48) --
	(101.63, 35.40) --
	(102.25, 35.31) --
	(102.87, 35.23) --
	(103.49, 35.15) --
	(104.10, 35.07) --
	(104.72, 34.99) --
	(105.34, 34.91) --
	(105.96, 34.84) --
	(106.58, 34.77) --
	(107.20, 34.70) --
	(107.82, 34.63) --
	(108.44, 33.46) --
	(109.06, 33.39) --
	(109.68, 33.39) --
	(110.30, 33.33) --
	(110.92, 33.26) --
	(111.54, 33.20) --
	(112.16, 33.13) --
	(112.78, 33.07) --
	(113.40, 33.00) --
	(114.02, 32.94) --
	(114.64, 32.88) --
	(115.26, 32.81) --
	(115.88, 32.75) --
	(116.50, 32.69) --
	(117.12, 32.63) --
	(117.74, 32.57) --
	(118.36, 32.52) --
	(118.98, 32.46) --
	(119.60, 32.40) --
	(120.22, 32.34) --
	(120.84, 32.29) --
	(121.46, 32.24) --
	(122.08, 32.18) --
	(122.69, 32.13) --
	(123.31, 32.08) --
	(123.93, 32.03) --
	(124.55, 31.98) --
	(125.17, 31.93) --
	(125.79, 31.89) --
	(126.41, 31.84) --
	(127.03, 31.80) --
	(127.65, 31.76) --
	(128.27, 31.72) --
	(128.89, 31.68) --
	(129.51, 31.65) --
	(130.13, 31.61) --
	(130.75, 31.58) --
	(131.37, 31.55) --
	(131.99, 31.53) --
	(132.61, 31.50) --
	(133.23, 31.48) --
	(133.85, 31.46) --
	(134.47, 31.44) --
	(135.09, 31.43) --
	(135.71, 31.42) --
	(136.33, 31.41) --
	(136.95, 31.41) --
	(137.57, 31.41) --
	(138.19, 31.42) --
	(138.81, 31.43) --
	(139.43, 31.44) --
	(140.05, 31.46) --
	(140.67, 31.48) --
	(141.28, 31.51) --
	(141.90, 31.54) --
	(142.52, 31.58) --
	(143.14, 31.63) --
	(143.76, 31.68) --
	(144.38, 31.73) --
	(145.00, 31.79) --
	(145.62, 31.86) --
	(146.24, 31.94) --
	(146.86, 32.02) --
	(147.48, 32.11) --
	(148.10, 32.20) --
	(148.72, 32.30) --
	(149.34, 32.40) --
	(149.96, 32.51) --
	(150.58, 32.63) --
	(151.20, 32.75) --
	(151.82, 32.88) --
	(152.44, 33.01) --
	(153.06, 33.15) --
	(153.68, 33.29) --
	(154.30, 33.44) --
	(154.92, 33.58) --
	(155.54, 33.74) --
	(156.16, 33.89) --
	(156.78, 34.05) --
	(157.40, 34.20) --
	(158.02, 34.36) --
	(158.64, 34.53) --
	(159.25, 34.69) --
	(159.87, 34.85) --
	(160.49, 35.01) --
	(161.11, 35.18) --
	(161.73, 35.34) --
	(162.35, 35.50) --
	(162.97, 35.66) --
	(163.59, 35.82) --
	(164.21, 35.98) --
	(164.83, 36.13) --
	(165.45, 36.29) --
	(166.07, 36.44) --
	(166.69, 36.59) --
	(167.31, 36.73) --
	(167.93, 36.88) --
	(168.55, 37.02) --
	(169.17, 37.16) --
	(169.79, 37.29) --
	(170.41, 37.42) --
	(171.03, 37.55) --
	(171.65, 37.68) --
	(172.27, 37.80) --
	(172.89, 37.91) --
	(173.51, 38.03) --
	(174.13, 38.14) --
	(174.75, 38.24) --
	(175.37, 38.34) --
	(175.99, 38.44) --
	(176.61, 38.53) --
	(177.23, 38.62) --
	(177.84, 38.70) --
	(178.46, 38.78) --
	(179.08, 38.86) --
	(179.70, 38.93) --
	(180.32, 38.99) --
	(180.94, 39.05) --
	(181.56, 39.10) --
	(182.18, 39.14) --
	(182.80, 39.18) --
	(183.42, 39.22) --
	(184.04, 39.25) --
	(184.66, 39.27) --
	(185.28, 39.28) --
	(185.90, 39.28) --
	(186.52, 39.28) --
	(187.14, 39.27) --
	(187.76, 39.26) --
	(188.38, 39.23) --
	(189.00, 39.19) --
	(189.62, 39.14) --
	(190.24, 39.09) --
	(190.86, 39.01) --
	(191.48, 38.93);

\path[draw=drawColor,line width= 1.2pt,dash pattern=on 1pt off 3pt ,line join=round,line cap=round] ( 24.79, 88.14) --
	( 25.41, 87.76) --
	( 26.03, 87.35) --
	( 26.65, 86.91) --
	( 27.27, 86.45) --
	( 27.89, 85.96) --
	( 28.51, 85.46) --
	( 29.13, 84.94) --
	( 29.74, 84.40) --
	( 30.36, 83.85) --
	( 30.98, 83.29) --
	( 31.60, 82.72) --
	( 32.22, 82.15) --
	( 32.84, 81.56) --
	( 33.46, 80.97) --
	( 34.08, 80.37) --
	( 34.70, 79.76) --
	( 35.32, 79.15) --
	( 35.94, 78.53) --
	( 36.56, 77.92) --
	( 37.18, 77.29) --
	( 37.80, 76.67) --
	( 38.42, 76.04) --
	( 39.04, 75.41) --
	( 39.66, 74.78) --
	( 40.28, 74.15) --
	( 40.90, 73.51) --
	( 41.52, 72.88) --
	( 42.14, 72.24) --
	( 42.76, 71.60) --
	( 43.38, 70.97) --
	( 44.00, 70.33) --
	( 44.62, 69.69) --
	( 45.24, 69.06) --
	( 45.86, 68.42) --
	( 46.48, 67.78) --
	( 47.10, 67.15) --
	( 47.72, 66.51) --
	( 48.33, 65.88) --
	( 48.95, 65.25) --
	( 49.57, 64.62) --
	( 50.19, 63.99) --
	( 50.81, 63.37) --
	( 51.43, 62.74) --
	( 52.05, 62.12) --
	( 52.67, 61.50) --
	( 53.29, 60.88) --
	( 53.91, 60.27) --
	( 54.53, 59.66) --
	( 55.15, 59.06) --
	( 55.77, 58.46) --
	( 56.39, 57.86) --
	( 57.01, 57.27) --
	( 57.63, 56.68) --
	( 58.25, 56.10) --
	( 58.87, 55.52) --
	( 59.49, 54.95) --
	( 60.11, 54.39) --
	( 60.73, 53.83) --
	( 61.35, 53.29) --
	( 61.97, 52.75) --
	( 62.59, 52.22) --
	( 63.21, 51.69) --
	( 63.83, 51.18) --
	( 64.45, 50.68) --
	( 65.07, 50.19) --
	( 65.69, 49.71) --
	( 66.31, 49.24) --
	( 66.92, 48.78) --
	( 67.54, 48.33) --
	( 68.16, 47.90) --
	( 68.78, 47.48) --
	( 69.40, 47.07) --
	( 70.02, 46.67) --
	( 70.64, 46.28) --
	( 71.26, 45.91) --
	( 71.88, 45.55) --
	( 72.50, 45.20) --
	( 73.12, 44.87) --
	( 73.74, 44.54) --
	( 74.36, 44.23) --
	( 74.98, 43.93) --
	( 75.60, 43.63) --
	( 76.22, 43.35) --
	( 76.84, 43.08) --
	( 77.46, 42.82) --
	( 78.08, 42.57) --
	( 78.70, 42.33) --
	( 79.32, 42.09) --
	( 79.94, 41.87) --
	( 80.56, 41.65) --
	( 81.18, 41.44) --
	( 81.80, 41.23) --
	( 82.42, 41.04) --
	( 83.04, 40.85) --
	( 83.66, 40.66) --
	( 84.28, 40.48) --
	( 84.90, 40.31) --
	( 85.51, 40.14) --
	( 86.13, 39.98) --
	( 86.75, 39.82) --
	( 87.37, 39.66) --
	( 87.99, 39.51) --
	( 88.61, 39.36) --
	( 89.23, 39.22) --
	( 89.85, 39.08) --
	( 90.47, 38.94) --
	( 91.09, 38.81) --
	( 91.71, 38.68) --
	( 92.33, 38.55) --
	( 92.95, 38.43) --
	( 93.57, 38.30) --
	( 94.19, 38.18) --
	( 94.81, 38.07) --
	( 95.43, 37.95) --
	( 96.05, 37.84) --
	( 96.67, 37.73) --
	( 97.29, 37.62) --
	( 97.91, 37.51) --
	( 98.53, 37.41) --
	( 99.15, 37.31) --
	( 99.77, 37.21) --
	(100.39, 37.11) --
	(101.01, 37.01) --
	(101.63, 36.91) --
	(102.25, 36.82) --
	(102.87, 36.73) --
	(103.49, 36.64) --
	(104.10, 36.55) --
	(104.72, 36.46) --
	(105.34, 36.38) --
	(105.96, 36.30) --
	(106.58, 36.21) --
	(107.20, 36.13) --
	(107.82, 36.06) --
	(108.44, 35.23) --
	(109.06, 35.15) --
	(109.68, 35.15) --
	(110.30, 35.08) --
	(110.92, 35.00) --
	(111.54, 34.93) --
	(112.16, 34.86) --
	(112.78, 34.79) --
	(113.40, 34.72) --
	(114.02, 34.65) --
	(114.64, 34.58) --
	(115.26, 34.51) --
	(115.88, 34.44) --
	(116.50, 34.37) --
	(117.12, 34.31) --
	(117.74, 34.24) --
	(118.36, 34.18) --
	(118.98, 34.12) --
	(119.60, 34.06) --
	(120.22, 33.99) --
	(120.84, 33.94) --
	(121.46, 33.88) --
	(122.08, 33.82) --
	(122.69, 33.76) --
	(123.31, 33.71) --
	(123.93, 33.66) --
	(124.55, 33.61) --
	(125.17, 33.56) --
	(125.79, 33.51) --
	(126.41, 33.46) --
	(127.03, 33.42) --
	(127.65, 33.37) --
	(128.27, 33.33) --
	(128.89, 33.29) --
	(129.51, 33.26) --
	(130.13, 33.22) --
	(130.75, 33.19) --
	(131.37, 33.16) --
	(131.99, 33.13) --
	(132.61, 33.11) --
	(133.23, 33.09) --
	(133.85, 33.07) --
	(134.47, 33.05) --
	(135.09, 33.04) --
	(135.71, 33.04) --
	(136.33, 33.03) --
	(136.95, 33.03) --
	(137.57, 33.04) --
	(138.19, 33.05) --
	(138.81, 33.06) --
	(139.43, 33.08) --
	(140.05, 33.10) --
	(140.67, 33.13) --
	(141.28, 33.17) --
	(141.90, 33.21) --
	(142.52, 33.26) --
	(143.14, 33.31) --
	(143.76, 33.37) --
	(144.38, 33.43) --
	(145.00, 33.50) --
	(145.62, 33.58) --
	(146.24, 33.67) --
	(146.86, 33.76) --
	(147.48, 33.86) --
	(148.10, 33.96) --
	(148.72, 34.07) --
	(149.34, 34.19) --
	(149.96, 34.31) --
	(150.58, 34.44) --
	(151.20, 34.57) --
	(151.82, 34.71) --
	(152.44, 34.85) --
	(153.06, 35.00) --
	(153.68, 35.15) --
	(154.30, 35.31) --
	(154.92, 35.47) --
	(155.54, 35.63) --
	(156.16, 35.80) --
	(156.78, 35.96) --
	(157.40, 36.13) --
	(158.02, 36.30) --
	(158.64, 36.47) --
	(159.25, 36.65) --
	(159.87, 36.82) --
	(160.49, 36.99) --
	(161.11, 37.16) --
	(161.73, 37.33) --
	(162.35, 37.50) --
	(162.97, 37.67) --
	(163.59, 37.84) --
	(164.21, 38.00) --
	(164.83, 38.17) --
	(165.45, 38.33) --
	(166.07, 38.49) --
	(166.69, 38.64) --
	(167.31, 38.80) --
	(167.93, 38.95) --
	(168.55, 39.09) --
	(169.17, 39.24) --
	(169.79, 39.38) --
	(170.41, 39.51) --
	(171.03, 39.65) --
	(171.65, 39.78) --
	(172.27, 39.90) --
	(172.89, 40.02) --
	(173.51, 40.14) --
	(174.13, 40.26) --
	(174.75, 40.36) --
	(175.37, 40.47) --
	(175.99, 40.57) --
	(176.61, 40.66) --
	(177.23, 40.75) --
	(177.84, 40.84) --
	(178.46, 40.92) --
	(179.08, 40.99) --
	(179.70, 41.06) --
	(180.32, 41.13) --
	(180.94, 41.18) --
	(181.56, 41.24) --
	(182.18, 41.28) --
	(182.80, 41.32) --
	(183.42, 41.36) --
	(184.04, 41.38) --
	(184.66, 41.40) --
	(185.28, 41.41) --
	(185.90, 41.41) --
	(186.52, 41.41) --
	(187.14, 41.40) --
	(187.76, 41.37) --
	(188.38, 41.34) --
	(189.00, 41.30) --
	(189.62, 41.25) --
	(190.24, 41.18) --
	(190.86, 41.10) --
	(191.48, 41.01);
\end{scope}
\end{tikzpicture}

%% file: Draft_36_AER.bib
@article{Easley1998,
author = {Easley, David and O'Hara, Maureen and Srinivas, P. S.},

isbn = {00221082},

journal = {Journal of Finance},
pmid = {400231},
title = {{Option volume and stock prices: Evidence on where informed traders trade}},
volume = {53},
number = {2},
pages = {431--465},
year = {1998}
}

@misc{Hu2018,
author = {Hu, Jianfeng},
booktitle = {Review of Finance},
title = {{Option Listing and Information Asymmetry}},
year = {2018}
}

@article{Jiang2005,
author = {Jiang, George J. and Tian, Yisong S.},

isbn = {5206213373},

journal = {Review of Financial Studies},
number = {4},
pages = {1305--1342},
pmid = {16514790},
title = {{The model-free implied volatility and its information content}},
volume = {18},
year = {2005}
}

@unpublished{Muravyev2016,
archivePrefix = {arXiv},
arxivId = {1111.2036},
author = {Muravyev, Dmitriy and Pearson, Neil D. and Pollet, Joshua Matthew},
booktitle = {SSRN},

eprint = {1111.2036},
isbn = {2041-8205},
title = {{Is There a Risk Premium in the Stock Lending Market? Evidence from Equity Options}},
year = {2016}
}

@article{kyle1985continuous,
  title={Continuous auctions and insider trading},
  author={Kyle, Albert S},
  journal={Econometrica},
  volume={53},
  number={6},
  pages={1315--1335},
  year={1985},
  publisher={JSTOR}
}

@article{ni2008volatility,
  title={Volatility information trading in the option market},
  author={Ni, Sophie X and Pan, Jun and Poteshman, Allen M},
  journal={Journal of Finance},
  volume={63},
  number={3},
  pages={1059--1091},
  year={2008},
  publisher={Wiley Online Library}
}

@article{pan2006information,
  title={The information in option volume for future stock prices},
  author={Pan, Jun and Poteshman, Allen M},
  journal={Review of Financial Studies},
  volume={19},
  number={3},
  pages={871--908},
  year={2006},
  publisher={Oxford University Press}
}

@article{augustin2019informed,
  title={Informed options trading prior to takeover announcements: Insider trading?},
  author={Augustin, Patrick and Brenner, Menachem and Subrahmanyam, Marti G},
  journal={Management Science},
  volume={65},
  number={12},
  pages={5697--5720},
  year={2019},
  publisher={INFORMS}
}

@article{back1993asymmetric,
  title={Asymmetric information and options},
  author={Back, Kerry},
  journal={Review of Financial Studies},
  volume={6},
  number={3},
  pages={435--472},
  year={1993},
  publisher={Oxford University Press}
}

@article{biais1994insider,
  title={Insider and liquidity trading in stock and options markets},
  author={Biais, Bruno and Hillion, Pierre},
  journal={Review of Financial Studies},
  volume={7},
  number={4},
  pages={743--780},
  year={1994},
  publisher={Oxford University Press}
}

@article{foster1996strategic,
  title={Strategic trading when agents forecast the forecasts of others},
  author={Foster, F Douglas and Viswanathan, S},
  journal={Journal of Finance},
  volume={51},
  number={4},
  pages={1437--1478},
  year={1996},
  publisher={Wiley Online Library}
}

@article{back2000imperfect,
  title={Imperfect competition among informed traders},
  author={Back, Kerry and Cao, C Henry and Willard, Gregory A},
  journal={Journal of Finance},
  volume={55},
  number={5},
  pages={2117--2155},
  year={2000},
  publisher={Wiley Online Library}
}

@article{rochet1994insider,
  title={Insider trading without normality},
  author={Rochet, Jean-Charles and Vila, Jean-Luc},
  journal={Review of Economic Studies},
  volume={61},
  number={1},
  pages={131--152},
  year={1994},
  publisher={Wiley-Blackwell}
}

@article{goyal2009cross,
  title={Cross-section of option returns and volatility},
  author={Goyal, Amit and Saretto, Alessio},
  journal={Journal of Financial Economics},
  volume={94},
  number={2},
  pages={310--326},
  year={2009},
  publisher={Elsevier}
}

@article{bali2013does,
  title={Does risk-neutral skewness predict the cross section of equity option portfolio returns?},
  author={Bali, Turan G and Murray, Scott},
  journal={Journal of Financial and Quantitative Analysis},
  volume={48},
  number={4},
  pages={1145--1171},
  year={2013},
  publisher={JSTOR}
}

@article{cao2005informational,
  title={Informational content of option volume prior to takeovers},
  author={Cao, Charles and Chen, Zhiwu and Griffin, John M},
  journal={Journal of Business},
  volume={78},
  number={3},
  pages={1073--1109},
  year={2005},
  publisher={JSTOR}
}

@article{arrow1954existence,
  title={Existence of an equilibrium for a competitive economy},
  author={Arrow, Kenneth J and Debreu, Gerard},
  journal={Econometrica},
  volume={22},
  number={3},
  pages={265--290},
  year={1954},
  publisher={JSTOR}
}

@article{Roll2009,
  title={Options trading activity and firm valuation},
  author={Roll, Richard and Schwartz, Eduardo and Subrahmanyam, Avanidhar},
  journal={Journal of Financial Economics},
  volume={94},
  number={3},
  pages={345--360},
  year={2009},
  publisher={Elsevier}
}

@article{grossman1980impossibility,
  title={On the impossibility of informationally efficient markets},
  author={Grossman, Sanford J and Stiglitz, Joseph E},
  journal={American Economic Review},
  volume={70},
  number={3},
  pages={393--408},
  year={1980},
  publisher={JSTOR}
}

@article{collin2021informed,
  title={Informed Trading in the Stock Market and Option-Price Discovery},
  author={Collin-Dufresne, Pierre and Fos, Vyacheslav and Muravyev, Dmitry},
  journal={Journal of Financial and Quantitative Analysis},
  volume={56},
  number={6},
  pages={1945--1984},
  year={2021},
  publisher={Cambridge University Press}
}

@article{zhan2022option,
  title={Option return predictability},
  author={Zhan, Xintong and Han, Bing and Cao, Jie and Tong, Qing},
  journal={Review of Financial Studies},
  volume={35},
  number={3},
  pages={1394--1442},
  year={2022}
}

@article{chabakauri2022multi,
  title={Multi-asset noisy rational expectations equilibrium with contingent claims},
  author={Chabakauri, Georgy and Yuan, Kathy and Zachariadis, Konstantinos E},
  journal={Review of Economic Studies},
  volume={89},
  number={5},
  pages={2445--2490},
  year={2022},
  publisher={Oxford University Press}
}

@article{black1973pricing,
  title={The pricing of options and corporate liabilities},
  author={Black, Fischer and Scholes, Myron},
  journal={Journal of Political Economy},
  volume={81},
  number={3},
  pages={637--654},
  year={1973},
  publisher={University of Chicago Press}
}

@article{cao2013cross,
  title={Cross section of option returns and idiosyncratic stock volatility},
  author={Cao, Jie and Han, Bing},
  journal={Journal of Financial Economics},
  volume={108},
  number={1},
  pages={231--249},
  doi={10.1016/j.jfineco.2012.11.010},
  year={2013},
  publisher={Elsevier}
}

@article{christoffersen2018illiquidity,
  title={Illiquidity premia in the equity options market},
  author={Christoffersen, Peter and Goyenko, Ruslan and Jacobs, Kris and Karoui, Mehdi},
  journal={Review of Financial Studies},
  volume={31},
  number={3},
  pages={811--851},
  year={2018},
  publisher={Oxford University Press}
}

@article{bates2000post,
  title={Post-'87 crash fears in the {S}\&{P} 500 futures option market},
  author={Bates, David S},
  journal={Journal of Econometrics},
  volume={94},
  number={1-2},
  pages={181--238},
  year={2000},
  publisher={Elsevier}
}

@article{ait2007maximum,
  title={Maximum likelihood estimation of stochastic volatility models},
  author={A{\"\i}t-Sahalia, Yacine and Kimmel, Robert},
  journal={Journal of Financial Economics},
  volume={83},
  number={2},
  pages={413--452},
  year={2007},
  publisher={Elsevier}
}

@article{christoffersen2010volatility,
  title={Volatility dynamics for the {S}\&{P}500: Evidence from realized volatility, daily returns, and option prices},
  author={Christoffersen, Peter and Jacobs, Kris and Mimouni, Karim},
  journal={Review of Financial Studies},
  volume={23},
  number={8},
  pages={3141--3189},
  year={2010},
  publisher={Oxford University Press}
}

@article{duffie2000transform,
  title={Transform analysis and asset pricing for affine jump-diffusions},
  author={Duffie, Darrell and Pan, Jun and Singleton, Kenneth},
  journal={Econometrica},
  volume={68},
  number={6},
  pages={1343--1376},
  year={2000},
  publisher={Wiley Online Library}
}

@article{heston1993closed,
  title={A closed-form solution for options with stochastic volatility with applications to bond and currency options},
  author={Heston, Steven L},
  journal={Review of Financial Studies},
  volume={6},
  number={2},
  pages={327--343},
  year={1993},
  publisher={Oxford University Press}
}

@article{bates1996jumps,
  title={Jumps and stochastic volatility: Exchange rate processes implicit in {D}eutsche {M}ark options},
  author={Bates, David S},
  journal={Review of Financial Studies},
  volume={9},
  number={1},
  pages={69--107},
  year={1996},
  publisher={Oxford University Press}
}

@article{britten2000option,
  title={Option prices, implied price processes, and stochastic volatility},
  author={Britten-Jones, Mark and Neuberger, Anthony},
  journal={Journal of Finance},
  volume={55},
  number={2},
  pages={839--866},
  year={2000},
  publisher={Wiley Online Library}
}

@article{ait2021implied,
  title={Implied stochastic volatility models},
  author={A{\"\i}t-Sahalia, Yacine and Li, Chenxu and Li, Chen Xu},
  journal={Review of Financial Studies},
  volume={34},
  number={1},
  pages={394--450},
  year={2021},
  publisher={Oxford University Press}
}

@article{carr2012explicit,
  title={Explicit constructions of martingales calibrated to given implied volatility smiles},
  author={Carr, Peter and Cousot, Laurent},
  journal={SIAM Journal on Financial Mathematics},
  volume={3},
  number={1},
  pages={182--214},
  year={2012},
  publisher={SIAM}
}

@article{lin1995trade,
  title={Trade size and components of the bid-ask spread},
  author={Lin, Ji-Chai and Sanger, Gary C and Booth, G Geoffrey},
  journal={Review of Financial Studies},
  volume={8},
  number={4},
  pages={1153--1183},
  year={1995},
  publisher={Oxford University Press}
}

@article{huang1997components,
  title={The components of the bid-ask spread: A general approach},
  author={Huang, Roger D and Stoll, Hans R},
  journal={Review of Financial Studies},
  volume={10},
  number={4},
  pages={995--1034},
  year={1997},
  publisher={Oxford University Press}
}

@article{glosten1988estimating,
  title={Estimating the components of the bid/ask spread},
  author={Glosten, Lawrence R and Harris, Lawrence E},
  journal={Journal of Financial Economics},
  volume={21},
  number={1},
  pages={123--142},
  year={1988},
  publisher={Elsevier}
}

@article{goyenko2009liquidity,
  title={Do liquidity measures measure liquidity?},
  author={Goyenko, Ruslan Y and Holden, Craig W and Trzcinka, Charles A},
  journal={Journal of Financial Economics},
  volume={92},
  number={2},
  pages={153--181},
  year={2009},
  publisher={Elsevier}
}

@article{makarov2020trading,
  title={Trading and arbitrage in cryptocurrency markets},
  author={Makarov, Igor and Schoar, Antoinette},
  journal={Journal of Financial Economics},
  volume={135},
  number={2},
  pages={293--319},
  year={2020},
  publisher={Elsevier}
}

@article{hendershott2011does,
  title={Does algorithmic trading improve liquidity?},
  author={Hendershott, Terrence and Jones, Charles M and Menkveld, Albert J},
  journal={Journal of Finance},
  volume={66},
  number={1},
  pages={1--33},
  year={2011},
  publisher={Wiley Online Library}
}

@article{berestycki2002asymptotics,
  title={Asymptotics and calibration of local volatility models},
  author={Berestycki, Henri and Busca, J and Florent, I},
  journal={Quantitative Finance},
  volume={2},
  number={1},
  pages={61},
  year={2002},
  publisher={IOP Publishing}
}

@article{driessen2009price,
  title={The price of correlation risk: Evidence from equity options},
  author={Driessen, Joost and Maenhout, Pascal J and Vilkov, Grigory},
  journal={Journal of Finance},
  volume={64},
  number={3},
  pages={1377--1406},
  year={2009},
  publisher={Wiley Online Library}
}

@article{coval2001expected,
  title={Expected option returns},
  author={Coval, Joshua D and Shumway, Tyler},
  journal={Journal of Finance},
  volume={56},
  number={3},
  pages={983--1009},
  year={2001},
  publisher={Wiley Online Library}
}

@article{admati1985noisy,
  title={A noisy rational expectations equilibrium for multi-asset securities markets},
  author={Admati, Anat R},
  journal={Econometrica},
  pages={629--657},
  year={1985},
  publisher={JSTOR}
}

@article{malamud2015noisy,
  title={Noisy {A}rrow-{D}ebreu equilibria},
  author={Malamud, Semyon},
  journal={Working Paper},
  year={2015}
}

@article{caballe1994imperfect,
  title={Imperfect competition in a multi-security market with risk neutrality},
  author={Caballe, Jordi and Krishnan, Murugappa},
  journal={Econometrica},
  pages={695--704},
  year={1994},
  publisher={JSTOR}
}

@article{dupire1994pricing,
  title={Pricing with a smile},
  author={Dupire, Bruno},
  journal={Risk},
  volume={7},
  number={1},
  pages={18--20},
  year={1994}
}

@book{hull2003options,
  title={Options, Futures and Other Derivatives},
  author={Hull, John C},
  year={2015},
  publisher={Pearson}
}

@book{vine2011options,
  title={Options: Trading Strategy and Risk Management},
  author={Vine, Simon},
  year={2011},
  publisher={John Wiley \& Sons}
}

@article{brennan1996information,
  title={Information, trade, and derivative securities},
  author={Brennan, Michael J and Cao, H Henry},
  journal={Review of Financial Studies},
  volume={9},
  number={1},
  pages={163--208},
  year={1996},
  publisher={Oxford University Press}
}

@book{back2010asset,
  title={Asset Pricing and Portfolio Choice Theory},
  author={Back, Kerry},
  year={2010},
  publisher={Oxford University Press}
}

@book{shiryaev1999essentials,
  title={Essentials of Stochastic Finance: Facts, Models, Theory},
  author={Shiryaev, Albert N},
  volume={3},
  year={1999},
  publisher={World scientific}
}

@article{merton1976option,
  title={Option pricing when underlying stock returns are discontinuous},
  author={Merton, Robert C},
  journal={Journal of Financial Economics},
  volume={3},
  number={1-2},
  pages={125--144},
  year={1976},
  publisher={Elsevier}
}

@article{bates1991crash,
  title={The crash of '87: was it expected? {T}he evidence from options markets},
  author={Bates, David S},
  journal={Journal of Finance},
  volume={46},
  number={3},
  pages={1009--1044},
  year={1991},
  publisher={Wiley Online Library}
}

@article{ross1976options,
  title={Options and efficiency},
  author={Ross, Stephen A},
  journal={Quarterly Journal of Economics},
  volume={90},
  number={1},
  pages={75--89},
  year={1976},
  publisher={MIT Press}
}

@article{Hasbrouck1995,
  author  = {Hasbrouck, Joel},
  title   = {One Security, Many Markets: Determining the Contributions to Price Discovery},
  journal = {Journal of Finance},
  year    = {1995},
  volume  = {50},
  number  = {4},
  pages   = {1175--1199},
  doi     = {10.1111/j.1540-6261.1995.tb04054.x}
}

@article{BollenWhaley2004,
  author  = {Bollen, Nicolas P. B. and Whaley, Robert E.},
  title   = {Does Net Buying Pressure Affect the Shape of Implied Volatility Functions?},
  journal = {Journal of Finance},
  year    = {2004},
  volume  = {59},
  number  = {2},
  pages   = {711--753},
  doi     = {10.1111/j.1540-6261.2004.00647.x}
}

@article{KaeckVanKervelSeeger2022,
  author  = {Kaeck, Andreas and van Kervel, Vincent and Seeger, Norman J.},
  title   = {Price impact versus bid--ask spreads in the index option market},
  journal = {Journal of Financial Markets},
  year    = {2022},
  volume  = {59},
  pages   = {100675},
  doi     = {10.1016/j.finmar.2021.100675}
}

@article{BakshiKapadiaMadan2003,
  author  = {Bakshi, Gurdip and Kapadia, Nikunj and Madan, Dilip},
  title   = {Stock Return Characteristics, Skew Laws, and the Differential Pricing of Individual Equity Options},
  journal = {Review of Financial Studies},
  year    = {2003},
  volume  = {16},
  number  = {1},
  pages   = {101--143},
  doi     = {10.1093/rfs/16.1.101}
}

@article{CarrWu2009,
  author  = {Carr, Peter and Wu, Liuren},
  title   = {Variance Risk Premiums},
  journal = {Review of Financial Studies},
  year    = {2009},
  volume  = {22},
  number  = {3},
  pages   = {1311--1341},
  doi     = {10.1093/rfs/hhn038}
}

@article{BollerslevTauchenZhou2009,
  author  = {Bollerslev, Tim and Tauchen, George and Zhou, Hao},
  title   = {Expected Stock Returns and Variance Risk Premia},
  journal = {Review of Financial Studies},
  year    = {2009},
  volume  = {22},
  number  = {11},
  pages   = {4463--4492},
  doi     = {10.1093/rfs/hhp008}
}

@article{ConradDittmarGhysels2013,
  author  = {Conrad, Jennifer and Dittmar, Robert F. and Ghysels, Eric},
  title   = {Ex Ante Skewness and Expected Stock Returns},
  journal = {Journal of Finance},
  year    = {2013},
  volume  = {68},
  number  = {1},
  pages   = {85--124},
  doi     = {10.1111/j.1540-6261.2012.01795.x}
}

@article{DennisMayhew2002,
  title   = {Risk-Neutral Skewness: Evidence from Stock Options},
  author  = {Dennis, Patrick and Mayhew, Stewart},
  journal = {Journal of Financial and Quantitative Analysis},
  year    = {2002},
  volume  = {37},
  number  = {3},
  pages   = {471--493},
  doi     = {10.2307/3594989}
}

@article{BakshiMadan2000,
  title   = {Spanning and Derivative-Security Valuation},
  author  = {Bakshi, Gurdip and Madan, Dilip},
  journal = {Journal of Financial Economics},
  year    = {2000},
  volume  = {55},
  number  = {2},
  pages   = {205--238},
  doi     = {10.1016/S0304-405X(99)00050-1}
}

@article{XingZhangZhao2010,
  title   = {What Does the Individual Option Volatility Smirk Tell Us About Future Equity Returns?},
  author  = {Xing, Yuhang and Zhang, Xiaoyan and Zhao, Rui},
  journal = {Journal of Financial and Quantitative Analysis},
  year    = {2010},
  volume  = {45},
  number  = {3},
  pages   = {641--662},
  doi     = {10.1017/S0022109010000220}
}

@article{CremersWeinbaum2010,
  title   = {Deviations from Put-Call Parity and Stock Return Predictability},
  author  = {Cremers, Martijn and Weinbaum, David},
  journal = {Journal of Financial and Quantitative Analysis},
  year    = {2010},
  volume  = {45},
  number  = {2},
  pages   = {335--367},
  doi     = {10.1017/S002210901000013X}
}

@article{BollerslevTodorov2011,
  title   = {Tails, Fears, and Risk Premia},
  author  = {Bollerslev, Tim and Todorov, Viktor},
  journal = {The Journal of Finance},
  year    = {2011},
  volume  = {66},
  number  = {6},
  pages   = {2165--2211},
  doi     = {10.1111/j.1540-6261.2011.01695.x}
}

@article{BollerslevTodorovXu2015,
  title   = {Tail Risk Premia and Return Predictability},
  author  = {Bollerslev, Tim and Todorov, Viktor and Xu, Lai},
  journal = {Journal of Financial Economics},
  year    = {2015},
  volume  = {118},
  number  = {1},
  pages   = {113--134},
  doi     = {10.1016/j.jfineco.2015.02.010}
}

@book{natenberg2014ovp,
  title        = {Option Volatility \& Pricing: Advanced Trading Strategies and Techniques},
  author       = {Natenberg, Sheldon},
  edition      = {2},
  year         = {2014},
  publisher    = {McGraw-Hill Education},
}

@book{mcmillan2012options,
  title        = {Options as a Strategic Investment},
  author       = {McMillan, Lawrence G.},
  edition      = {5},
  year         = {2012},
  publisher    = {New York Institute of Finance},
}

@book{overby2007optionsplaybook,
  title        = {The Options Playbook},
  author       = {Overby, Brian},
  year         = {2007},
  publisher    = {TradeKing},
}

@article{KozhanNeubergerSchneider2013,
  title        = {The Skew Risk Premium in the Equity Index Market},
  author       = {Kozhan, Roman and Neuberger, Anthony and Schneider, Paul},
  journal      = {The Review of Financial Studies},
  volume       = {26},
  number       = {9},
  pages        = {2174--2203},
  year         = {2013},
  doi          = {10.1093/rfs/hht039},
}

@article{gonzalo1995estimation,
  title   = {Estimation of Common Long-Memory Components in Cointegrated Systems},
  author  = {Gonzalo, Jes{\'u}s and Granger, Clive W. J.},
  journal = {Journal of Business \& Economic Statistics},
  volume  = {13},
  number  = {1},
  pages   = {27--35},
  year    = {1995}
}

@article{chakravarty2004informed,
  title   = {Informed Trading in Stock and Option Markets},
  author  = {Chakravarty, Sugato and Gulen, Huseyin and Mayhew, Stewart},
  journal = {Journal of Finance},
  volume  = {59},
  number  = {3},
  pages   = {1235--1257},
  year    = {2004},
  doi     = {10.1111/j.1540-6261.2004.00661.x}
}

@article{holden1992longlived,
  title   = {Long-Lived Private Information and Imperfect Competition},
  author  = {Holden, Craig W. and Subrahmanyam, Avanidhar},
  journal = {Journal of Finance},
  volume  = {47},
  number  = {1},
  pages   = {247--270},
  year    = {1992},
  doi     = {10.1111/j.1540-6261.1992.tb03985.x}
}

@article{GarbadeSilber1983,
  author  = {Garbade, Kenneth D. and Silber, William L.},
  title   = {Price Movements and Price Discovery in Futures and Cash Markets},
  journal = {Review of Economics and Statistics},
  year    = {1983},
  volume  = {65},
  number  = {2},
  pages   = {289--297},
  doi     = {10.2307/1924495}
}

@article{Chan1992,
  author  = {Chan, Kalok},
  title   = {A Further Analysis of the Lead--Lag Relationship Between the Cash Market and Stock Index Futures Market},
  journal = {Review of Financial Studies},
  year    = {1992},
  volume  = {5},
  number  = {1},
  pages   = {123--152},
  doi     = {10.1093/rfs/5.1.123}
}

@article{BlancoBrennanMarsh2005,
  author  = {Blanco, Roberto and Brennan, Simon and Marsh, Ian W.},
  title   = {An Empirical Analysis of the Dynamic Relation between Investment-Grade Bonds and Credit Default Swaps},
  journal = {Journal of Finance},
  year    = {2005},
  volume  = {60},
  number  = {5},
  pages   = {2255--2281},
  doi     = {10.1111/j.1540-6261.2005.00798.x}
}

@article{LongstaffMithalNeis2005,
  author  = {Longstaff, Francis A. and Mithal, Sanjay and Neis, Eric},
  title   = {Corporate Yield Spreads: Default Risk or Liquidity? New Evidence from the Credit Default Swap Market},
  journal = {Journal of Finance},
  year    = {2005},
  volume  = {60},
  number  = {5},
  pages   = {2213--2253},
  doi     = {10.1111/j.1540-6261.2005.00797.x}
}

@article{AcharyaJohnson2007,
  author  = {Acharya, Viral V. and Johnson, Timothy C.},
  title   = {Insider Trading in Credit Derivatives},
  journal = {Journal of Financial Economics},
  year    = {2007},
  volume  = {84},
  number  = {1},
  pages   = {110--141},
  doi     = {10.1016/j.jfineco.2006.05.003}
}

@article{kellertsengmath,
  title   = {An Infinite-Dimensional Insider Trading Game},
  author  = {Christian Keller and Michael C. Tseng},
  journal = {arXiv preprint},
  year    = {2026}
}
